\documentclass[twocolumn,aps,pra,superscriptaddress,amsmath,amssym,floatfix,nofootinbib,longbibliography]{revtex4-1}

\usepackage{graphicx}
\usepackage{bm}
\usepackage{color}
\usepackage[usenames,dvipsnames]{xcolor}
\usepackage{tikz}
\usetikzlibrary{%
    decorations.pathreplacing,%
    decorations.pathmorphing,%
    arrows
}
\usepackage{amssymb}
\usepackage{enumerate}
\usepackage{comment}
\usepackage[normalem]{ulem}

\def\be{\begin{equation}}
\def\ee{\end{equation}}

\newcommand{\ket}[1]{\left\vert #1 \right\rangle}
\newcommand{\bra}[1]{\left\langle #1 \right\vert}

\newcommand{\abs}[1]{\left| #1 \right|}
\newcommand{\blue}[1]{{\color{blue}#1}}

\usepackage[colorlinks=true,linkcolor=MidnightBlue,citecolor=blue,filecolor=green,urlcolor=PineGreen]{hyperref}
\definecolor{ceruleanblue}{rgb}{0.16, 0.32, 0.75}
\definecolor{darkspringgreen}{rgb}{0.09, 0.45, 0.27}
\definecolor{deeplilac}{rgb}{0.6, 0.33, 0.73}
\definecolor{goldenyellow}{rgb}{1.0, 0.87, 0.0}
\definecolor{cobalt}{rgb}{0.0, 0.28, 0.67}
\definecolor{selectiveyellow}{rgb}{1.0, 0.73, 0.0}
\definecolor{turquoiseblue}{rgb}{0.0, 1.0, 0.94}
\definecolor{vividviolet}{rgb}{0.62, 0.0, 1.0}
\definecolor{neongreen}{rgb}{0.22, 0.88, 0.08}
\definecolor{codegreen}{rgb}{0,0.6,0}
\definecolor{codegray}{rgb}{0.5,0.5,0.5}
\definecolor{codepurple}{rgb}{0.58,0,0.82}
\definecolor{backcolour}{rgb}{0.95,0.95,0.95}
\definecolor{huntergreen}{rgb}{0.21, 0.37, 0.23}
\definecolor{lavenderpurple}{rgb}{0.59, 0.48, 0.71}
\definecolor{coquelicot}{rgb}{1.0, 0.22, 0.0}
\definecolor{crimsonglory}{rgb}{0.75, 0.0, 0.2}
\definecolor{deeppink}{rgb}{1.0, 0.08, 0.58}
\definecolor{electricviolet}{rgb}{0.56, 0.0, 1.0}
\definecolor{electricgreen}{rgb}{0.0, 1.0, 0.0}
\definecolor{mint}{rgb}{0.24, 0.71, 0.54}
\definecolor{dodgerblue}{rgb}{0.12, 0.56, 1.0}
\definecolor{lincolngreen}{rgb}{0.11, 0.35, 0.02}
\definecolor{persianblue}{rgb}{0.11, 0.22, 0.73}
\definecolor{patriarch}{rgb}{0.5, 0.0, 0.5}
\definecolor{amaranth}{rgb}{0.9, 0.17, 0.31}

\begin{document}
\title{%Remote Two--Qubit Entanglement %Genesis 
%by Joint Homodyne Detection of Fluorescence 
%Optimizing entanglement yield: Feedback control based on measurements of two qubits' spontaneous emission
Entanglement--Preserving Limit Cycles \\ from Sequential Quantum Measurements and Feedback
}
\author{Philippe Lewalle} 
\email{plewalle@ur.rochester.edu}
\affiliation{Department of Physics and Astronomy, University of Rochester, Rochester, NY 14627, USA}
\affiliation{Center for Coherence and Quantum Optics, University of Rochester, Rochester, NY 14627, USA}
\author{Cyril Elouard} 
\affiliation{Department of Physics and Astronomy, University of Rochester, Rochester, NY 14627, USA}
\affiliation{Center for Coherence and Quantum Optics, University of Rochester, Rochester, NY 14627, USA}
\author{Andrew N. Jordan} 
\affiliation{Department of Physics and Astronomy, University of Rochester, Rochester, NY 14627, USA}
\affiliation{Center for Coherence and Quantum Optics, University of Rochester, Rochester, NY 14627, USA}
\affiliation{Institute for Quantum Studies, Chapman University, Orange, CA 92866, USA}
\date{\today}

\begin{abstract}
Entanglement generation and preservation is a key task in quantum information processing, and a variety of protocols exist to entangle remote qubits via measurement of their spontaneous emission. 
We here propose feedback methods, based on monitoring the fluorescence of two qubits and using only local $\pi$--pulses for control, to increase the yield and/or lifetime of entangled two--qubit states. 
Specifically, we describe a protocol based on photodetection of spontaneous emission (i.e.~using quantum jump trajectories) which allows for entanglement preservation via measurement undoing, creating a limit cycle around a Bell states. 
We then demonstrate that a similar modification can be made to a recent feedback scheme based on homodyne measurement (i.e.~using diffusive quantum trajectories), [L.~S.~Martin and K.~B.~Whaley, arXiv:1912.00067] in order to increase the lifetime of the entanglement it creates. 
Our schemes are most effective for high measurement efficiencies, and the impact of less-than-ideal measurement efficiency is quantified.  
The method we describe here combines proven techniques in a novel way, complementing existing protocols, and offering a pathway towards generating and protecting entangled states so that they may be used in various applications on demand. 

\end{abstract}

\maketitle

\section{Introduction}

Entanglement is one of the key features of quantum systems which allows for potential information--processing advantages, over those possible in purely classical systems.
An unmonitored spontaneous emission process leads to decoherence and loss of entanglement \cite{YuEberly_2004}.
On the other hand, measurement of such decay channels via photodetection has been proven to be an effective means of generating entanglement \cite{Cabrillo1998, Bose1999, Plenio1999, BarrettKok, DuanKimble2003, Browne2003, Simon2003, Lim2005, Moehring2007, Maunz2009, Hofmann2012, Hanson2013Heralded, HansonLoopholeFree, Ohm2017}.
Such processes can be realized with more general time--continuous measurements \cite{BarrettKok, Mintert2005, Carvalho2007, Viviescas2010, Mascarenhas2010, Mascarenhas2011, Carvalho2011, Santos2012, FlorTeach2019, LongFlor2019, Leigh2019}, in which the entanglement generation is tracked by the same process that creates it. 

\par Advances in continuous quantum measurement (stochastic quantum trajectories) in general \cite{BookCarmichael, BookPercival, BookGardiner, BookWiseman, BookBarchielli, BookJacobs, Wiseman1996Review, Brun2001Teach, Jacobs2006}, 
% , Wiseman1996Review, Gisin1984, Barchielli1986, Caves1987, Diosi1988, Gisin1992-1, Gisin1992-2, Wiseman1993prl, Wiseman1993pra1, Wiseman1993pra2, Brun2001Teach, Dziarmaga2004, Jacobs2006, Gough2004, Gambetta2008, Santos2011, Gough2012, Murch2013, Jordan2013rev, Chantasri2013, Weber2014,  Chantasri2015, Rouchon2015, Korotkov2016, Leigh2016, Gross_2018, Gough2018, Gough2018_JMP, Gough2019-1, Gough2019-2, Crowder2019
have been consistently connected to the development of Hamiltonian feedback protocols, conditioned on the real--time measurement record, which aim to implement useful quantum control tasks \cite{Zhang2017, Wiseman1994, Ahn2002, *Ahn2003, *Ahn2003-2, *Ahn2003-3,  Sarovar2004, Mancini2007, Carvalho2007-2, Carvalho2008, Hill2008, Serafini2010, Vijay2012, Balouchi_2014, Rheda2014, Szigeti2014,  Hsu2016,  Magazzu2018, Zhang2018, Gourgy2018,  Minev2018, Martin2019, Leigh2019, Cardona2019, Mohseninia2019}. %Doherty1999, Doherty2000, Jacobs2003, vHandel2005, vHandel2005b, Mirrahimi2005, Combes2006, Wiseman_2006, Combes2008, Jacobs2008, Shabani2008, Mabuchi_2009, Combes2010, Combes2010b, Schirmer2010, Teo_2014, Taylor2017, Jiang2019,
This pattern can be seen in the work most directly related to ours: study of the quantum trajectories from monitoring a single qubit's spontaneous emission has led from theory \cite{Bolund2014, Jordan2015flor, FlorTeach2019} to experiments without \cite{Campagne-Ibarcq2016, naghiloo2015fluores, Mahdi2016, Tan2017, Ficheux2018}, and then with  \cite{PCI-2016-2, Mahdi2017Qtherm}, feedback. 
Theory without \cite{Viviescas2010, Santos2012, LongFlor2019}, and with \cite{BarrettKok, Mascarenhas2010, Leigh2019}, feedback has been developed in the two qubit case that the present work adds to. 

\par Our proposal here involves supplementing existing measurement and feedback schemes \cite{LongFlor2019, Leigh2019}, based on monitoring two qubits via their natural decay channel, with fast $\pi$--pulses. We show that this allows us to trap the two--qubit state in limit cycles close to a Bell state. 
%In this way, we are able to utilize the entanglement creation capability of continuous fluorescence measurements, while selectively eliminating the factors in those dynamics which can also erase that entanglement.
It is worth noting that we consider ``continuous'' measurement that relies on finite time-steps $\Delta t$ (i.e.~finite detector integration time leading the measurement record at each step), such that fast operations can be interjected so as to effectively take place ``between'' measurements.
While this is a reasonable regime to consider on real devices, it marks a mathematical departure from approaches to continuous measurements and feedback that are defined strictly in the time--continuum limit (where $\Delta t$ becomes an infinitesimal $dt$).
We also point out that most existing schemes which address the task of interest rely on additional resources, such as ancillary qubits or additional transitions for storing quantum information. 
While there are potential advantages to such approaches, ours requires \emph{only} the two qubits and feedback based on local operations and classical communication (LOCC).
The use of LOCC for feedback implies that the measurements are the only non--local element in our scheme, and must therefore be entirely responsible for entanglement generation; operations local to each qubit cannot increase the concurrence of the two--qubit state at all \cite{Plenio07,Horodecki09}. 
The role of the feedback is to allow measurements to better generate entanglement, or prevent subsequent measurements from decreasing the entanglement; our scheme leads to significant gains in entanglement yield and lifetime. 

\par The functioning of our control scheme brings to mind some other topics in the quantum measurement and control literature. 
First, the use of fast $\pi$--pulses to effectively reverse decoherence processes has its roots in spin--echo techniques \cite{HahnSpinEcho}; more recently this has been generalized into ``bang--bang'' (BB) type control schemes (which may themselves be viewed as a subset of dynamical decoupling protocols) \cite{ViolaBangBang, ViolaDecoupling, Viola1999, Viola2000, Byrd2002, Viola2003, Byrd2003, Viola2004, Byrd2004, Facchi2004, Facchi2005, Khodjasteh2005, Morton2006, Pryadko2009, Damodarakurup2009, Wang2012, Xu2012, Bhole2016}. 
While there has been work which combines dynamical decoupling or BB control with other quantum error correction methods \cite{Byrd2002, Byrd2003, Byrd2004}, or with measurement via the quantum Zeno effect \cite{Facchi2004, Facchi2005}, few past works interject fast BB--like controls inbetween measurement or other types of controls \cite{Ticozzi2006, Gong2013, Sun2010}. 
Second, we will see that the way we use our BB--like feedback, especially in conjunction with photodetection, is effectively equivalent to a measurement reversal procedure \cite{Jordan2006Reversal, JordanCPReversal, Katz2008, Kim2009, Sun2010,  Korotkov2010, Kim2012, Korotkov2012}.

\par We will proceed as follows: We first consider jump trajectories from ideal photodetection measurements in Sec.~\ref{sec-PDfeed-basic}. We demonstrate a simple feedback procedure based on fast $\pi$--pulses, which allows us to preserve virtually all concurrence generated by our measurements for arbitrarily long times.
Next we develop the corresponding procedure in the homodyne case \cite{Viviescas2010, LongFlor2019}, building on the recent scheme by Martin and Whaley \cite{Leigh2019} (which is, in turn, connected to our recent works \cite{FlorTeach2019, LongFlor2019}). 
The existing scheme implements local unitary feedback operations, and allows for deterministic generation of a Bell state based on ideal operation in the time--continuum limit. 
We exit the time--continuum assumption, and add $\pi$--pulse based BB--like control atop the local feedback rotations derived in Ref.~\cite{Leigh2019}.
This is shown to again lead to a stable limit cycle about a Bell state, which may preserve the entanglement generated by the Hamiltonian control indefinitely.  
In Sec.~\ref{sec-etaimpact} we re-consider each of the above schemes, assuming that we have inefficient measurements (but still an otherwise ideal setup). We perform a numerical analysis to quantify how the performance of our schemes degrade when state purity is gradually lost due to accumulated inefficient measurements.
Conclusions and outlook are presented in Sec.~\ref{sec-conclusions}.

\begin{figure}
    \centering
    \begin{tikzpicture}
    [wave3/.style={decorate,decoration={snake,post length=0.2cm,amplitude=0.15cm,segment length=0.2cm}}]
    % background box
    \filldraw[fill = cobalt!60!white, draw = Plum!10!white, rounded corners = 0.5cm] (0,0) rectangle (8,4);
    % qubits and cavities
    \filldraw[fill = black!50, color = black!50] (3.18,1.92) -- (2.85,1.5) -- (3.18,1.08) -- cycle;
    \filldraw[fill = black!50, color = black!50] (3.18,3.42) -- (2.85,3) -- (3.18,2.58) -- cycle;
    \filldraw[fill = black!50, color = black!50] (4.2,3.6) -- (3.6,3.6) -- (4.2,3.0) -- cycle;
    \filldraw[fill = black!50, color = black!50] (4.2,0.9) -- (3.6,0.9) -- (4.2,1.5) -- cycle;
    \filldraw[even odd rule,inner color=selectiveyellow ,outer color=black!50,color = black!50] (3.6,1.5) circle (0.6);
    \filldraw[even odd rule,inner color=selectiveyellow ,outer color=black!50,color = black!50] (3.6,3) circle (0.6);
    \draw[selectiveyellow,wave3,line width = 0.05cm,->] (3.6,3) -- (4.2,3.6);
    \draw[selectiveyellow,wave3,line width = 0.05cm,->] (3.6,1.5) -- (4.2,0.9);
    \draw[white,line width = 0.05cm] (3.35,1.65) -- (3.85,1.65);
    \draw[white,line width = 0.05cm] (3.35,1.35) -- (3.85,1.35);
    \draw[white,line width = 0.05cm] (3.35,3.15) -- (3.85,3.15);
    \draw[white,line width = 0.05cm] (3.35,2.85) -- (3.85,2.85);
    % inefficiency splitters
    %\draw[draw = white, left color = vividviolet!70!turquoiseblue, right color = vividviolet!50!white] (4.65,3.1) -- (4.9,3.35) -- (5.15,3.1) -- (4.9,2.85) -- cycle;
    %\draw[white] (4.65,3.1) -- (5.15,3.1);
    %\draw[draw = white, left color = vividviolet!70!turquoiseblue, right color = vividviolet!50!white] (4.65,1.4) -- (4.9,1.65) -- (5.15,1.4) -- (4.9,1.15) -- cycle;
    %\draw[white] (4.65,1.4) -- (5.15,1.4);
    \draw[draw = white, left color = vividviolet!70!turquoiseblue, right color = vividviolet!50!white] (6,2.75) -- (6.25,3) -- (6.5,2.75) -- (6.25,2.5) -- cycle;
    \draw[white] (6.25,2.5) -- (6.25,3);
    \draw[draw = white, left color = vividviolet!70!turquoiseblue, right color = vividviolet!50!white] (6,1.75) -- (6.25,2) -- (6.5,1.75) -- (6.25,1.5) -- cycle;
    \draw[white] (6.25,1.5) -- (6.25,2);
    % signal path and main splitter
    \draw[selectiveyellow,line width = 0.05cm,->] (4.2,3.6) -- (4.3,3.7) -- (5.5,2.5);
    \draw[selectiveyellow,line width = 0.05cm,->] (4.2,0.9) -- (4.3,0.8) -- (5.5,2.0);
    \draw[turquoiseblue!50,line width = 0.05cm] (4.1,3.75) -- (4.5,3.75);
    \draw[turquoiseblue!50,line width = 0.05cm] (4.1,0.75) -- (4.5,0.75);
    \draw[left color = turquoiseblue, right color = Cerulean, draw = white] (5.25,2.25)-- (5.75,2.75) -- (6.25,2.25) -- (5.75,1.75) -- cycle;
    \draw[white] (5.25,2.25) -- (6.25,2.25);
    \draw[selectiveyellow, dotted, line width = 0.05cm] (5.5,2.5) -- (6,2);
    \draw[selectiveyellow, dotted, line width = 0.05cm] (5.5,2) -- (6,2.5);
    \draw[selectiveyellow,line width = 0.05cm,->] (6,2.5) -- (6.5,3);
    \draw[selectiveyellow,line width = 0.05cm,->] (6,2) -- (6.5,1.5);
    % detectors and cables
    \draw[fill = black, rounded corners = 0.05cm] (6.7,2.8) arc[radius = 0.27, start angle = -45, end angle = 135] --  cycle;
    \draw[fill = black, rounded corners = 0.05cm] (6.7,1.7) arc[radius = 0.27, start angle = 45, end angle = -135] --  cycle;
    \draw[black,line width = 0.05cm,rounded corners = 0.3 cm,->] (6.6,1.4) -- (7,1) -- (6,0.5) -- (1.5,0.5) -- (1,1) -- (1.0,1.5);
    \draw[black,line width = 0.05cm,rounded corners = 0.3 cm,->] (6.6,3.1) -- (7,3.5) -- (7.3,3) -- (7.3,1) -- (6.2,0.3) -- (1.3,0.3) -- (0.7,0.8) -- (0.7,1.5);
    % computer and feedback
    \draw[fill = black!20!white, rounded corners = 0.05cm] (0.5,1.6) -- (1.2,1.6) -- (1.1,1.8) -- (0.6,1.8) -- cycle;
    \draw[fill = black!20!white, rounded corners = 0.05cm] (0.5,1.9) -- (1.2,1.9) -- (1.2,2.4) -- (0.5,2.4) -- cycle;
    \draw[fill = white, draw = white, rounded corners = 0.05cm] (0.6,2) -- (1.1,2) -- (1.1,2.3) -- (0.6,2.3) -- cycle;
    \draw[line width = 0.05cm, rounded corners = 0.1cm] (1.3,2.2) -- (1.5,2.2) -- (1.5,3) -- (2,3);
    \draw[line width = 0.05cm, rounded corners = 0.1cm] (1.3,2.0) -- (1.5,2.0) -- (1.5,1.5) -- (2,1.5);
    \draw[fill = black, rounded corners = 0.05cm] (1.7,1.45) rectangle (2,1.55);
    \draw[fill = black, rounded corners = 0.05cm] (1.7,2.95) rectangle (2,3.05);
    \draw[red,wave3,line width = 0.05cm,->] (2,1.5) -- (2.8,1.5);
    \draw[red,wave3,line width = 0.05cm,->] (2,3) -- (2.8,3);
    % inefficiency extra paths
    %\draw[black,opacity = 0.3,line width = 0.05cm] (4.2,2.4) -- (4.9,3.1);
    %\draw[black,opacity = 0.3,line width = 0.05cm] (4.2,2.1) -- (4.9,1.4);
    %\draw[selectiveyellow, opacity = 0.5, line width = 0.05cm,->] (4.9,3.1) -- (5.2,3.4);
    %\draw[selectiveyellow, opacity = 0.5, line width = 0.05cm,->] (4.9,1.4) -- (5.2,1.1);
    \draw[black,opacity = 0.3,line width = 0.05cm] (6.6,2.4) -- (6.25,2.75);
    \draw[black,opacity = 0.3,line width = 0.05cm] (6.6,2.1) -- (6.25,1.75);
    \draw[selectiveyellow, opacity = 0.5, line width = 0.05cm,->] (6.25,2.75) -- (5.95,3.05);
    \draw[selectiveyellow, opacity = 0.5, line width = 0.05cm,->] (6.25,1.75) -- (5.95,1.45);
    \end{tikzpicture} \\ \begin{picture}(0,0)
    \put(-13,82){\color{white} A}
    \put(-13,61){\color{white} B}
    \put(-50,81){\color{red} $\hat{F}_A$}
    \put(-50,61){\color{red} $\hat{F}_B$}
    \put(33,82){\tiny \color{selectiveyellow} $1$}
    \put(33,64){\tiny \color{selectiveyellow} $2$}
    \put(73.5,96.5){\tiny \color{selectiveyellow} $4$}
    \put(73.5,49){\tiny \color{selectiveyellow} $3$}
    \put(18,73){\tiny \color{white} 50/50}
    %\put(0,73){\tiny vac.}
    %\put(21,87){\color{vividviolet} \footnotesize $\eta_1$}
    %\put(21,61){\color{vividviolet} \footnotesize $\eta_2$}
    %\put(34,110){\color{selectiveyellow} $a_{1,\ell}^\dag$}
    %\put(34,34){\color{selectiveyellow} $a_{2,\ell}^\dag$}
    \put(74,85){\color{vividviolet} \footnotesize $\eta_4$}
    \put(74,63){\color{vividviolet} \footnotesize $\eta_3$}
    \put(45,103){\color{selectiveyellow} $a_{4,\ell}^\dag$}
    \put(45,45){\color{selectiveyellow} $a_{3,\ell}^\dag$}
    \put(72,73){\tiny vac.}
    \put(-108,110){(a)}
    \end{picture} \vspace{-10pt} \\
    %%%%%%%%%%%%%%%%%%%%%%%%%%%%%%%%%%%%%%%%%%
    %%%%%%%%%%%%%%%%%%%%%%%%%%%%%%%%%%%%%%%%%%
    \begin{tikzpicture}
    [wave3/.style={decorate,decoration={snake,post length=0.2cm,amplitude=0.15cm,segment length=0.2cm}}]
    % background box
    \filldraw[fill = cobalt!60!white, draw = Plum!10!white, rounded corners = 0.5cm] (0,0.25) rectangle (8,4.25);
    % qubits and cavities
    \filldraw[fill = black!50, color = black!50] (3.18,1.92) -- (2.85,1.5) -- (3.18,1.08) -- cycle;
    \filldraw[fill = black!50, color = black!50] (3.18,3.42) -- (2.85,3) -- (3.18,2.58) -- cycle;
    \filldraw[fill = black!50, color = black!50] (4.2,3.6) -- (3.6,3.6) -- (4.2,3.0) -- cycle;
    \filldraw[fill = black!50, color = black!50] (4.2,0.9) -- (3.6,0.9) -- (4.2,1.5) -- cycle;
    \filldraw[even odd rule,inner color=selectiveyellow ,outer color=black!50,color = black!50] (3.6,1.5) circle (0.6);
    \filldraw[even odd rule,inner color=selectiveyellow ,outer color=black!50,color = black!50] (3.6,3) circle (0.6);
    \draw[selectiveyellow,wave3,line width = 0.05cm,->] (3.6,3) -- (4.2,3.6);
    \draw[selectiveyellow,wave3,line width = 0.05cm,->] (3.6,1.5) -- (4.2,0.9);
    \draw[white,line width = 0.05cm] (3.35,1.65) -- (3.85,1.65);
    \draw[white,line width = 0.05cm] (3.35,1.35) -- (3.85,1.35);
    \draw[white,line width = 0.05cm] (3.35,3.15) -- (3.85,3.15);
    \draw[white,line width = 0.05cm] (3.35,2.85) -- (3.85,2.85);
    % inefficiency splitters
    %\draw[draw = white, left color = vividviolet!70!turquoiseblue, right color = vividviolet!50!white] (4.65,3.1) -- (4.9,3.35) -- (5.15,3.1) -- (4.9,2.85) -- cycle;
    %\draw[white] (4.65,3.1) -- (5.15,3.1);
    %\draw[draw = white, left color = vividviolet!70!turquoiseblue, right color = vividviolet!50!white] (4.65,1.4) -- (4.9,1.65) -- (5.15,1.4) -- (4.9,1.15) -- cycle;
    %\draw[white] (4.65,1.4) -- (5.15,1.4);
    \draw[draw = white, left color = vividviolet!70!turquoiseblue, right color = vividviolet!50!white] (6,2.75) -- (6.25,3) -- (6.5,2.75) -- (6.25,2.5) -- cycle;
    \draw[white] (6.25,2.5) -- (6.25,3);
    \draw[draw = white, left color = vividviolet!70!turquoiseblue, right color = vividviolet!50!white] (6,1.75) -- (6.25,2) -- (6.5,1.75) -- (6.25,1.5) -- cycle;
    \draw[white] (6.25,1.5) -- (6.25,2);
    % signal path and main splitter
    \draw[selectiveyellow,line width = 0.05cm,->] (4.2,3.6) -- (4.3,3.7) -- (5.5,2.5);
    \draw[selectiveyellow,line width = 0.05cm,->] (4.2,0.9) -- (4.3,0.8) -- (5.5,2.0);
    \draw[turquoiseblue!50,line width = 0.05cm] (4.1,3.75) -- (4.5,3.75);
    \draw[turquoiseblue!50,line width = 0.05cm] (4.1,0.75) -- (4.5,0.75);
    \draw[left color = turquoiseblue, right color = Cerulean, draw = white] (5.25,2.25)-- (5.75,2.75) -- (6.25,2.25) -- (5.75,1.75) -- cycle;
    \draw[white] (5.25,2.25) -- (6.25,2.25);
    \draw[selectiveyellow, dotted, line width = 0.05cm] (5.5,2.5) -- (6,2);
    \draw[selectiveyellow, dotted, line width = 0.05cm] (5.5,2) -- (6,2.5);
    \draw[selectiveyellow,line width = 0.05cm,->] (6,2.5) -- (6.5,3);
    \draw[selectiveyellow,line width = 0.05cm,->] (6,2) -- (6.5,1.5);
    % detectors and cables
    \draw[turquoiseblue,line width = 0.05cm] (6.625,1.625) -- (6.375,1.375);
    \draw[turquoiseblue,line width = 0.05cm] (6.625,2.875) -- (6.375,3.125);
    \draw[left color = turquoiseblue, right color = Cerulean, draw = white] (6.5,3.25) -- (6.75,3.5) -- (7,3.25) -- (6.75,3) -- cycle;
    \draw[white] (6.25,2.5) -- (6.25,3);
    \draw[left color = turquoiseblue, right color = Cerulean, draw = white] (6.5,1.25) -- (6.75,1.5) -- (7,1.25) -- (6.75,1) -- cycle;
    \draw[white] (6.75,1) -- (6.75,1.5);
    \draw[white] (6.75,3) -- (6.75,3.5);
    \draw[selectiveyellow,line width = 0.05cm,->] (7.6,2.4) -- (6.5,3.5);
    \draw[selectiveyellow,line width = 0.05cm,->] (7.6,2.1) -- (6.5,1);
    \draw[selectiveyellow,line width = 0.05cm,->] (6.5,3) -- (7,3.5);
    \draw[selectiveyellow,line width = 0.05cm,->] (6.5,1.5) -- (7,1);
    \draw[fill = black, rounded corners = 0.05cm] (7.15,3.35) arc[radius = 0.2, start angle = -45, end angle = 135] --  cycle;
    \draw[fill = black, rounded corners = 0.05cm] (7.15,1.15) arc[radius = 0.2, start angle = 45, end angle = -135] --  cycle;
    \draw[fill = black, rounded corners = 0.05cm] (6.63,3.63) arc[radius = 0.2, start angle = 45, end angle = 225] --  cycle;
    \draw[fill = black, rounded corners = 0.05cm] (6.63,0.87) arc[radius = 0.2, start angle = -45, end angle = -225] --  cycle;
    \draw[line width = 0.05cm] (7,3.5) arc[radius = 0.35, start angle=-45, end angle = 225]; 
    \draw[line width = 0.05cm] (7,1) arc[radius = 0.35, start angle=45, end angle = -225];
    \draw[line width = 0.05cm, rounded corners = 0.3cm, ->] (6.4,0.7) -- (6.0,0.7) -- (5.5,0.5) -- (0.85,0.5) -- (0.85,1.5);
    \draw[line width = 0.05cm, rounded corners = 0.3cm, ->] (6.4,3.8) -- (6.0,3.8) -- (5.5,4) -- (0.85,4) -- (0.85,2.5);
    % computer and feedback
    \draw[fill = black!20!white, rounded corners = 0.05cm] (0.5,1.6) -- (1.2,1.6) -- (1.1,1.8) -- (0.6,1.8) -- cycle;
    \draw[fill = black!20!white, rounded corners = 0.05cm] (0.5,1.9) -- (1.2,1.9) -- (1.2,2.4) -- (0.5,2.4) -- cycle;
    \draw[fill = white, draw = white, rounded corners = 0.05cm] (0.6,2) -- (1.1,2) -- (1.1,2.3) -- (0.6,2.3) -- cycle;
    \draw[line width = 0.05cm, rounded corners = 0.1cm] (1.3,2.2) -- (1.5,2.2) -- (1.5,3) -- (2,3);
    \draw[line width = 0.05cm, rounded corners = 0.1cm] (1.3,2.0) -- (1.5,2.0) -- (1.5,1.5) -- (2,1.5);
    \draw[fill = black, rounded corners = 0.05cm] (1.7,1.45) rectangle (2,1.55);
    \draw[fill = black, rounded corners = 0.05cm] (1.7,2.95) rectangle (2,3.05);
    \draw[red,wave3,line width = 0.05cm,->] (2,1.5) -- (2.8,1.5);
    \draw[red,wave3,line width = 0.05cm,->] (2,3) -- (2.8,3);
    % inefficiency extra paths
    %\draw[black,opacity = 0.3,line width = 0.05cm] (4.2,2.4) -- (4.9,3.1);
    %\draw[black,opacity = 0.3,line width = 0.05cm] (4.2,2.1) -- (4.9,1.4);
    %\draw[selectiveyellow, opacity = 0.5, line width = 0.05cm,->] (4.9,3.1) -- (5.2,3.4);
    %\draw[selectiveyellow, opacity = 0.5, line width = 0.05cm,->] (4.9,1.4) -- (5.2,1.1);
    \draw[black,opacity = 0.3,line width = 0.05cm] (6.6,2.4) -- (6.25,2.75);
    \draw[black,opacity = 0.3,line width = 0.05cm] (6.6,2.1) -- (6.25,1.75);
    \draw[selectiveyellow, opacity = 0.5, line width = 0.05cm,->] (6.25,2.75) -- (5.95,3.05);
    \draw[selectiveyellow, opacity = 0.5, line width = 0.05cm,->] (6.25,1.75) -- (5.95,1.45);
    \end{tikzpicture} \\ \begin{picture}(0,0)(0,7)
    \put(-13,82){\color{white} A}
    \put(-13,61){\color{white} B}
    \put(-50,81){\color{red} $\hat{F}_A$}
    \put(-50,61){\color{red} $\hat{F}_B$}
    \put(-50,105){\color{red} $\hat{\mathcal{U}}_A$}
    \put(-50,39){\color{red} $\hat{\mathcal{U}}_B$}
    \put(33,82){\tiny \color{selectiveyellow} $1$}
    \put(33,64){\tiny \color{selectiveyellow} $2$}
    \put(66,76.5){\tiny \color{selectiveyellow} $4$}
    \put(66,69.5){\tiny \color{selectiveyellow} $3$}
    \put(18,73){\tiny \color{white} 50/50}
    %\put(0,73){\tiny vac.}
    %\put(21,87){\color{vividviolet} \footnotesize $\eta_1$}
    %\put(21,61){\color{vividviolet} \footnotesize $\eta_2$}
    %\put(34,110){\color{selectiveyellow} $a_{1,\ell}^\dag$}
    %\put(34,34){\color{selectiveyellow} $a_{2,\ell}^\dag$}
    \put(74,85){\color{vividviolet} \footnotesize $\eta_4$}
    \put(74,63){\color{vividviolet} \footnotesize $\eta_3$}
    \put(85,85){\color{turquoiseblue} \footnotesize $\varphi$}
    \put(85,63){\color{turquoiseblue} \footnotesize $\phi$}
    \put(45,103){\color{selectiveyellow} $a_{4,\ell}^\dag$}
    \put(45,45){\color{selectiveyellow} $a_{3,\ell}^\dag$}
    \put(72,73){\tiny vac.}
    \put(100,73){\tiny \color{selectiveyellow} LO}
    \put(76,118){$r_4$}
    \put(76,27){$r_3$}
    \put(-106,118){(b)}
    \end{picture} \vspace{-10pt}
    \caption{We illustrate an apparatus for creating and preserving entanglement between qubits $A$ and $B$, using measurements of spontaneous emission and feedback based on those measurements. Panel (a) shows a device based on photocounting measurements, whereas panel (b) shows a corresponding device based on homodyne detection. In either case, cavities and transmission lines capture spontaneous emission and route it to measurement devices. The emitted signals from each qubit are mixed on a 50/50 beamsplitter and then monitored continuously, with a measurement result acquired every integration interval $\Delta t$. We suppose that $\Delta t \ll T_1$. Feedback control is exerted by fast $\pi$--pulses (fast compared to both $T_1$ and $\Delta t$), which quickly flip either qubit $A$ ($\hat{F}_A$) and/or qubit $B$ ($\hat{F}_B$) may or may not be implemented at the end of each detector integration time-step, conditioned on the measurement outcome. Additional qubit rotations $\hat{\mathcal{U}}_A$ and $\hat{\mathcal{U}}_B$ are used in the homodyne case (b). The cavities must be engineered such that the photons implementing the $\pi$-pulses, or other rotations, do not couple to the output modes which lead to the measurement devices. The purple beam-splitters model photon losses, where the incoming signal scatters according to the transformation $\hat{a}_i^\dag \rightarrow \sqrt{\eta_i}\hat{a}_{i,s}^\dag+\sqrt{1-\eta_i}\hat{a}_{i,\ell}^\dag$; there is a probability $\eta_i$ that the signal is transmitted, but a probability $1-\eta_i$ that it is reflected into the lost mode corresponding to $a_{i,\ell}^\dag$. Perfect measurement efficiency ($\eta_i = 1$) corresponds to lossless transmission from qubits to detectors. For a more comprehensive treatment of the measurements in this scheme, see \cite{LongFlor2019}. %Values of $\eta$ less than one model the loss of some signal photons (i.e.~inefficient measurements); outcomes in the unmeasured output modes may then be nonzero, and must be traced out, leading to mixed states. We first develop our schemes in the ideal case $\eta_3 = 1 = \eta_4$, presented in Secs.~\ref{sec-PDfeed-basic} and \ref{sec-HomFeed-basic}; we then consider the more realistic case $\eta_3 = \eta = \eta_4$ for $\eta < 1$ in Sec.~\ref{sec-etaimpact}. A more comprehensive treatment of all elements of these schemes, except for the feedback, can be found in \cite{LongFlor2019}.
    } 
    \label{fig-exp}
\end{figure}

\section{Photodetection--based feedback: Concurrence Preservation via Measurement Undoing \label{sec-PDfeed-basic}}

We begin with the case of jump trajectories, obtained from photodetection of two qubits' spontaneous emission, as per the device illustrated in Fig.~\ref{fig-exp}(a).
It will be helpful to recapitulate a few of our previous results \cite{LongFlor2019}, which will prove key to the scheme we now construct.
Firstly, with the two--qubit state initialized in $\ket{ee}$, two clicks are expected over the course of an experiment, absent any re--excitation of either qubit after it decays; 
the first click heralds the generation of a Bell state $\ket{\Psi^\pm}=(\ket{eg}\pm\ket{ge})/\sqrt{2}$ between the emitters, while the second click eliminates the entanglement, generating the state $\ket{gg}$.
Secondly, Bell states of the form $\ket{\Phi^\pm} = (\ket{ee}\pm\ket{gg})/\sqrt{2}$ hold their entanglement longer on average than the states $\ket{\Psi^\pm}$ under fluorescence and photodetection; this is because one click heralds complete disentanglement for a state $\ket{\Psi^\pm}$, whereas a state $\ket{\Phi^\pm}$ requires either two clicks or a long (compared to $T_1$) wait time to asymptotically disentangle the qubits.

While these even and odd parity Bell states behave differently, a $\pi$--rotation on a single qubit is all that is required to change from one type to the other.
Mathematically, we say that flipping qubit A and leaving qubit B alone can be represented by the unitary operation $\hat{F}_A = i\hat{\sigma}_y^A\otimes\openone^B$,
such that $\hat{F}_A \ket{\Psi^\pm} \propto \ket{\Phi^\mp}$ up to a global phase factor.
A feedback scheme for entanglement creation is thus easily identified: Starting from $\ket{ee}$ we wait for a click which heralds the creation of a state $\ket{\Psi^\pm}$; when that happens, we immediately flip one of the two qubits (e.g.~by the operation $\hat{F}_A$) to obtain the more--robust $\ket{\Phi^\mp}$ state instead. 
If we measure a single photon emission after obtaining a state of the type $\ket{\Phi^\pm}$, this subsequent click just takes us back to $\ket{\Psi^\pm}$ (which can again be immediately reset to $\ket{\Phi^\mp}$ by flipping one qubit). 

Between two clicks, the evolution of the two qubit system still degrades entanglement, such that additional pulses are needed to fully preserve state $\ket{\Phi^\pm}$. Consider evolution of a state of form $\mathsf{a}\ket{ee}+\mathsf{d}\ket{gg}$ under measurement dynamics for a step of duration $\Delta t$, in which neither detector receives a photon (the result of the majority of the individual measurements, for $\Delta t \ll T_1$). The Kraus operator implementing the resulting state update \cite{LongFlor2019} is
\be 
\hat{\mathcal{M}}_{00}= 
\left( \begin{array}{cccc}
1-\epsilon & 0 & 0 & 0\\
0 & \sqrt{1-\epsilon} & 0 & 0 \\
0 & 0 & \sqrt{1-\epsilon} & 0 \\
0 & 0 & 0 & 1
\end{array} \right),
\ee 

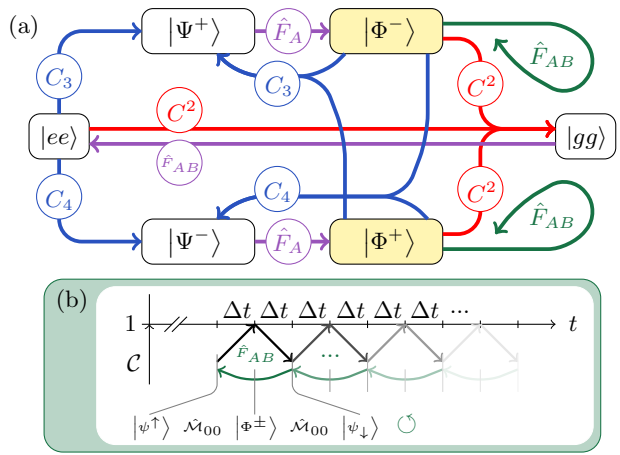
\begin{figure}
\begin{tikzpicture}
\filldraw[color = darkspringgreen!30,draw = darkspringgreen,rounded corners = 0.25cm] (0.2,-2.5) rectangle (7.6,-0.2);
\filldraw[color = white, draw = white,rounded corners = 0.25cm] (0.9,-2.4) rectangle (7.5,-0.3);
\filldraw[color=white, draw = black, rounded corners = 0.15cm] (0,1.4) rectangle (0.8,2);
\filldraw[color=white, draw = black, rounded corners = 0.15cm] (1.5,0) rectangle (3,0.6);
\filldraw[color=goldenyellow!30, draw = black, rounded corners = 0.15cm] (4,0) rectangle (5.5,0.6);
\filldraw[color=white, draw = black, rounded corners = 0.15cm] (7,1.4) rectangle (7.8,2);
\filldraw[color=white, draw = black, rounded corners = 0.15cm] (1.5,2.8) rectangle (3,3.4);
\filldraw[color=goldenyellow!30, draw = black, rounded corners = 0.15cm] (4,2.8) rectangle (5.5,3.4);
\draw[color = red,->, line width = 0.05cm] (0.8,1.8) -- (7,1.8);
\draw[color = red,->,rounded corners = 0.3cm, line width = 0.05cm] (5.5,3.0) -- (6,3.0) -- (6,1.8) -- (7,1.8);
\draw[color = red,->,rounded corners = 0.3cm, line width = 0.05cm] (5.5,0.4) -- (6,0.4) -- (6,1.8) -- (7,1.8);
\draw[color = deeplilac,->, line width = 0.05cm] (7,1.6) -- (0.8,1.6);
\draw[color = darkspringgreen,rounded corners = 0.3cm,->, line width = 0.05cm] (5.5,3.2) -- (7.2,3.2) -- (7.6,2.7) -- (7.2,2.2) -- (6.5,2.6) -- (6.2,3);
\draw[color = darkspringgreen,rounded corners = 0.3cm,->, line width = 0.05cm] (5.5,0.2) -- (7.2,0.2) -- (7.6,0.7) -- (7.2,1.2) -- (6.5,0.8) -- (6.2,0.4);
\draw[color = deeplilac,->, line width = 0.05cm] (3,3.1) -- (4,3.1);
\draw[color = deeplilac,->, line width = 0.05cm] (3,0.3) -- (4,0.3);
\draw[color = ceruleanblue,->,line width = 0.05cm, rounded corners = 0.3cm] (0.4,2) -- (0.4,3.1) -- (1.5,3.1);
\draw[color = ceruleanblue,->,line width = 0.05cm, rounded corners = 0.3cm] (0.4,1.4) -- (0.4,0.3) -- (1.5,0.3);
\draw[color = ceruleanblue,->,line width = 0.05cm, rounded corners = 0.3cm] (4.2,2.8) -- (3.9,2.5) -- (2.8,2.5) -- (2.5,2.8);
\draw[color = ceruleanblue,->,line width = 0.05cm, rounded corners = 0.3cm] (4.2,0.6) -- (4.2,1.7) -- (3.9,2.5) -- (2.8,2.5) -- (2.5,2.8);
\draw[color = ceruleanblue,->,line width = 0.05cm, rounded corners = 0.3cm] (5.3,0.6) -- (5,0.9) -- (2.8,0.9) -- (2.5,0.6);
\draw[color = ceruleanblue,->,line width = 0.05cm, rounded corners = 0.3cm] (5.3,2.8) -- (5.3,1.7) -- (5,0.9) -- (2.8,0.9) -- (2.5,0.6);
\filldraw[color = white, draw = ceruleanblue] (0.4,2.5) circle (0.3);
\filldraw[color = white, draw = ceruleanblue] (0.4,0.9) circle (0.3);
\filldraw[color = white, draw = ceruleanblue] (3.3,2.4) circle (0.3);
\filldraw[color = white, draw = ceruleanblue] (3.3,1.0) circle (0.3);
\filldraw[color = white, draw = deeplilac] (3.45,3.1) circle (0.3);
\filldraw[color = white, draw = deeplilac] (3.45,0.3) circle (0.3);
\filldraw[color = white, draw = deeplilac] (2,1.35) circle (0.3);
\filldraw[color = white, draw = red] (2,2.05) circle (0.3);
\filldraw[color = white, draw = red] (6,2.4) circle (0.3);
\filldraw[color = white, draw = red] (6,1.0) circle (0.3);
%%%%%%%%%%%%%%%%%%%%%%%%%%%%%%%%%%%%%%%%%%%%%%%%%%%%%%%%%%%
\draw[->] (2.0,-0.8) -- (7.0,-0.8);
\draw (1.9,-0.9) -- (2.1,-0.7);
\draw (1.8,-0.9) -- (2.0,-0.7);
\draw (1.5,-0.8) -- (1.9,-0.8);
\draw[->] (1.6,-1.6) -- (1.6,-0.8);
\draw (1.6,-0.8) -- (1.6, -0.4);
\draw (2.5,-0.85) -- (2.5,-0.75);
\draw (3,-0.85) -- (3,-0.75);
\draw (3.5,-0.85) -- (3.5,-0.75);
\draw (4,-0.85) -- (4,-0.75);
\draw (4.5,-0.85) -- (4.5,-0.75);
\draw (5,-0.85) -- (5,-0.75);
\draw (5.5,-0.85) -- (5.5,-0.75);
\draw (6,-0.85) -- (6,-0.75);
\draw (6.5,-0.85) -- (6.5,-0.75);
\draw[->,line width = 0.03cm] (2.5,-1.3) -- (3.0,-0.8);
\draw[->,line width = 0.03cm] (3.0,-0.8) -- (3.5,-1.3);
\draw[->,line width = 0.03cm,darkspringgreen,rounded corners = 0.3cm] (3.5,-1.4) -- (3.0,-1.6) -- (2.5,-1.4);
\draw[black!70,->,line width = 0.03cm] (3.5,-1.3) -- (4.0,-0.8);
\draw[black!70,->,line width = 0.03cm] (4.0,-0.8) -- (4.5,-1.3);
\draw[->,line width = 0.03cm,darkspringgreen!70,rounded corners = 0.3cm] (4.5,-1.4) -- (4.0,-1.6) -- (3.5,-1.4);
\draw[black!40,->,line width = 0.03cm] (4.5,-1.3) -- (5.0,-0.8);
\draw[black!40,->,line width = 0.03cm] (5.0,-0.8) -- (5.5,-1.3);
\draw[->,line width = 0.03cm,darkspringgreen!40,rounded corners = 0.3cm] (5.5,-1.4) -- (5.0,-1.6) -- (4.5,-1.4);
\draw[black!10,->,line width = 0.03cm] (5.5,-1.3) -- (6.0,-0.8);
\draw[black!10,->,line width = 0.03cm] (6.0,-0.8) -- (6.5,-1.3);
\draw[->,line width = 0.03cm,darkspringgreen!10,rounded corners = 0.3cm] (6.5,-1.4) -- (6.0,-1.6) -- (5.5,-1.4);
\draw[black!50,rounded corners = 0.1cm] (2.5,-1.2) -- (2.5,-1.7) -- (1.6,-2);
\draw[black!50] (3,-1.4) -- (3,-2);
\draw[black!50,rounded corners = 0.1cm] (3.5,-1.2) -- (3.5,-1.7) -- (4.4,-2);
\draw[black!40] (4,-1.4) -- (4,-1.7);
\draw[black!30] (4.5,-1.2) -- (4.5,-1.7);
\draw[black!20] (5,-1.4) -- (5,-1.7);
\draw[black!15] (5.5,-1.2) -- (5.5,-1.7);
\draw[black!10] (6,-1.4) -- (6,-1.7);
\draw[black!5] (6.5,-1.2) -- (6.5,-1.7);
\end{tikzpicture}\begin{picture}(0,0)(230,-71)
\put(0,88){(a)}
\put(12,46){$\ket{ee}$}
\put(211,46){$\ket{gg}$}
\put(60,6){$\ket{\Psi^-}$}
\put(134,6){$\ket{\Phi^+}$}
\put(60,86){$\ket{\Psi^+}$}
\put(134,86){$\ket{\Phi^-}$}
\put(197,76){\color{darkspringgreen} $\hat{F}_{AB}$}
\put(197,15){\color{darkspringgreen} $\hat{F}_{AB}$}
\put(173.5,64){\color{red} $C^2$}
\put(173.5,24){\color{red} $C^2$}
\put(60,54.5){\color{red} $C^2$}
\put(57.5,37){\tiny \color{deeplilac} $\hat{F}_{AB}$}
\put(100,85){\color{deeplilac} $\hat{F}_A$}
\put(100,5.5){\color{deeplilac} $\hat{F}_A$}
\put(97,65){\color{ceruleanblue} $C_3$}
\put(96.5,25){\color{ceruleanblue} $C_4$}
\put(14,68){\color{ceruleanblue} $C_3$}
\put(14,22){\color{ceruleanblue} $C_4$}
\put(18,-16){(b)}
\put(45,-40){$\mathcal{C}$}
\put(44,-25){1}
\put(81,-20){$\Delta t$}
\put(95,-20){$\Delta t$}
\put(109.5,-20){$\Delta t$}
\put(124,-20){$\Delta t$}
\put(138,-20){$\Delta t$}
\put(152,-20){$\Delta t$}
\put(167,-18){...}
\put(212,-25){$t$}
\put(86,-35){\color{darkspringgreen}\tiny $\hat{F}_{AB}$}
\put(118,-35){\color{darkspringgreen} ...}
\put(47,-63){{\tiny $\ket{\psi^\uparrow} \:\: \hat{\mathcal{M}}_{00} \:\: \ket{\Phi^\pm} \:\: \hat{\mathcal{M}}_{00} \:\: \ket{\psi_\downarrow}\:\:$} {\normalsize\color{darkspringgreen}$\boldsymbol{\circlearrowleft}$}}
\end{picture}
\caption{We lay out a flowchart (a) describing our feedback procedure. 
We begin with the separable state $\ket{ee}$, and see a rapid rise in average concurrence as the first clicks (either at port 3 or 4, denoted by {\color{ceruleanblue} $C_3$} or {\color{ceruleanblue} $C_4$}, respectively) put our qubits in the $\ket{\Psi^\pm}$ Bell states. 
As described in the main text and in our previous work \cite{LongFlor2019}, the $\ket{\Phi^\pm}$ Bell states are more robust against disentanglement, however; we consequently flip qubit $A$ with a $\pi$-pulse ({\color{deeplilac} $\hat{F}_A$}) immediately after the first click heralds entanglement, such that we take $\ket{\Psi^\pm} \rightarrow \ket{\Phi^\mp}$ (neglecting any global phase factors). 
Single clicks then send us back to the $\ket{\Psi^\pm}$ Bell states, rather than to the separable state $\ket{gg}$. 
When no detector click is received, states $\ket{\Phi^\pm}$ gradually lose concurrence as amplitude shifts from $\ket{ee}$ to $\ket{gg}$. 
By flipping both qubits ({\color{darkspringgreen} $\hat{F}_{AB}$}) between these no--click measurements, we implement a state--recycling scheme, however. This may be understood as introducing limit cycle in the concurrence $\mathcal{C}$, using the fast $\pi$--pulses {\color{darkspringgreen}$\hat{F}_{AB}$}, as illustrated in (b). It can also be understood in terms of a measurement reversal: 
If the effect of the null measurement is described by $\hat{\mathcal{M}}_{00}\ket{\Phi^\pm} \rightarrow \ket{\psi_\downarrow}$, then flipping both qubits leads to a state $\ket{\psi^\uparrow}$ (i.e.~$\hat{F}_{AB} \ket{\psi_\downarrow} \rightarrow \ket{\psi^\uparrow}$) , which is the same as $\ket{\psi_\downarrow}$ except that the amplitudes on $\ket{ee}$ and $\ket{gg}$ are swapped.
This change is substantial, because the next no--click measurement then undoes the first, i.e.~$\hat{\mathcal{M}}_{00}\ket{\psi^\uparrow}\rightarrow \ket{\Phi^\pm}$, thereby resetting the state in a way that traps the concurrence in a cycle near $\mathcal{C}=1$.
The net effect of this scheme is that once we are in the cycle about the $\ket{\Phi^\pm}$ state, only a double--click {\color{red} $C^2$}, in which both qubits emit in the same timestep, can completely disentangle them. 
In the event of such a rare double--click, we simply flip both qubits ({\color{deeplilac} $\hat{F}_{AB}$}), and thereby restart the whole scheme from $\ket{ee}$. The concurrence yield of this scheme is shown in Fig.~\ref{fig-ConcIdealFB}. }
    \label{fig-flowchart}
\end{figure}
\begin{figure}
    \centering
    \includegraphics[width=\columnwidth,trim = {5 0 30 25}, clip]{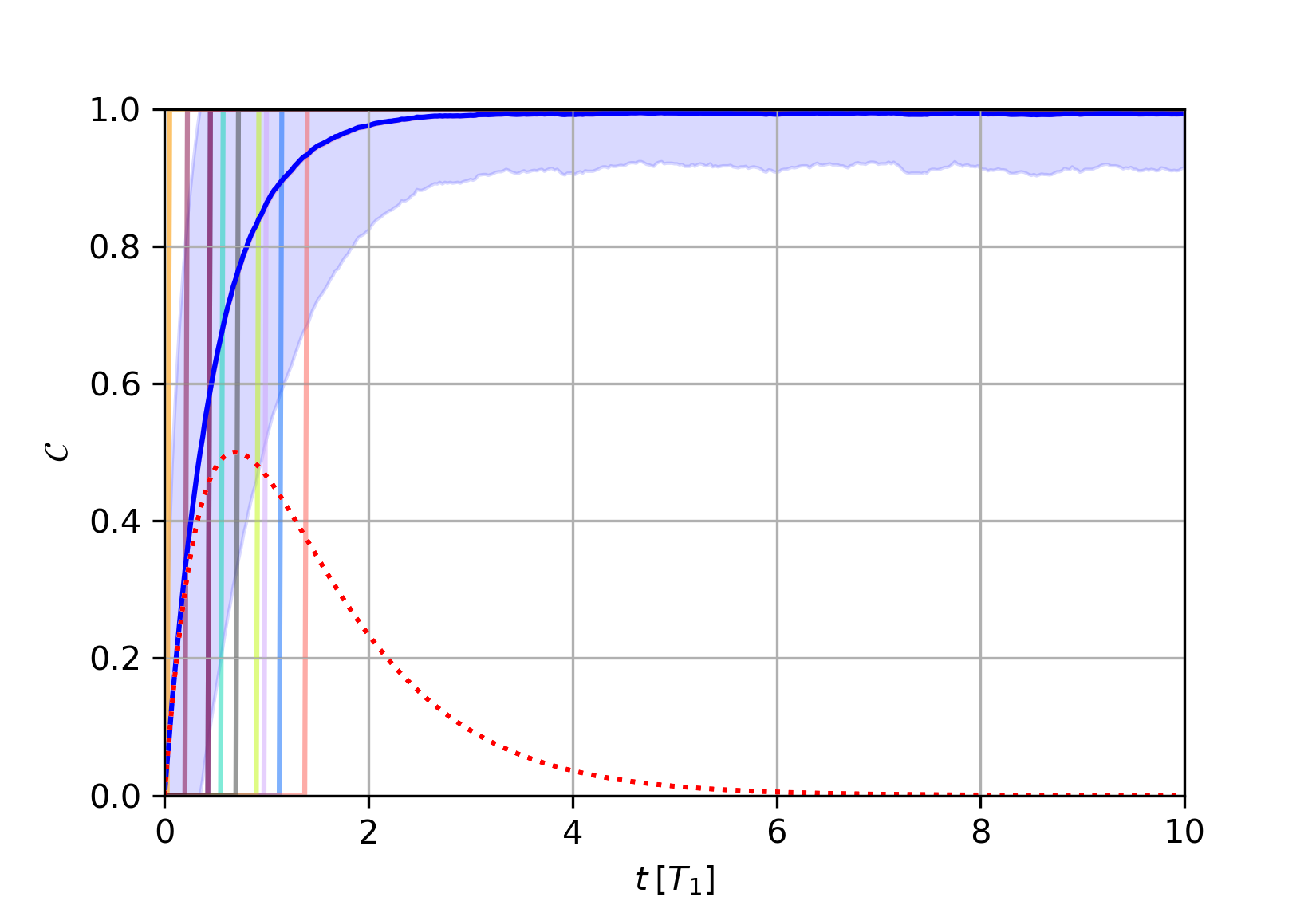}
    \caption{We show the concurrence $\mathcal{C}$ as a function of time, obtained via the feedback scheme described in Fig.~\ref{fig-flowchart} and the main text. The concurrence of individual jump trajectories are shown (background, multiple colors), as is the average concurrence over an ensemble of trajectories (dark blue, surrounded by a pale envelope of $\pm$ one standard deviation). Idealizations implicit in this simulation include 1) capture and detection efficiency are perfect, 2) no environmental channels apart from the decay channel we measure exist, and that our $\pi$-pulses are 3) free of errors and 4) implemented instantaneously as a measurement timestep completes and its outcome is recorded. We see that within 2--3 $T_1$, we are able to drive the average concurrence to $\mathcal{C}\gtrsim0.99$ and maintain it there indefinitely with our protocol. Most individual trajectories achieve $\mathcal{C} = 1$ exactly, but at a random time (because jumps occur stochastically). The dotted red curve shows the average concurrence from the measurement alone, without feedback.} 
    \label{fig-ConcIdealFB}
\end{figure}


\noindent
where $\epsilon \equiv \gamma \: \Delta t$, and $\epsilon$ should be assumed small (i.e.~measurements are performed on timescale which is fast compared to $T_1$). 
Repeated evolution of this type gradually causes the concurrence to decay, as the amplitude in $\ket{gg}$ grows relative to that in $\ket{ee}$ (with every step $\Delta t$ in which no photons are received, our supposed probability of ultimately getting the outcome $\ket{gg}$ instead of $\ket{ee}$ increases). Suppose however that upon receiving a no--click result, we flip both qubits, according to the operation 
\be
\hat{F}_{AB} = (i\hat{\sigma}_y^A\otimes\openone^B)\cdot(i\openone^A\otimes\hat{\sigma}_y^B) = \hat{F}_A\cdot \hat{F}_B.
\ee
We find that after a second step of measurement without a click at the detectors, the two--qubit state is unchanged, i.e.~we find
\be \label{pd-bell-recycle}
\frac{\hat{\mathcal{M}}_{00} \hat{F}_{AB} \hat{\mathcal{M}}_{00} \ket{\Phi^\pm}}{\left| \hat{\mathcal{M}}_{00} \hat{F}_{AB} \hat{\mathcal{M}}_{00} \ket{\Phi^\pm} \right|} \propto \ket{\Phi^\pm}
\ee
up to a global phase factor.
Effectively, if we flip the slightly larger amplitude from $\ket{gg}$ back to $\ket{ee}$, the next step of no--click evolution will simply undo the previous one \cite{Sun2010, Jordan2006Reversal, JordanCPReversal, Katz2008, Kim2009, Korotkov2010, Kim2012, Korotkov2012}; thus we can effectively ``recycle'' the $\ket{\Phi^\pm}$ states indefinitely during a stretch of no--click measurement outcomes by quickly flipping both qubits after every other such measurement. 
%In the language of measurement reversal (or measurement undoing) \cite{Jordan2006Reversal, JordanCPReversal, Katz2008, Kim2009, Korotkov2010, Kim2012, Korotkov2012}, controlling each qubit with fast double flips ensures that each measurement $\hat{\mathcal{M}}_{00}$ reverses (or undoes) the previous one. 
The utility of flipping operations for reversing entanglement decay due to a damping channel has been noted before, by Sun et al.~\cite{Sun2010}. 
The measurement reversal succeeds most of the time, because the outcome corresponding to $\hat{\mathcal{M}}_{00}$ occurs with probability $O(1)$, whereas results involving one or two clicks occur with probabilities $O(\epsilon)$ or $O(\epsilon^2)$, respectively \cite{LongFlor2019}.
Only the double click, which is the rarest of these options, disentangles the qubits. 
The recycling operation we have described actually works on any state, because applying the total scheme twice, as per
\be \label{mapping+2}
\hat{\mathcal{M}}_{00} \hat{F}_{AB} \hat{\mathcal{M}}_{00}\hat{\mathcal{M}}_{00} \hat{F}_{AB} \hat{\mathcal{M}}_{00} \propto \openone,
\ee
amounts to an identity operation.
Therefore, the procedure can be seen as a general measurement reversal, analogous to the superconducting phase experimental results of Ref.~\cite{Kim2012}. 
Our procedure can effectively freeze the state evolution \emph{between} click events into a small limit cycle (of size $\sim \Delta t$) around any desired state; the application of primary interest here is stabilization of the Bell states $\ket{\Phi^\pm}$, but one could imagine other uses as well.
%Once the qubits are entangled, the present scheme can, under ideal conditions, keep both qubits entangled indefinitely under all circumstances except a rare ``double emission'' event in which both qubits emit photons in the same timestep $\Delta t$.
A flowchart in Fig.~\ref{fig-flowchart} represents the entire feedback process we have just described (including the correction of jumps due to a single emission), and the behavior of the concurrence, obtained from numerical simulation of trajectories under the measurement and feedback protocol, is shown in Fig.~\ref{fig-ConcIdealFB}. 

\par We may more--formally frame the state evolution of the recycling scheme between clicks as an iterative map, such that 
\be 
\ket{\psi_{k+1}} = \frac{\hat{\mathcal{M}}_{00} \hat{F}_{AB} \hat{\mathcal{M}}_{00} \ket{\psi_k}}{\left| \hat{\mathcal{M}}_{00} \hat{F}_{AB} \hat{\mathcal{M}}_{00} \ket{\psi_k} \right|}.
\ee
It is then straightforward to verify that to $O(\Delta t)$, the concurrence $\mathcal{C}$ is unchanged over one step step of the recycling (which covers a total evolution time of $2 \Delta t$), i.e.
\be
\dot{\mathcal{C}} \approx \frac{\mathcal{C}_{k+1}-\mathcal{C}_k}{2\Delta t} = 0.
\ee
This implies that all states are at a fixed point in this iterative mapping of the concurrence, and that therefore the preservation sits at the border between stability and instability \cite{BookOtt, BookStrogatz}; in other words, any errors which may occur as the scheme progresses are simply preserved, without being either suppressed or amplified. 
%What we have just described is an example of a measurement reversal, or measurement un--doing; in that language \cite{Jordan2006Reversal, JordanCPReversal, Katz2008, Kim2009, Korotkov2010, Kim2012, Korotkov2012}, each measurement $\mathcal{M}_{00}$ undoes or reverses the previous one, once the two--qubit flip control is added.
%The measurement reversal occurs most of the time, because the outcome corresponding to $\mathcal{M}_{00}$ occurs with probability $O(1)$, whereas results involving one or two clicks occur with probabilities $O(\epsilon)$ or $O(\epsilon^2)$, respectively \cite{LongFlor2019}.

%%%%%%%%%%%%%%%%%%%%%%%%%%%%%%%%%%%%%%%%%%%%%%%%%%%%%%%%%%%%%%%%%%%%%%%%%%%%%%%%%%%%%%%%%%%%%%%%%%%
%%%%%%%%%%%%%%%%%%%%%%%%%%%%%%%%%%%%%%%%%%%%%%%%%%%%%%%%%%%%%%%%%%%%%%%%%%%%%%%%%%%%%%%%%%%%%%%%%%%
%%%%%%%%%%%%%%%%%%%%%%%%%%%%%%%%%%%%%%%%%%%%%%%%%%%%%%%%%%%%%%%%%%%%%%%%%%%%%%%%%%%%%%%%%%%%%%%%%%%
\section{Adapting the Recycling Scheme to Homodyne--based Feedback \label{sec-HomFeed-basic}}

There has been considerable work on the entangling properties of continuous homodyne measurements as well \cite{LongFlor2019, Carvalho2007, Viviescas2010, Mascarenhas2011}. Martin and Whaley recently derived a feedback scheme based on such measurements which deterministically generates a Bell state in a finite time \cite{Leigh2019}. 
We will summarize their scheme using the notation of our previous works \cite{LongFlor2019}, and then show that the same principles used above can be applied to this case too, i.e.~we will demonstrate that adding fast $\pi$--pulses into the continuous measurement \cite{LongFlor2019} and Hamiltonian feedback protocol \cite{Leigh2019} will allow us to stabilize the entangled state once it is created, instead of having it decay away.

Homodyne detection of fluorescence monitoring quadratures $90^\circ$ out of phase, instead of photodetection (see Fig.~\ref{fig-exp}(b)), generates diffusive quantum trajectories and entangles the emitting qubits to the same degree as photodetection, on average \cite{LongFlor2019, Viviescas2010, Mascarenhas2011}.
The Kraus operator representing a measurement of the quadrature $\phi = 0$ at port 3, and $\varphi = 90^\circ$ at port 4 may be written $\hat{\mathcal{M}}_{\text{hom}}\propto$
\be \label{M-hom}
\ \left( \begin{array}{cccc}
1-\epsilon & 0 & 0 & 0 \\
\sqrt{\epsilon(1-\epsilon)} (X -i Y) & \sqrt{1-\epsilon} & 0 & 0 \\
\sqrt{\epsilon(1-\epsilon)} (X +i Y) & 0 & \sqrt{1-\epsilon} & 0 \\
\epsilon (X^2 + Y^2-1) & 
\sqrt{\epsilon}(X + iY) & \sqrt{\epsilon}(X - i Y) & 1
\end{array} \right),
\ee 
where $X = r_3 \sqrt{\Delta t/2}$ is the outcome of the measurement at port 3, $Y = r_4 \sqrt{\Delta t/2}$ is the outcome of the measurement at port 4, and $\epsilon = \gamma\:\Delta t$ \cite{LongFlor2019, FlorTeach2019}.
Martin and Whaley have recently derived the local/separable unitary feedback operation
\be \label{homfeedU}
\hat{\mathcal{U}} = \exp\left[i \:\Delta t\sqrt{\frac{\gamma}{2}} \:\frac{\mathsf{a} \left( r_3 (\hat{\sigma}_y^A+\hat{\sigma}_y^B) + r_4(\hat{\sigma}_x^B-\hat{\sigma}_x^A) \right)}{\mathsf{a}+\sqrt{1-\mathsf{a}^2}}  \right],
\ee
which acts on a state of the type \be \label{a-state} \ket{\psi} = \mathsf{a} \ket{ee} - \text{sgn}(\mathsf{a})\sqrt{1-\mathsf{a}^2}\ket{gg}, \ee (for real $\mathsf{a}$). 
In the ideal case, one completely cancels the measurement noise by applying this operation, generating deterministic dynamics that are optimal (within continuously--applied Hamiltonian protocols using LOCC, and restricted to states of the type \eqref{a-state}) for driving the system towards an entangled state $\ket{\Psi^-}$.

\par Note that for the choice $\phi = 0$ and $\varphi = 90^\circ$, the measurement records may be written in terms of a signal, and noise term modeled with a Wiener increment $dW$, according to
\begin{subequations} \be 
r_3 = \sqrt{\tfrac{\gamma}{2}}\left\langle \sigma_x^A + \sigma_x^B \right\rangle +  \frac{dW_3}{dt},
\ee \be 
r_4 = \sqrt{\tfrac{\gamma}{2}}\left\langle \sigma_y^A - \sigma_y^B \right\rangle +  \frac{dW_4}{dt}.
\ee \end{subequations}
For a state of the type \eqref{a-state}, we find that $\left\langle \sigma_x^A + \sigma_x^B \right\rangle = 0 = \left\langle \sigma_y^A - \sigma_y^B \right\rangle$, such that the measurements are effectively of the  ``no--knowledge'' type, which are generally useful for cancelling noise \cite{Szigeti2014}. 
The utility of feedback preserving such a condition in the process of entanglement generation, which is related to the concept of a decoherence--free subspace, has been demonstrated for different types of measurements \cite{Hill2008} (e.g.~dispersive measurements). 
These ideas can be helpfully connected to properties of our present scheme: 
First, the feedback protocol (ideally executed) ensures that $r_3$ and $r_4$ are pure noise, which is closely related to the feedback ensuring the state remains of the form \eqref{a-state}. 
Second, the readouts $r$ scale like $dW/dt$ in the time--continuum limit.
%, such that It\^{o} calculus requires that we take $r^2\rightarrow 1/dt$. 
%An equation of motion, effectively of the It\^{o} type, can then be obtained by writing $\ket{\psi(t+\Delta t)}=\hat{\mathcal{U}}\hat{\mathcal{M}}_{\text{hom}}\ket{\psi(t)}/|\hat{\mathcal{U}}\hat{\mathcal{M}}_{\text{hom}}\ket{\psi(t)}|$ for $\ket{\psi}$ as in \eqref{a-state}, expanding the RHS (written in terms of $r_3$ and $r_4$) to $O(\Delta t^2)$, taking $r_3^2 \rightarrow 1/\Delta t \leftarrow r_4^2$, and \emph{only then} truncating the subsequent expressions to $O(\Delta t)$.
An equation of motion can then be obtained by writing $\ket{\psi(t+\Delta t)}=\hat{\mathcal{U}}\hat{\mathcal{M}}_{\text{hom}}\ket{\psi(t)}/|\hat{\mathcal{U}}\hat{\mathcal{M}}_{\text{hom}}\ket{\psi(t)}|$ for $\ket{\psi}$ as in \eqref{a-state} and expanding the RHS (written in terms of $r_3$ and $r_4$) to $O(\Delta t)$, applying It\^{o}'s lemma $r_{3,4}^2 \to 1/\Delta t$. 

\begin{figure}
    \centering
    \includegraphics[width=\columnwidth,trim = {5 0 30 25}, clip]{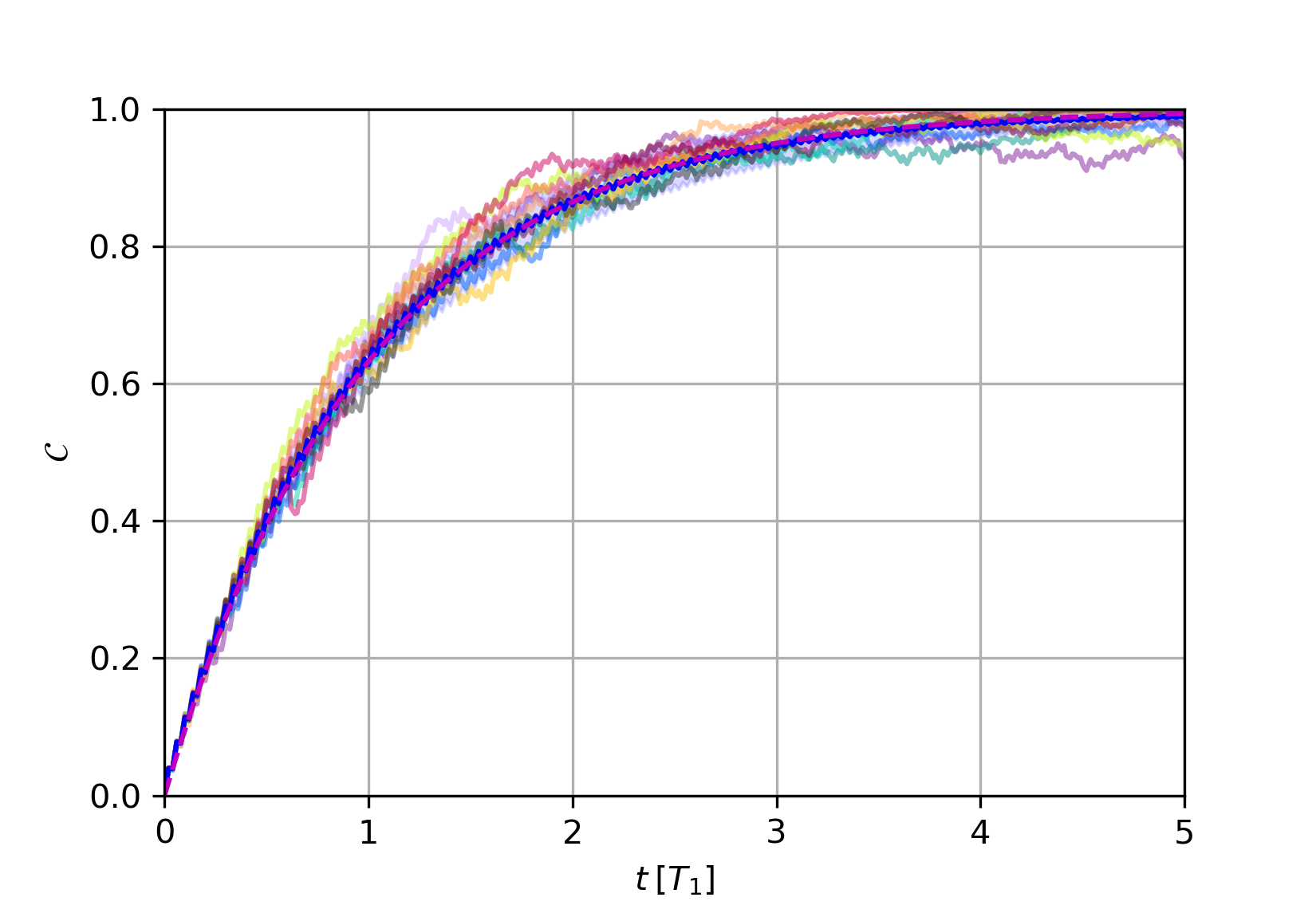}\\
    \includegraphics[width=\columnwidth,trim = {5 0 30 25}, clip]{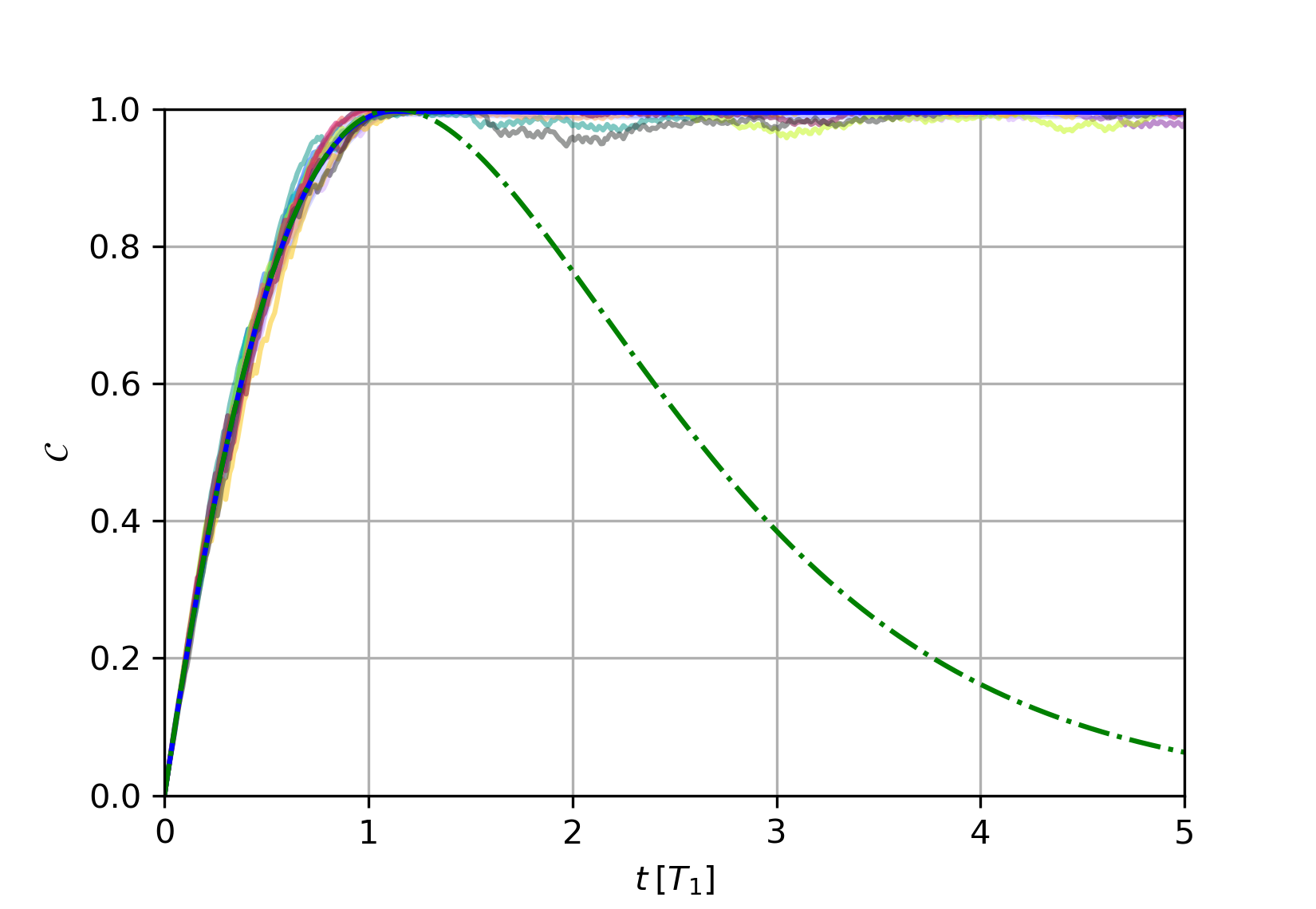}\\
    \begin{picture}(0,0)
    \put(-85,330){(a)}
    \put(-85,160){(b)}
    \end{picture} \vspace{-10pt}
    \caption{We apply the homodyne feedback scheme of \cite{Leigh2019}, using the initial state $\ket{ee}$, and add $\pi$--pulses between every other cycle of measurement and feedback. In (a), we apply our $\pi$--pulse modification over the entire time evolution; while this is not as effective as the ideal case shown in (b), it serves to demonstrate the stability of our modified scheme.
    In (b), we only add the $\pi$--pulses after the time $t_e =  (\pi/4 + \ln\sqrt{2})T_1 \approx 1.13 T_1$ at which maximum entanglement is achieved by the scheme of \cite{Leigh2019} alone. 
    %In other words, we begin with the state $\ket{ee}$ at $t=0$, and apply $...\hat{\mathcal{U}}\mathcal{M}\ket{ee}$ every $\Delta t$ until $t_e$. After $t_e$, we modify this procedure in the spirit of our photodetection recycling scheme, such that we interject $\pi$--pulses every other measurement cycle, i.e.~we do $... \hat{\mathcal{U}}\mathcal{M}F_{AB}\hat{\mathcal{U}}\mathcal{M}\hat{\mathcal{U}}\mathcal{M}F_{AB}\hat{\mathcal{U}}\mathcal{M}\ket{\psi(t_e)}$ (where $4\Delta t$ of evolution are explicitly written). 
    We see that as in the photocounting case, this procedure creates an entanglement--preserving limit cycle. The above simulation assumes that the $\hat{F}_{AB}$ are applied instantaneously, and uses $\Delta t = 0.01 T_1$ for the measurement and feedback. This homodyne scheme attains $\mathcal{C} = 1$ exactly, instead of approach $\mathcal{C} \approx 1$ asymptotically (as in Fig.~\ref{fig-ConcIdealFB}). The approach to $\mathcal{C}=1$ occurs about twice as fast in this homodyne case, as compared with the photodetection case. The analytic solution from \cite{Leigh2019}, without the additional flipping operations, is plotted in dash--dotted green for reference. Deviations from perfectly deterministic dynamics are due to the effects of finite $\Delta t$; we see that these non--idealities have virtually no impact on our ability to preserve concurrence. Up to the jagged ``teeth'' from the finite $\Delta t$, the average concurrence in panel (a) is in good agreement with the analytical solution \eqref{diffCsol}, shown in dashed magenta. }
    \label{fig-homfeed}
\end{figure}
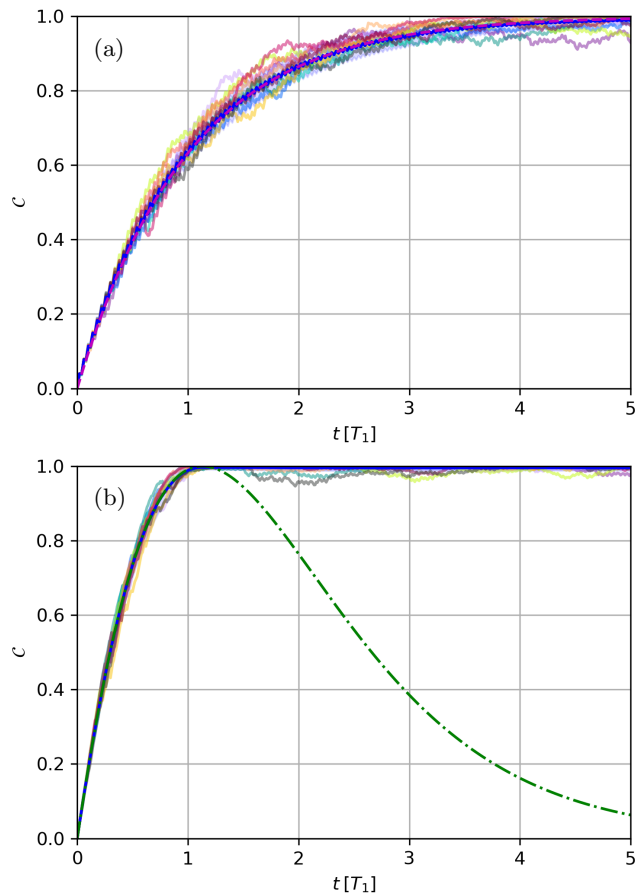

The result can be written as an iterative update
\be \label{map-a}
\mathsf{a}_{k+1} = \mathsf{a}_k - \epsilon \frac{ \mathsf{a}_k\: \text{sgn}(\mathsf{a}_k) \sqrt{1- \mathsf{a}_k^2}}{ \mathsf{a}_k + \text{sgn}(\mathsf{a}_k)\sqrt{1- \mathsf{a}_k^2 }},\quad \text{or}
\ee
\be \label{map-ad}
\mathsf{a}_{k+1} = \mathsf{a}_k+ \epsilon \frac{\mathsf{a}_k \mathsf{d}_k}{\mathsf{a}_k-\mathsf{d}_k},\quad \mathsf{d}_{k+1} = \mathsf{d}_k- \epsilon \frac{\mathsf{a}_k^2}{\mathsf{a}_k-\mathsf{d}_k}
\ee
(where the latter uses $-\text{sgn}(\mathsf{a}_k)\sqrt{1-\mathsf{a}_k}\rightarrow \mathsf{d}_k$).
In the time--continuum limit, these can be written instead as differential equations
\be \label{diffeq-a}
\dot{\mathsf{a}} = - \gamma \frac{\mathsf{a}\: \text{sgn}(\mathsf{a})\sqrt{1-\mathsf{a}^2}}{\mathsf{a}+\text{sgn}(\mathsf{a})\sqrt{1-\mathsf{a}^2}}, \quad\text{or}
\ee
\be \label{diffeq-ad}
\dot{\mathsf{a}} = \gamma \frac{\mathsf{a}\: \mathsf{d}}{\mathsf{a}-\mathsf{d}},\quad \dot{\mathsf{d}} = - \gamma \frac{\mathsf{a}^2}{\mathsf{a}-\mathsf{d}}.
\ee
The expression \eqref{diffeq-a} or \eqref{diffeq-ad} is entirely equivalent to the equation derived in \cite{Leigh2019}, there written instead in terms of the concurrence $\mathcal{C}$, according to
\be \label{leighC}
\dot{\mathcal{C}} = \bigg\lbrace \begin{array}{ll} 
\gamma \left(1- \mathcal{C} + \sqrt{1-\mathcal{C}^2}\right) & \text{ for }|\mathsf{a}| > |\mathsf{d}| \\
\gamma \left(1- \mathcal{C} - \sqrt{1-\mathcal{C}^2}\right) & \text{ for }|\mathsf{a}| < |\mathsf{d}|
\end{array}.
\ee
The solution for the case $|\mathsf{a}|>|\mathsf{d}|$ leads to a concurrence which rises to $\mathcal{C}=1$ (the state is $\ket{\Phi^-}$, with $\mathsf{a} = 1/\sqrt{2} = - \mathsf{d}$), which then switches over to the decaying solution associated with the case $\abs{\mathsf{a}}<\abs{\mathsf{d}}$, as amplitude continues to shift from $\ket{ee}$ over to $\ket{gg}$ (see the green dash--dotted curve in Fig.~\ref{fig-homfeed}(b)).

\begin{figure*}
\hspace{0.75cm}\begin{tabular}{cc}
\includegraphics[width=.45\textwidth,trim = {40 29 0 30},clip]{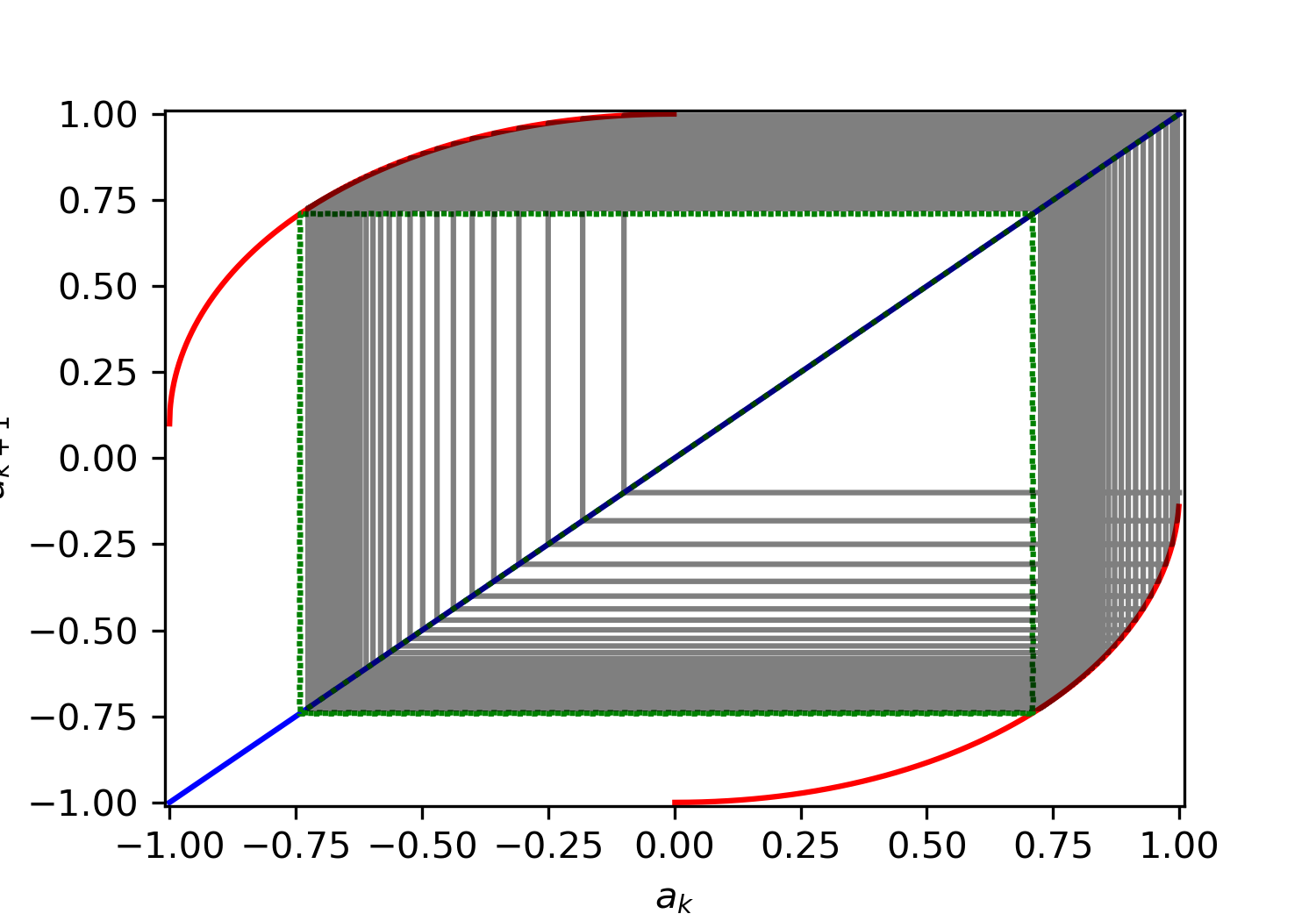}&
\includegraphics[width=.45\textwidth,trim = {40 29 0 30},clip]{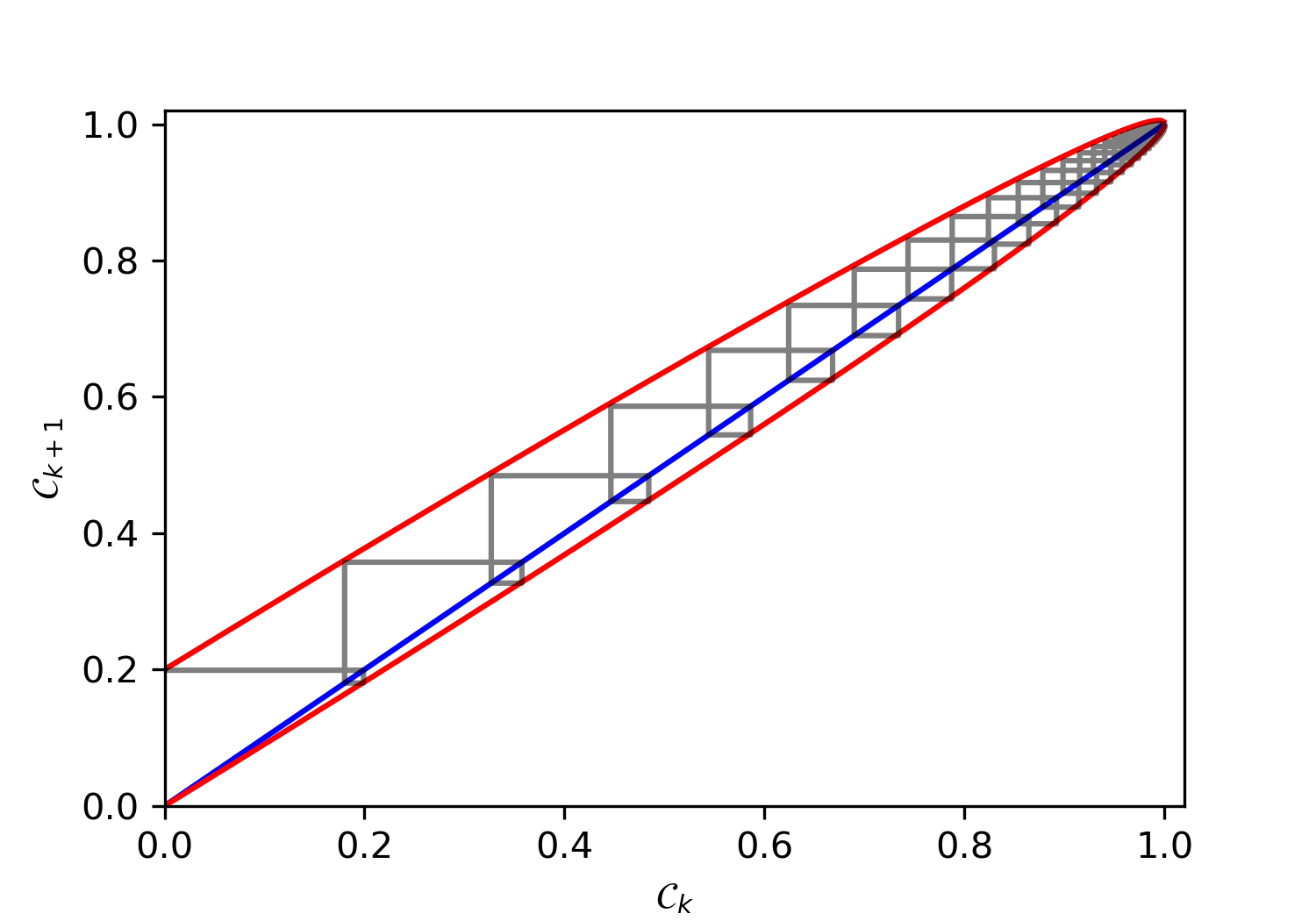} \\
\includegraphics[width=.45\textwidth,trim = {40 29 0 30},clip]{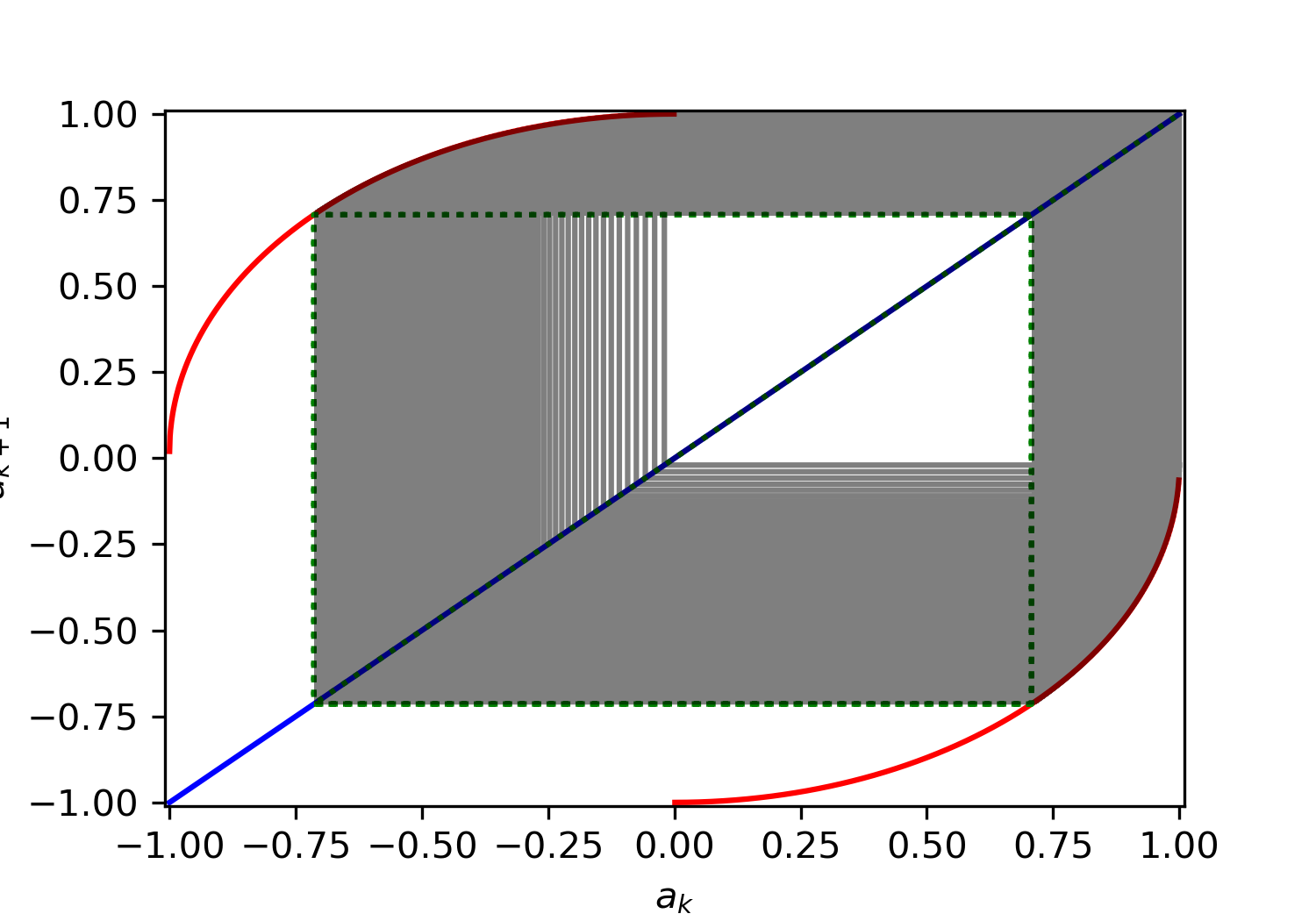} &
\includegraphics[width=.45\textwidth,trim = {40 29 0 30},clip]{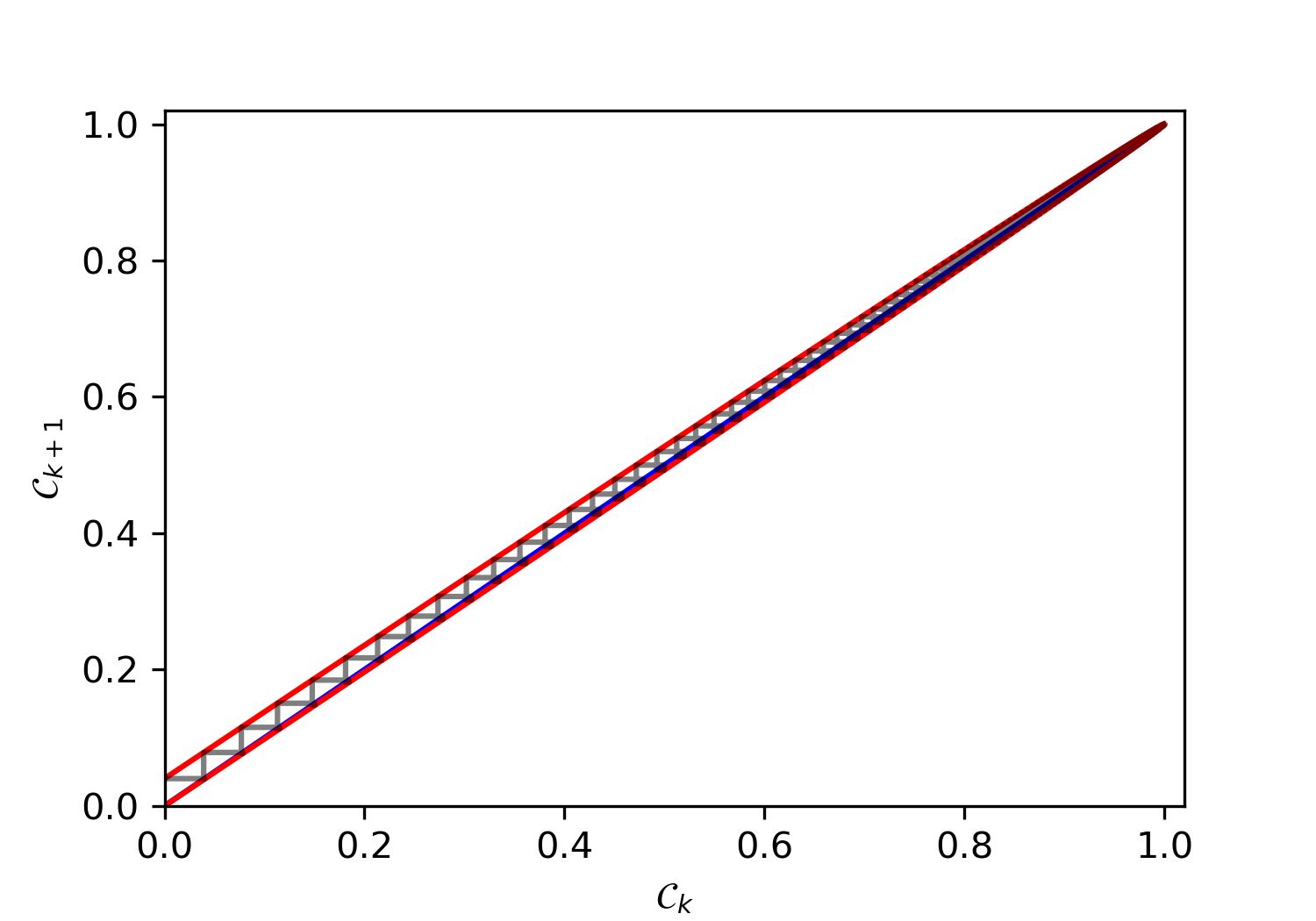}
\end{tabular} \\ \hspace{0.75cm}
\begin{picture}(0,0)(-4,0)
\put(-223,270){(a)}
\put(10,270){(b)}
\put(-223,128){(c)}
\put(10,128){(d)}
\put(-240,278){1}
\put(-255,260){$\mathsf{a}_{k+1}$}
\put(-240,213){0}
\put(-243,148){-1}
\put(-240,136){1}
\put(-255,118){$\mathsf{a}_{k+1}$}
\put(-240,71){0}
\put(-243,6){-1}
\put(-7,278){1}
\put(-20,260){$\mathcal{C}_{k+1}$}
\put(-7,148){0}
\put(-7,136){1}
\put(-20,118){$\mathcal{C}_{k+1}$}
\put(-7,6){0}
\put(-235,-4){-1}
\put(-132.5,-4){0}
\put(-45,8){$\mathsf{a}_k$}
\put(-45,150){$\mathsf{a}_k$}
\put(-33,-4){1}
\put(1,-4){0}
\put(185,8){$\mathcal{C}_k$}
\put(185,150){$\mathcal{C}_k$}
\put(197,-4){1}
\end{picture} \vspace{3pt}
\caption{We generate cobweb plots for the mapping \eqref{homflipcycle}, expressed as one--dimensional mappings either in terms of the coefficient on $\ket{ee}$, i.e.~$\mathsf{a}_{k+1} = f(\mathsf{a}_k)$ (see (a) and (c), where $f(\mathsf{a}_k)$ is from \eqref{mapflip-ad}), or in terms of the concurrence, i.e.~$\mathcal{C}_{k+1} = g(\mathcal{C}_k)$ (see (b) and (d), where $g(\mathcal{C}_k)$ is \eqref{Ck1}). All of the plots shown are initialized at $\mathsf{a}_0 = 1$ (and therefore $\mathcal{C}_0 = 0$). We use $\epsilon = 0.1$ in plots (a) and (b); this is about the largest $\epsilon$ can get before our approximations to $O(\epsilon)$ fall apart entirely; they are included here because it is easier to visualize how the mapping works when simplified to this coarse--grained level. We reduce $\epsilon$ to $0.02$ in plots (c) and (d), in order to show how the plots scale into the regime where our scheme is actually intended to operate, and our approximations are more appropriate. The dotted green box in plots (a) and (c) show the Bell state to which the scheme converges, where $\mathsf{a}$ and $\mathsf{d}$ simply alternate between $1/\sqrt{2}$ and $-1/\sqrt{2}$ (the state there is always $\ket{\Phi^-}$, up to a global sign). }
\label{fig-cobweb}
\end{figure*}
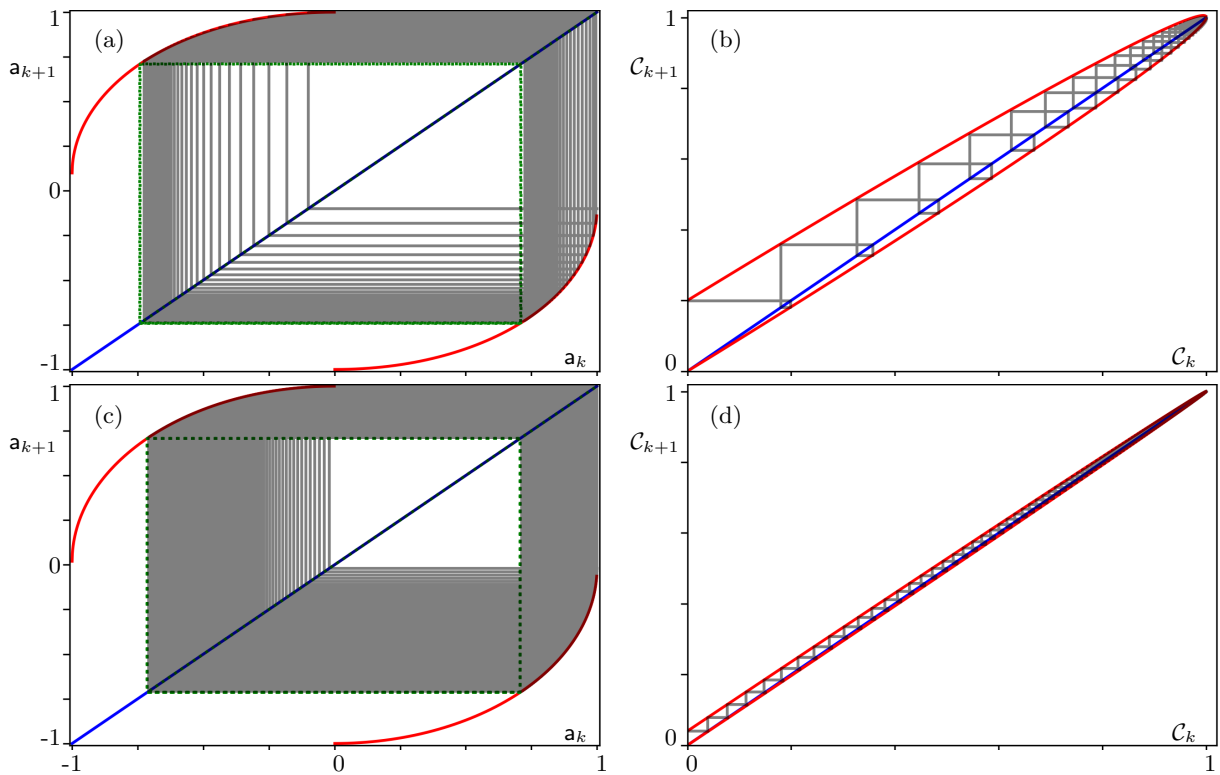

\par We are now in a position to formally consider our proposed modification, where we again interject fast flips $\hat{F}_{AB}$ of both qubits in between the measurements and Hamiltonian feedback just described.
In the photodetection case, we saw that the addition of operations $\hat{F}_{AB}$ allowed us to turn decay of the concurrence into a limit cycle in which successive measurements undid each other.
The idea now is similar: In order to stabilize the concurrence, we wish to trap the system in a limit cycle which alternates between the solution of growing concurrence and that of decaying concurrence \eqref{leighC}, instead of having the $|\mathsf{a}|<|\mathsf{d}|$ solution take over and eat away at the entanglement the moment we have generated a Bell state.
%{\color{blue}\sout{We will shortly be able to show that the behavior of our solution is effectively equivalent if we 1) interject flips $F_{AB}$ after every timestep of evolution under $\hat{\mathcal{U}}\hat{\mathcal{M}}_{\text{hom}}$, or 2) if we only interject $F_{AB}$ after every other timestep;
%while doing flipping operations twice as often may not be practical in a real device, the mathematics are simpler, so we will develop what follows from that perspective, bearing in mind that the effectiveness of our scheme is not appreciably affected either way.}[I find this paragraph rather confusing and not very useful for the following. In addition we do not come back to it later as promised. Why not saying that pulses are applied every step from the beginning?]}

\par Interjecting a flipping operation between every detector timestep (including the measurement and the immediate feedback \eqref{homfeedU}) may be described by the state update
\be \label{homflipcycle}
\ket{\psi(t+\Delta t)} = \frac{\hat{F}_{AB} \hat{\mathcal{U}} \hat{\mathcal{M}}_{\text{hom}} \ket{\psi(t)}}{|\hat{F}_{AB} \hat{\mathcal{U}} \hat{\mathcal{M}}_{\text{hom}} \ket{\psi(t)}|},
\ee
and we will assume $\ket{\psi}$ is of the form $\mathsf{a}\ket{ee} + \mathsf{d}\ket{gg}$, where $\mathsf{a}$ and $\mathsf{d}$ are assumed to be real \emph{and} to have opposite signs (as above).
The addition of $\hat{F}_{AB}$ interchanges the amplitudes on $\ket{ee}$ and $\ket{gg}$, such that we may make a slight modification to \eqref{map-ad}, which now reads
\be \label{mapflip-ad}
\mathsf{a}_{k+1} = \mathsf{d}_k- \epsilon \frac{\mathsf{a}_k^2}{\mathsf{a}_k-\mathsf{d}_k},\quad \mathsf{d}_{k+1} = \mathsf{a}_k+ \epsilon \frac{\mathsf{a}_k \mathsf{d}_k}{\mathsf{a}_k-\mathsf{d}_k}.
\ee
Equivalently, the flips result in alternation between the cases $|\mathsf{a}|>|\mathsf{d}|$ or $|\mathsf{a}|<|\mathsf{d}|$ every $\Delta t$, such that the concurrence will rise in one step, and then fall the next.
The concurrence is defined as $\mathcal{C}_k = - 2 \mathsf{a}_k \mathsf{d}_k$. 
Concatenating two steps of evolution in the concurrence allows us to quantify the net effect of our scheme.
We find that to $O(\epsilon)$, we have 
\be \label{Ck1}
\mathcal{C}_{k+1} = -2 \mathsf{a}_{k+1} \mathsf{d}_{k+1} = \mathcal{C}_k(1-\epsilon) + 2 \:\epsilon\: \mathsf{a}_k^2,
\ee
which may be repeated to find
\be \label{Ck2}
\mathcal{C}_{k+2} = \mathcal{C}_k - 2\: \epsilon \:\mathcal{C}_k +2 \:\epsilon.
\ee
The aggregate evolution across two cycles of this process is well--described by 
\be \label{diffC}
\dot{\mathcal{C}} \approx \frac{\mathcal{C}_{k+2} - \mathcal{C}_k}{2 \Delta t} \quad\rightarrow\quad \dot{\mathcal{C}} = \gamma (1-\mathcal{C}),
\ee
which should be understood as the average evolution across a rising and falling step. 
As the feedback here ensures near deterministic dynamics for small $\epsilon$, this average evolution can be taken as representative of the behavior of all trajectories. 
The solution to the continuous version of this equation, e.g.~for the \emph{least favorable} case $\mathcal{C}_0 = 0$ (no initial concurrence), is
\be \label{diffCsol}
\mathcal{C}(t) = 1 - e^{-\gamma t}.
\ee
The actual process matches this idealized solution up to small ``teeth'', reflecting the individual steps of alternating growth and decay for finite $\Delta t$. 
This is illustrated Fig.~\ref{fig-homfeed}(a); note that in simulation to generate this figure, we use the operator $F_{AB}$ after every \emph{other} application of $\hat{\mathcal{U}}\hat{\mathcal{M}}_{\text{hom}}$, rather than between every cycle of measurement and Hamiltonian feedback. 
Using the flips half as often doubles the size of the ``teeth'', but they remain bound about the idealized solution we have just derived\footnote{Strictly speaking, the flips can be spaced many more steps apart; this comes at the cost of increasing the size of the limit cycle about the Bell state, but with little other change to how our system functions. The effect of decreasing or increasing the frequency of flips in the photodetection case is similar.}.
%We do this following the same logic as in the photodetection case. %; there we applied flips after every other null measurement to make successive measurements undo each other. 
% Near $\ket{\Phi^\pm}$, there is no harm in applying flips every cycle rather than every other, but it is not strictly necessary either.
We have done our homodyne derivations above with the flips every cycle for mathematical simplicity.%, but performing flips half as often is adequate. 

\par Many of the properties of \eqref{diffC} are highly desirable.
First we see that the mapping of interest has a single \emph{stable} fixed point at $\mathcal{C} = 1$; this arises because solutions to \eqref{leighC} grow faster (when $\abs{\mathsf{a}}>\abs{\mathsf{d}}$) than they decay (when $\abs{\mathsf{d}}>\abs{\mathsf{a}}$) for $\mathcal{C} < 1$, such that the mapping \eqref{Ck2} \emph{always} yields a net gain in entanglement.
That net gain is greater when the entanglement is smaller.
Ideally, one does not begin to interject joint $\pi$-pulses $\hat{F}_{AB}$ while $|\mathsf{a}| > |\mathsf{d}|$, but rather waits for the Bell state to be created by the scheme of \cite{Leigh2019} alone, and only then turns on the extra controls (see Fig.~\ref{fig-homfeed}(b)). 
While the stability of our flipping scheme never allows a net decrease in concurrence, and can be used to generate entanglement, it truly excels at preserving concurrence after the Hamiltonian feedback has operated on its own to generate it. 
%The stability of our modified scheme means that it is ultimately quite robust against errors in timing the start of flipping operations; indeed, the \emph{worst case} solution \eqref{diffCsol} we have for high--efficiency measurements and high--fidelity feedback operations asymptotically approaches $\mathcal{C} =1$ for long times. 
%See Fig.~\ref{fig-homfeed}(a) for a direct comparison and further comments.
The use of a finite time-step means that the Hamiltonian portion of the feedback \eqref{homfeedU} from \cite{Leigh2019} does not operate perfectly, and small deviations from deterministic dynamics occur; however the scheme is still stable, as evidenced by the numerical simulations in Fig.~\ref{fig-homfeed}.
All of the properties of the discrete mappings incorporating our flipping operations can be visualized in the cobweb plots Fig.~\ref{fig-cobweb}. 
These require that we recast our equations into one--dimensional mappings, which can be obtained from \eqref{mapflip-ad} and \eqref{Ck1} using the substitutions $\mathsf{d}_k \rightarrow - \text{sgn}(\mathsf{a}_k) \sqrt{1-\mathsf{a}_k^2}$, or $\mathsf{a}_k^2 \rightarrow \tfrac{1}{2} \pm \tfrac{1}{2}\sqrt{1-\mathcal{C}_k^2}$, respectively; the operation $\hat{F}_{AB}$ in each cycle causes the sign in the latter expression to alternate with every iteration, which is effectively averaged over in obtaining \eqref{Ck2}.

It is possible to recast the derivation above in terms of a different parameterization of the two--qubit state. 
Let us define $(\mathsf{a}, \mathsf{d}) = (\cos \theta, -\sin \theta)$, with $\theta \in [0,\pi/2]$. 
In the case of continuous feedback only, we find the equation for $\theta$ given by,
\begin{equation}
{\dot \theta} = \gamma \frac{\cos \theta}{\cos \theta + \sin \theta}.    
\end{equation}
Starting at $\theta=0$, this equation has a solution of 
\begin{equation}
    e^{-\theta} \cos \theta = e^{-\gamma t},
\end{equation}
which is transcendental.
In the case of adding the fast $\pi$-pulses, we find the equation for $\theta$ given by,
\begin{equation}
{\dot \theta} = \gamma \frac{\cos \theta - \sin \theta}{\cos \theta + \sin \theta}.    
\end{equation}
This equation has a solution
\begin{equation}
    \cos \theta - \sin \theta = e^{-\gamma t/2},
\end{equation}
which can equivalently be expressed by
\begin{equation}
\cos^2 \theta = \frac{1}{2} + \frac{1}{2} \sqrt{1 - (1 - e^{-\gamma t})^2},    
\end{equation}
with $\sin^2\theta = 1 - \cos^2 \theta$, and consistent with the statement $\mathsf{a}_k^2 \rightarrow \tfrac{1}{2} \pm \tfrac{1}{2}\sqrt{1-\mathcal{C}_k^2}$ in conjunction with the solution \eqref{diffCsol}.

\par We briefly summarize what has been presented so far: We have demonstrated that feedback based on qubit flips, %\blue{\sout{, as in similar BB control protocols,}} 
and utilized in conjunction with measurements of qubits' spontaneous emission, is able to protect the qubits' concurrence against the monitored $T_1$ decay processes.
The regime in which we operate is one where the measurement intervals (detector integration intervals) are much shorter (perhaps 2 orders of magnitude smaller) than the $T_1$ time of the qubits, and the qubit flips are executed at least one order of magnitude faster than that.
For example, in superconducting qubits, $T_1 \approx 50\,\mu\mathrm{s}$, $\Delta t$ can be as short as $20\,\mathrm{ns}$, while $t_\pi \approx 5\,\mathrm{ns}$.
We have shown that fast $\pi$--pulses form the basis of a good control strategy for entanglement preservation in such scenarios, either in conjunction with photodetection, or as a supplement to existing Hamiltonian feedback \cite{Leigh2019} based on homodyne detection instead; in either case, the addition of fast BB--like $\pi$--pulses allows us to trap the two--qubit dynamics in an arbitrarily small limit cycle about a fixed point at a Bell state.

%%%%%%%%%%%%%%%%%%%%%%%%%%%%%%%%%%%%%%%%%%%%%%%%%%%%%%%%%%%%%%%%%%%%%%%%%%%%%%%%%%%%%%%%%%%%%%%%%%%
%%%%%%%%%%%%%%%%%%%%%%%%%%%%%%%%%%%%%%%%%%%%%%%%%%%%%%%%%%%%%%%%%%%%%%%%%%%%%%%%%%%%%%%%%%%%%%%%%%%
%%%%%%%%%%%%%%%%%%%%%%%%%%%%%%%%%%%%%%%%%%%%%%%%%%%%%%%%%%%%%%%%%%%%%%%%%%%%%%%%%%%%%%%%%%%%%%%%%%%
\section{Impact of Measurement Inefficiency \label{sec-etaimpact}}

Our discussion so far has focused on establishing the utility and dynamical properties of our proposed scheme with an ideal apparatus. 
Several of the assumptions implicit in the idealized analysis are however never fully achieved in practice. For example, it is difficult to make measurements with near--unit efficiency, to implement feedback operations without some processing delay time, and to implement feedback operations with perfect fidelity. 
Any of these factors should be expected to degrade the performance of any feedback control protocol relative to the ideal case. 
We will here focus on analyzing the impact of of measurement inefficiency. 
Including finite detector efficiency generically introduces mixed states as some of the signal is lost, increasing the complexity of the equations describing the state evolution.
As such, our program now is to study the inefficient case, for both the photodetection-- and homodyne--based schemes discussed above, using numerical simulation. 
Our aim here is not to find the best possible modification to our feedback scheme for the more realistic case of inefficient measurements, but simply to quantify the effect of inefficiency on the simple $\pi$--pulse--based strategies we have proposed above.

\par Measurement inefficiency may be modeled by using an ideal detector, but with a lossy channel in front of it. In other words, it is possible to model measurement inefficiency by introducing some finite probability that photons arriving at the ideal detector are instead diverted into some lost channel. 
This is illustrated in Fig.~\ref{fig-exp} by the unbalanced (purple) beam-splitters in channels 3 and 4, which allow photons to transmit to the detector with probability $\eta_3$ or $\eta_4$, but otherwise reflect them into a channel in which they are irretrievably lost.
We briefly review the formal model of such a situation to Appendix~\ref{sec-kraus-review}, and discuss it in much greater detail in Refs.~\cite{FlorTeach2019, LongFlor2019}.
The ideal case we treated above is that for which $\eta_3 = 1 = \eta_4$, and we are now generalizing to the case where we allow $\eta_3 < 1$ and $\eta_4 < 1$. 

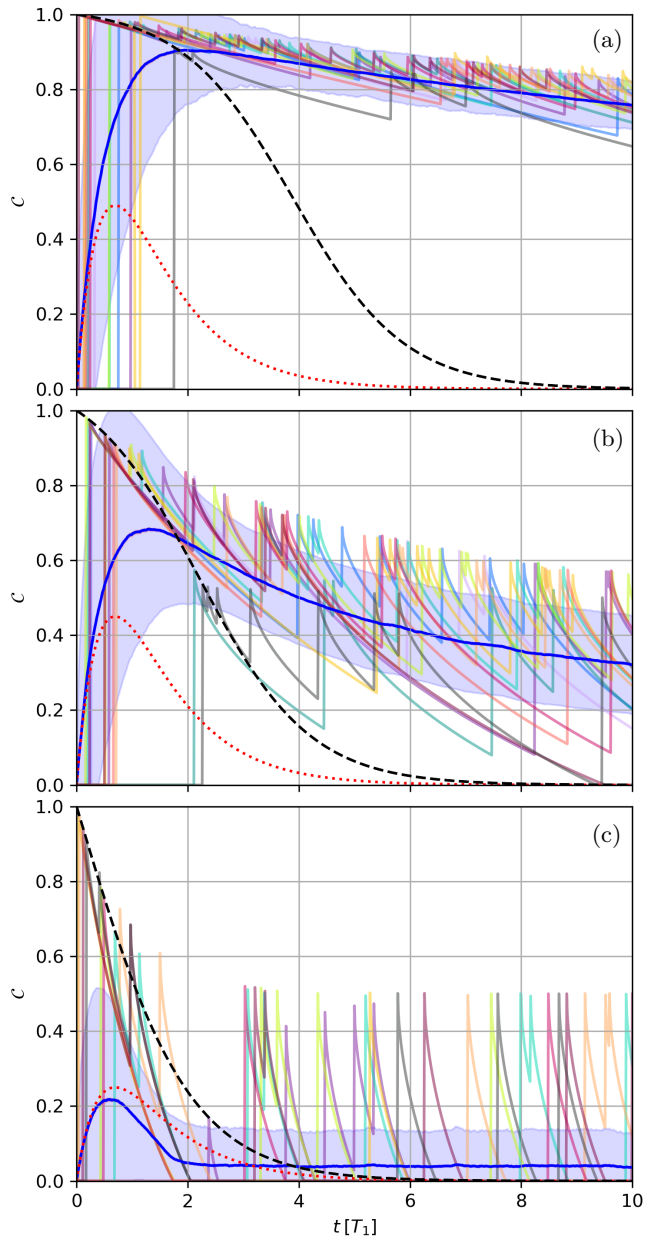
\begin{figure}
    \centering
    \includegraphics[width = .99\columnwidth,trim = {10 28 30 28pt},clip]{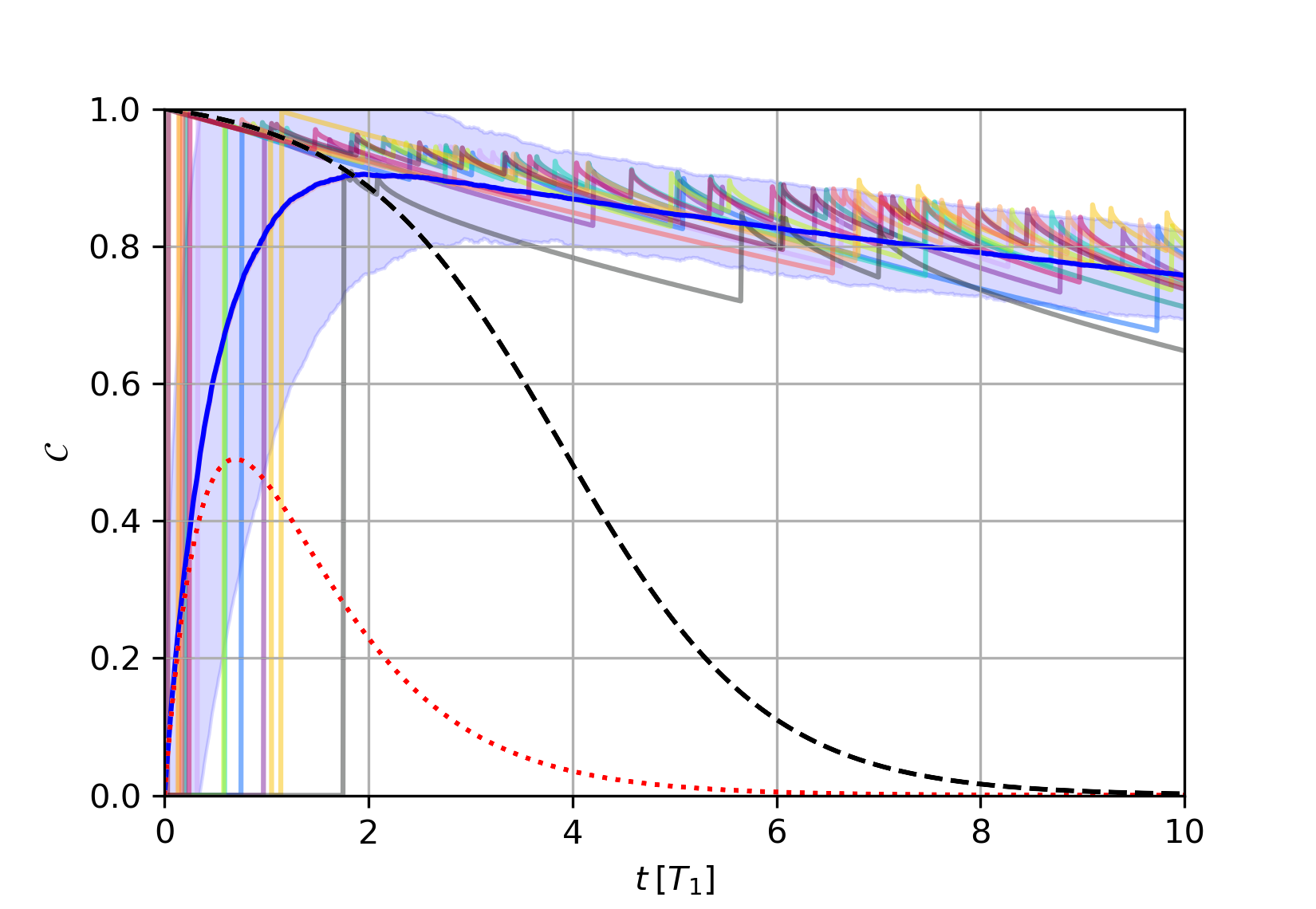} \\
    \includegraphics[width = .99\columnwidth,trim = {10 28 30 28pt},clip]{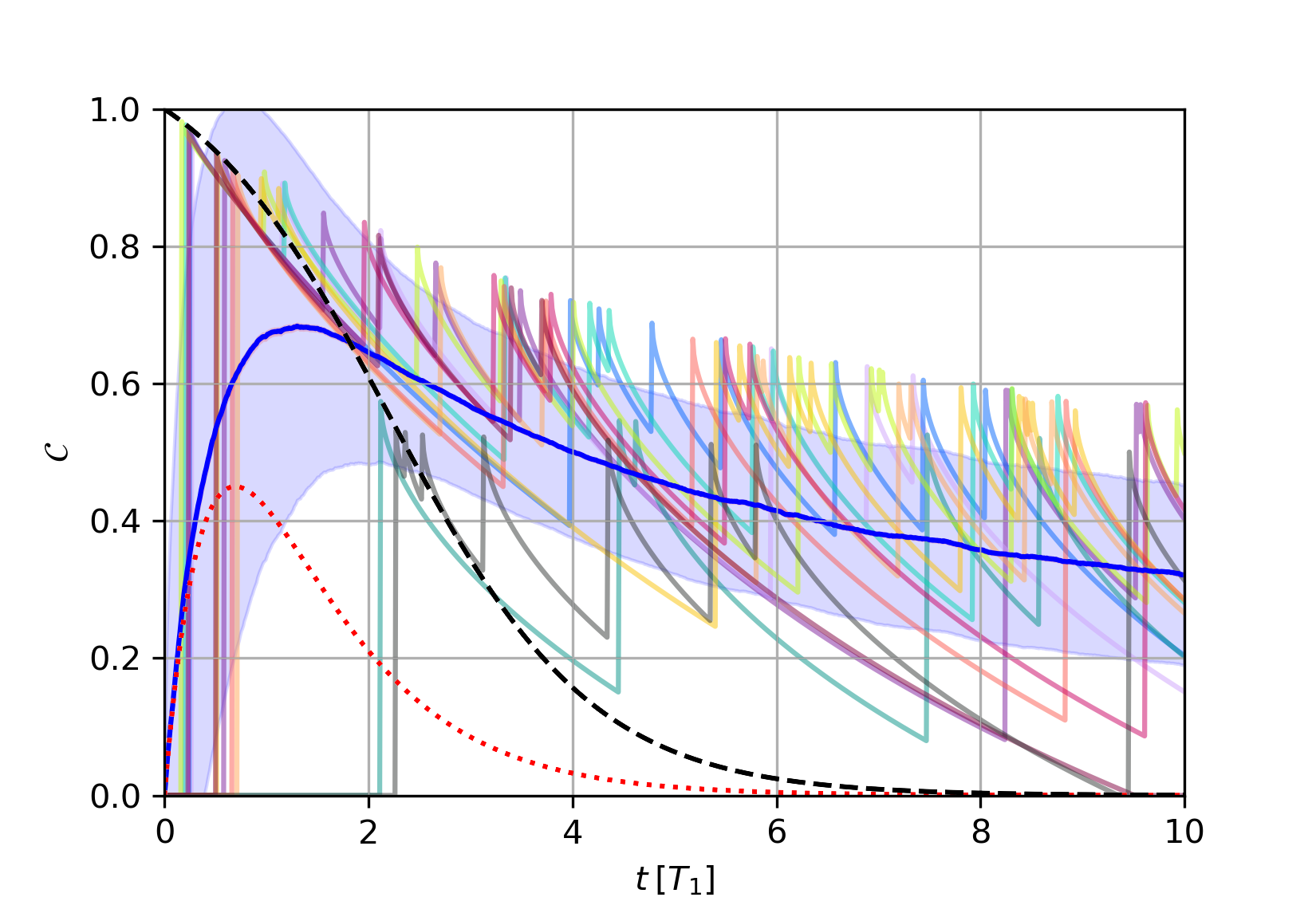} \\
    \includegraphics[width = .99\columnwidth,trim = {10 0 30 28pt},clip]{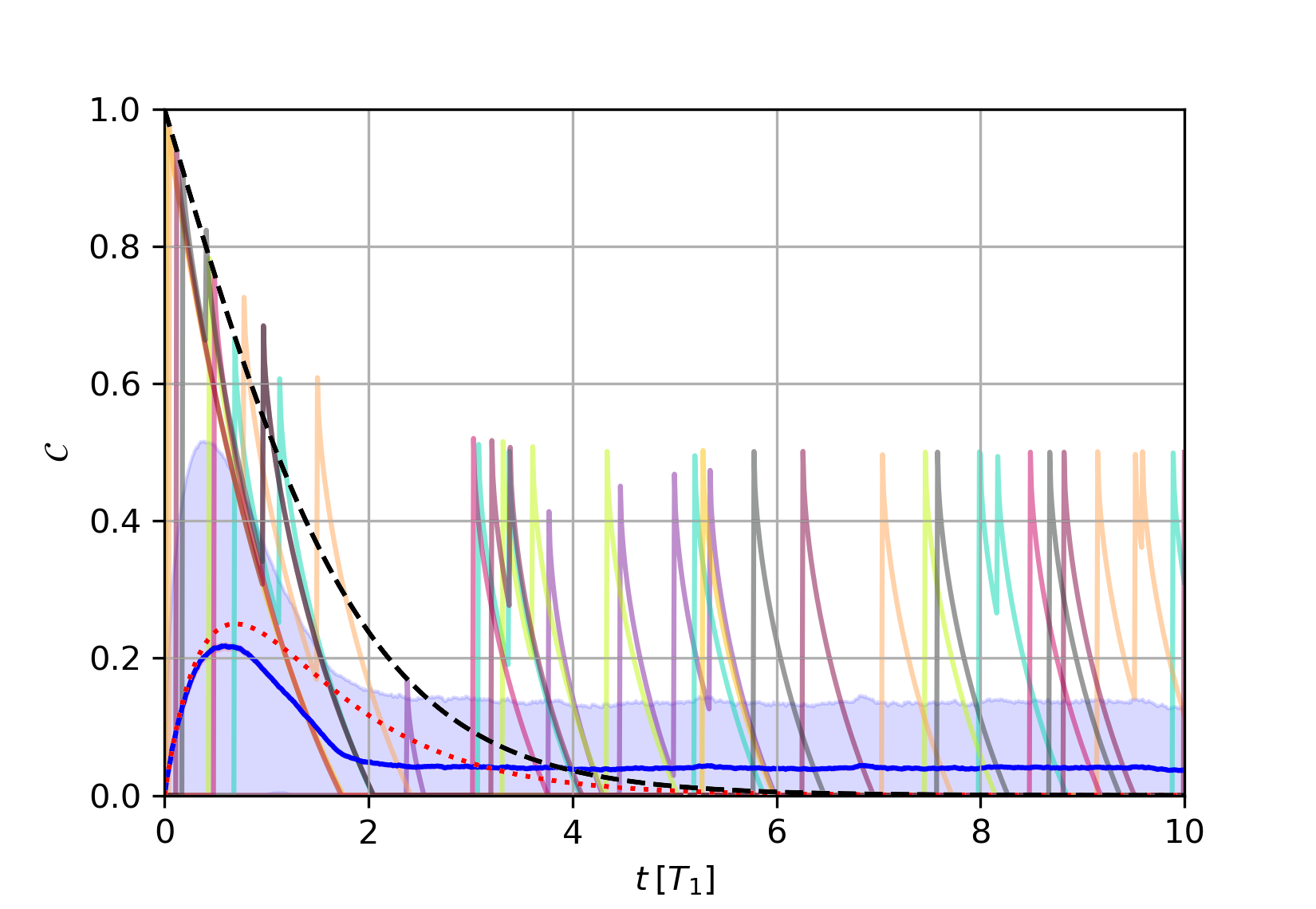} \\
    \begin{picture}(0,0)
    \put(100,463){(a)}
    \put(100,312){(b)}
    \put(100,162){(c)}
    \end{picture}\vspace{-20pt}
    \caption{We plot the evolution of the concurrence $\mathcal{C}$ for trajectories arising from inefficient photodetection, and including BB--like feedback and measurement reversal as described in the main text; these plots should be compared with Fig.~\ref{fig-ConcIdealFB}, which illustrates the corresponding process under ideal circumstances. We here use symmetric ($\eta_3 = \eta = \eta_4$) measurement efficiencies $\eta = 0.98$ (a), $\eta = 0.90$ (b), and $\eta = 0.50$ (c). We see that for measurement efficiencies close to the ideal, e.g.~as in (a) and (b), the average concurrence with feedback always exceeds that without (well approximated by $\bar{\mathcal{C}}(t) = 2 \eta e^{-\gamma t}(1-e^{-\gamma t})$ \cite{LongFlor2019}, shown in dotted red). Even in (c), where this is no longer true, the ability to maintain any concurrence at long times is still advantageous compared with doing nothing. The upper bound \eqref{CmaxPDeta} on the concurrence derived in \cite{LongFlor2019} and shown in dashed black, for the case without feedback, shows the the extent to which degradation in the measurement efficiency affects the ability to generate entanglement to begin with, and provides another useful reference against which our feedback may be compared.}
    \label{fig-PDetaFB}
\end{figure}

%%%%%%%%%%%%%%%%%%%%%%%%%%%%%%%%%%%%%%%%%%%%%%%%%%%%%%%%%%%%%%%%%%%%%%%%%%%%%%%%%%%%%%%%%%%%%%%%%%%
\subsection{Inefficient Photodetection}

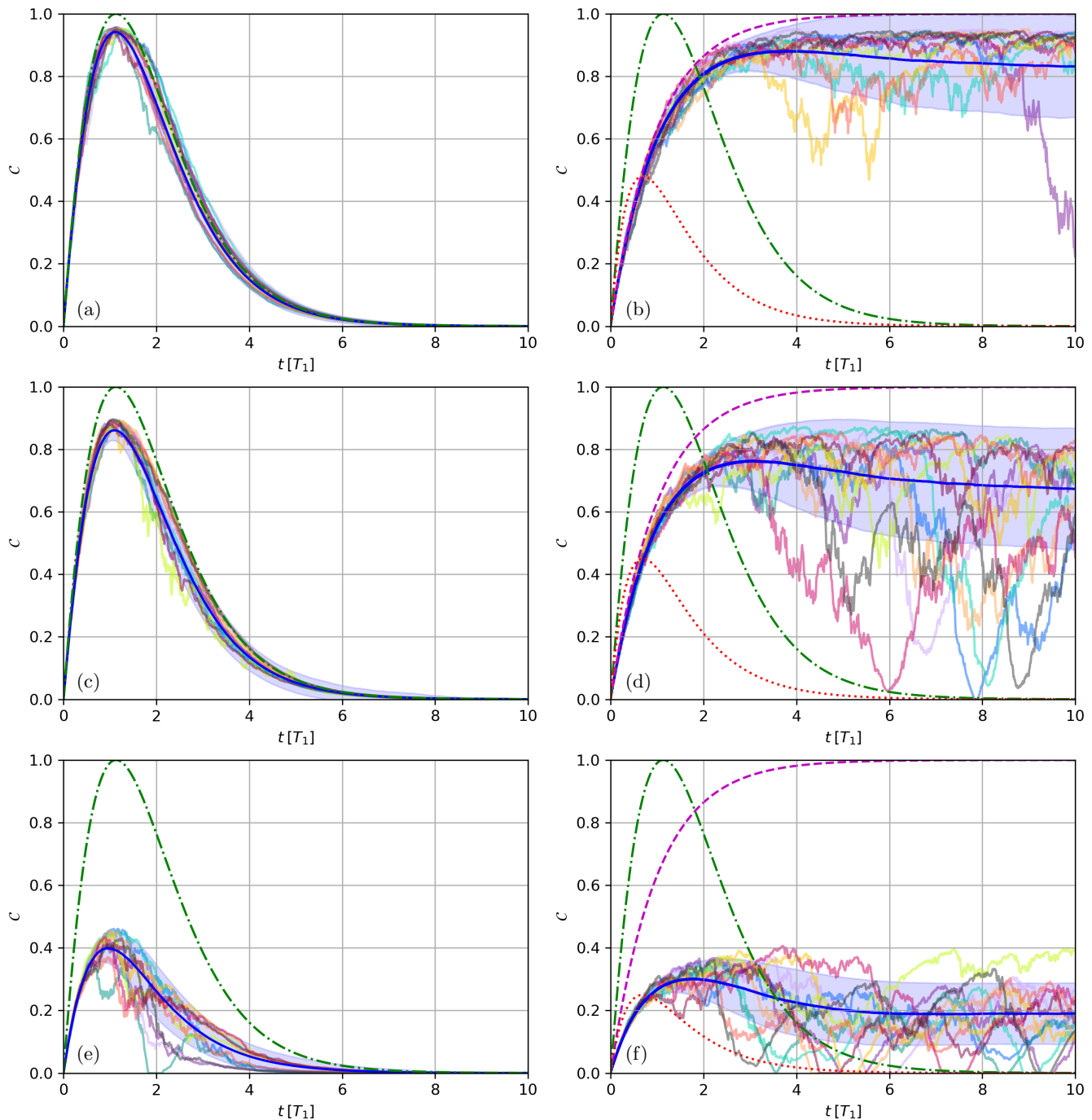
\begin{figure*}
\begin{tabular}{cc}
\includegraphics[width = .49\textwidth,trim = {10 3 30 28}, clip]{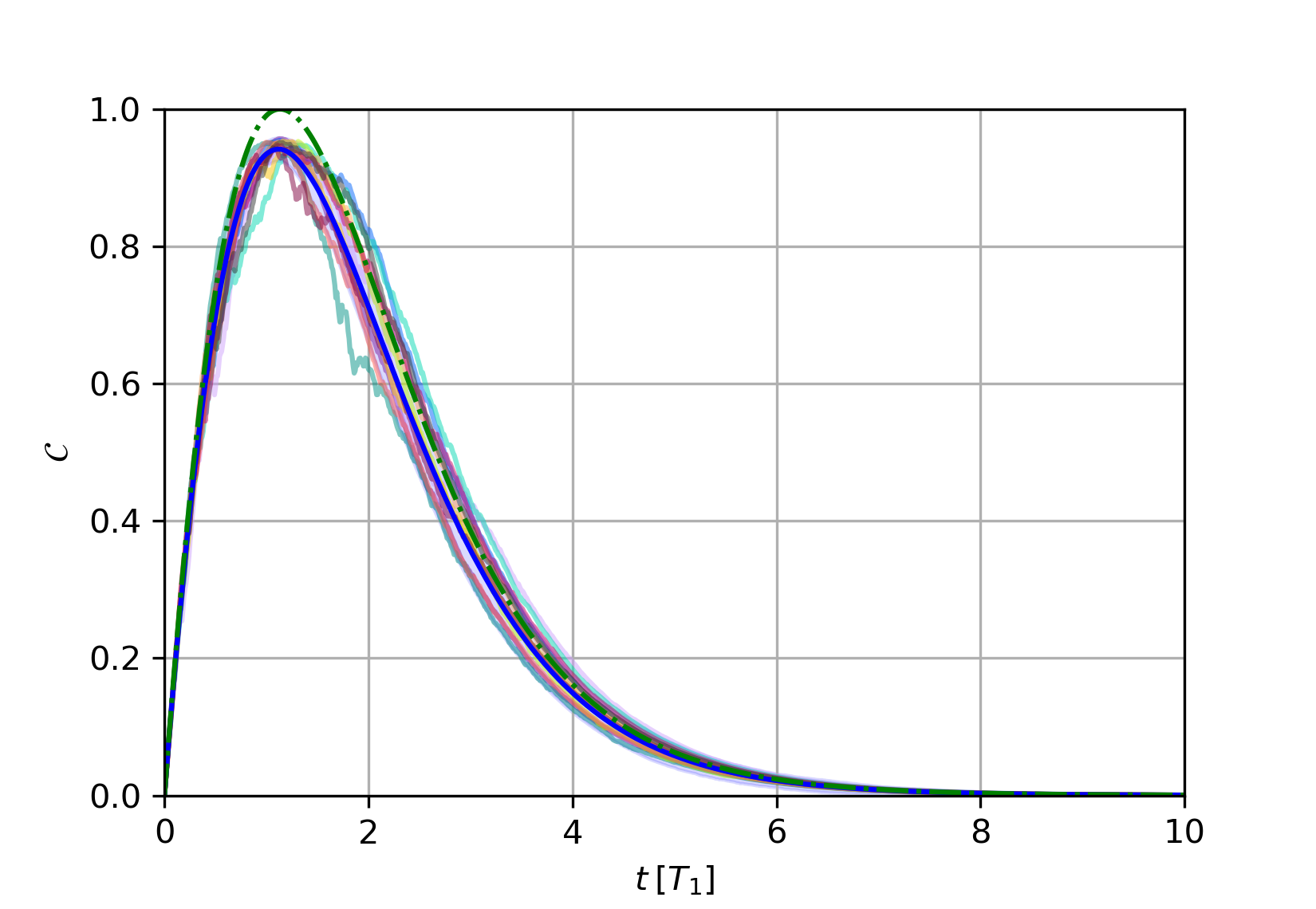} &
\includegraphics[width = .49\textwidth,trim = {10 3 30 28}, clip]{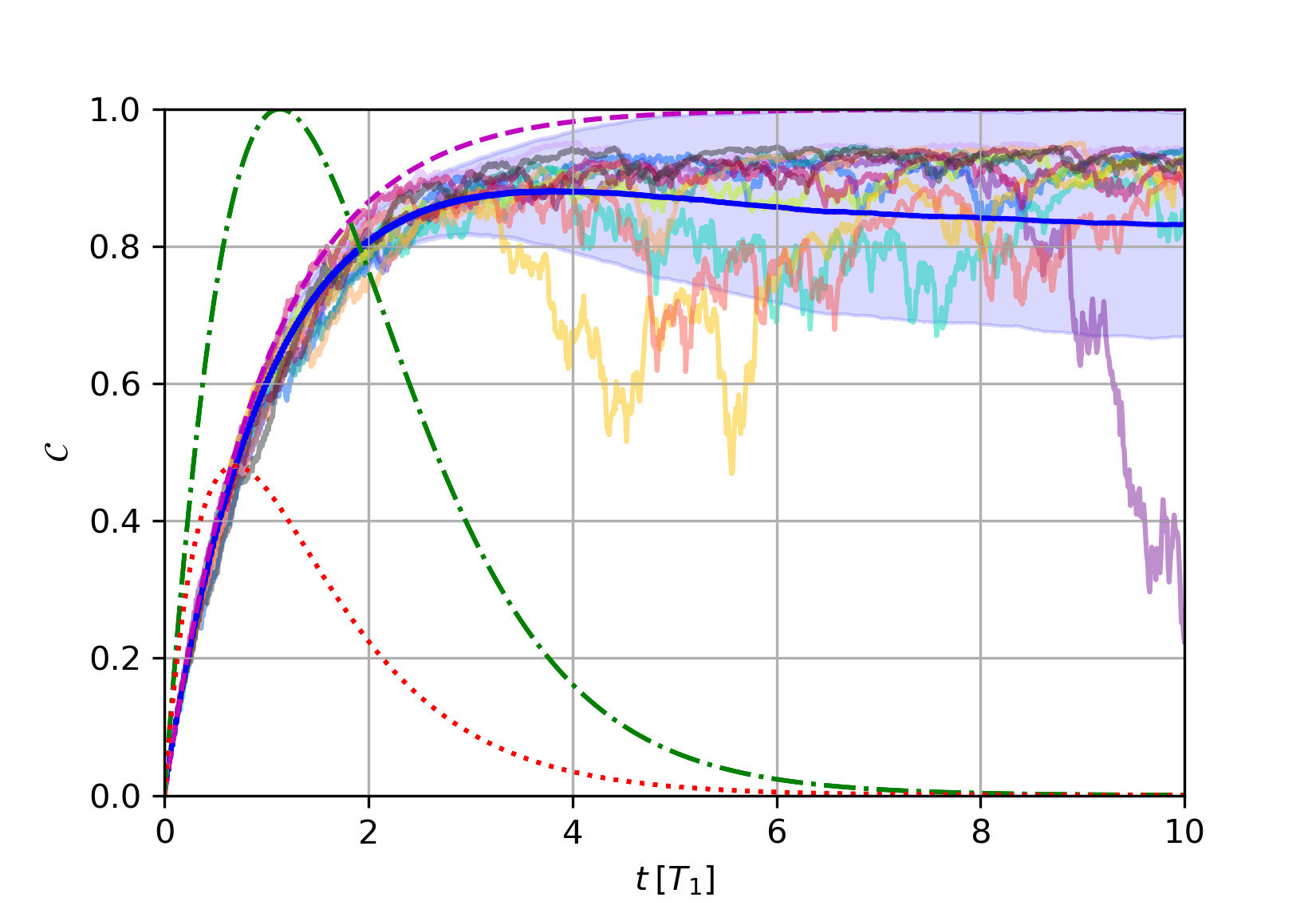} \\
\includegraphics[width = .49\textwidth,trim = {10 3 30 28}, clip]{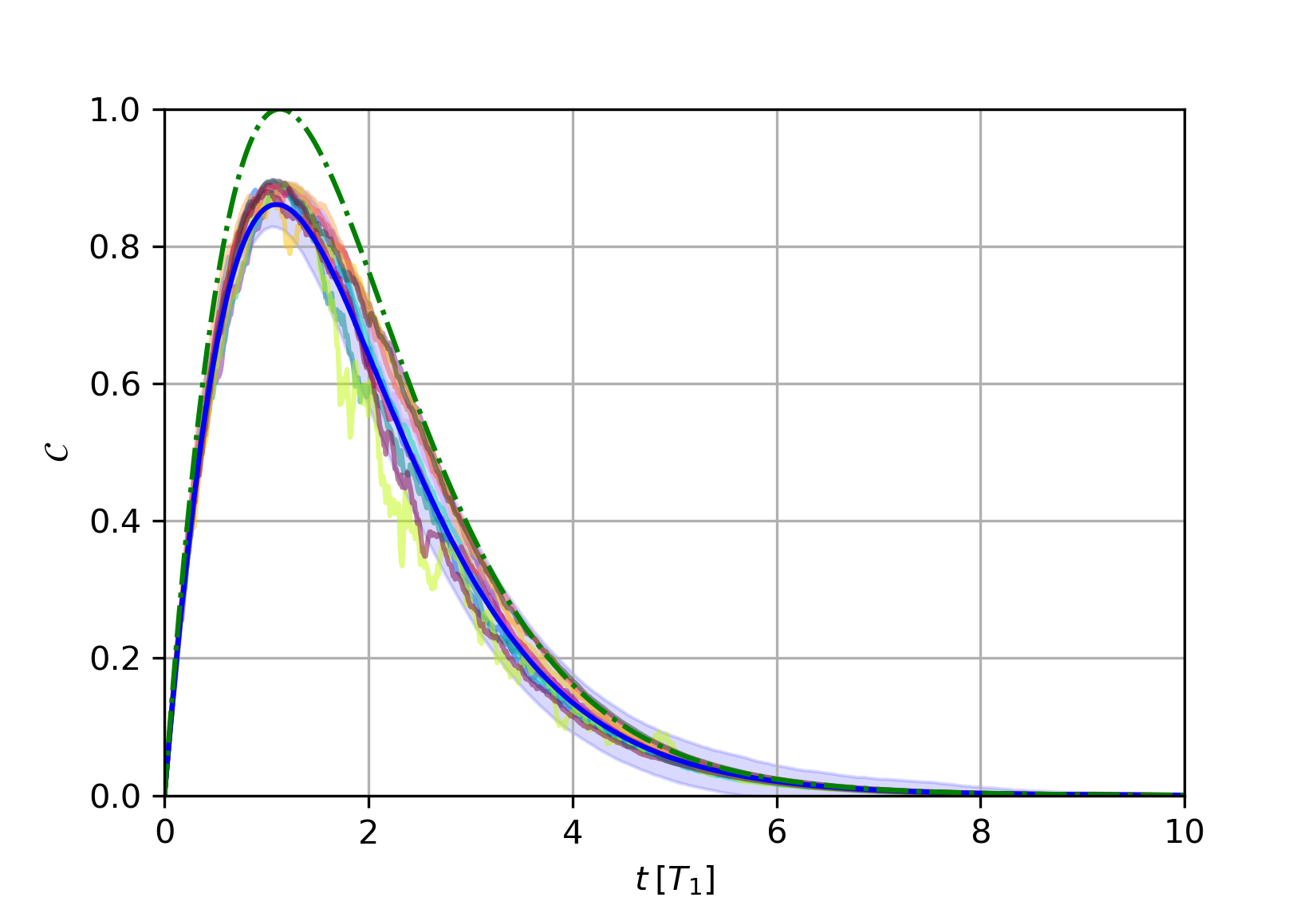} &
\includegraphics[width = .49\textwidth,trim = {10 3 30 28}, clip]{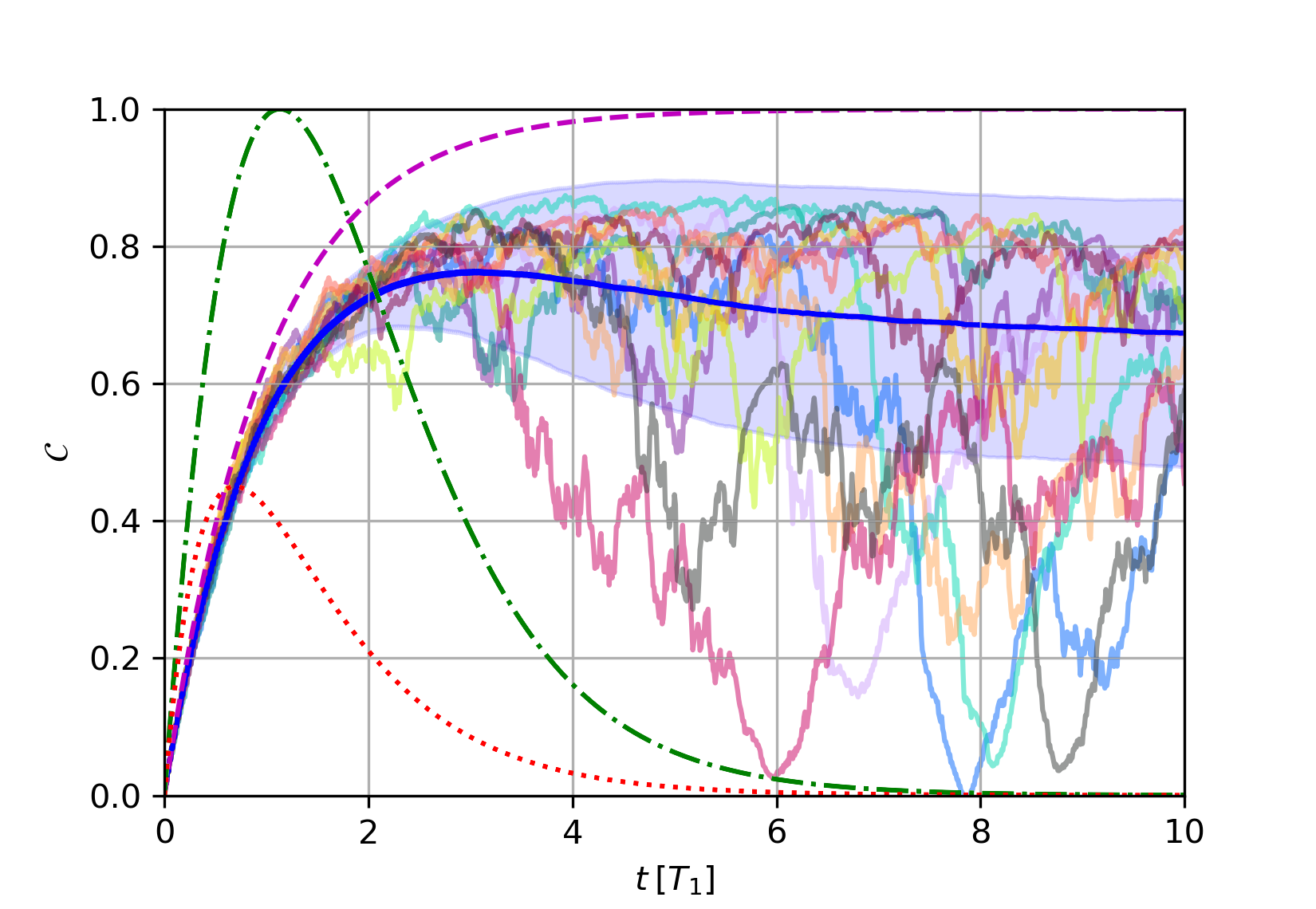} \\
\includegraphics[width = .49\textwidth,trim = {10 3 30 28}, clip]{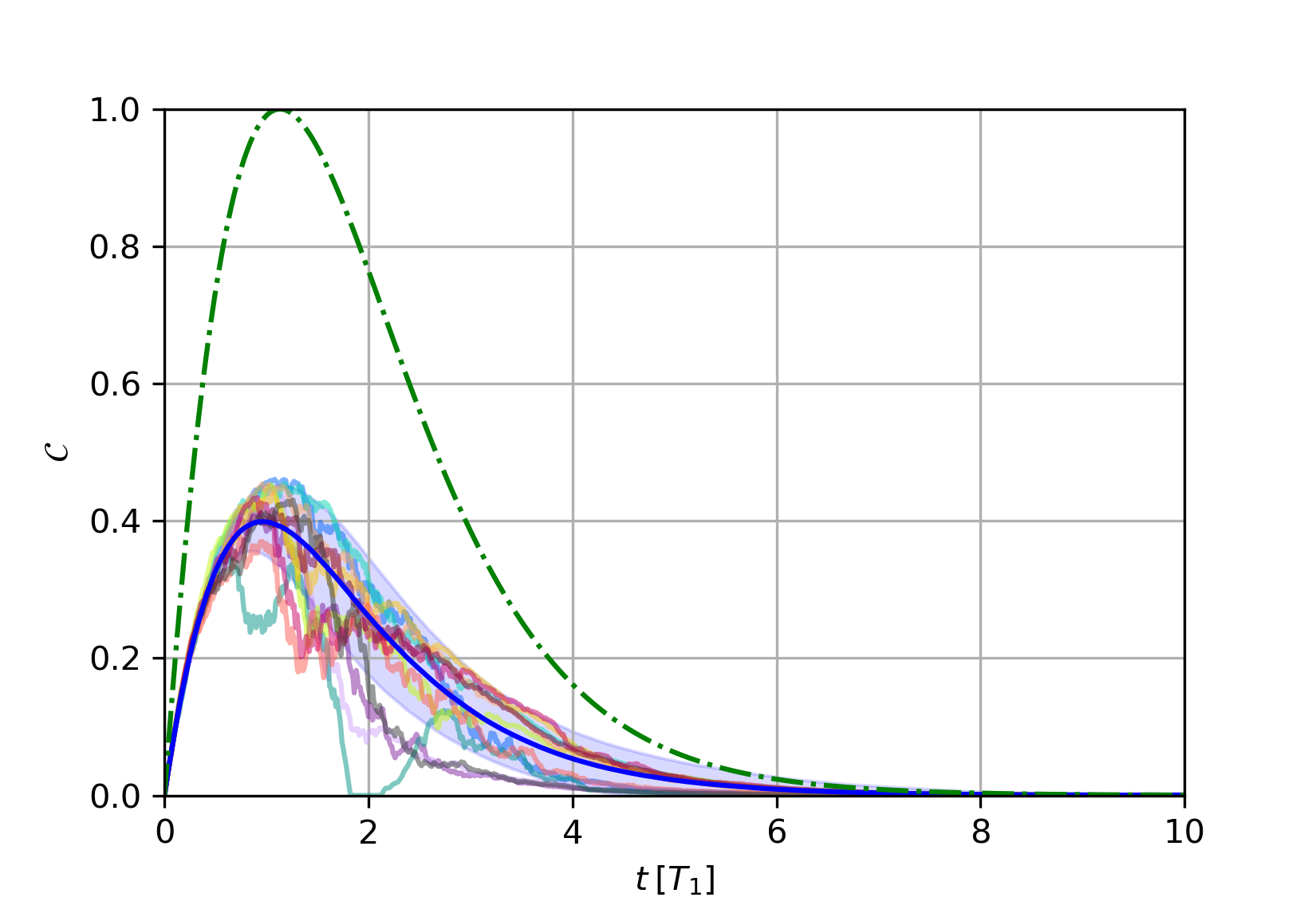} &
\includegraphics[width = .49\textwidth,trim = {10 3 30 28}, clip]{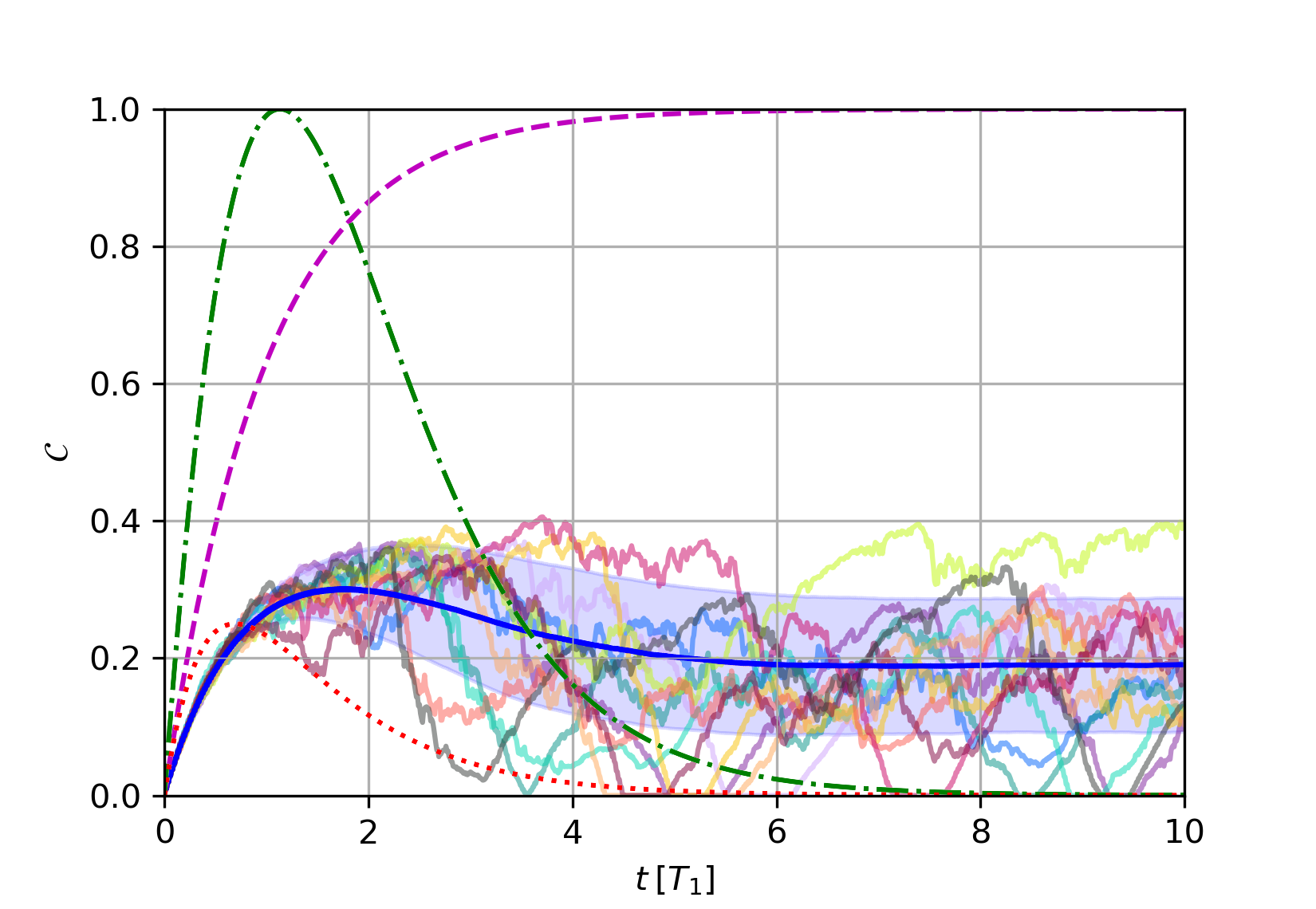}
\end{tabular} \\ 
\begin{picture}(0,0)
\put(-217,378){(a)}
\put(38,378){(b)}
\put(-217,205){(c)}
\put(38,205){(d)}
\put(-217,32){(e)}
\put(38,32){(f)}
\end{picture} \vspace{-10pt}
\caption{We simulate inefficient \blue{homodyne} measurements with the feedback process \eqref{homfeedU-eta}, both alone (a,c,e), and with added $\pi$-pulses on both qubits every $\Delta t$ (b,d,f). We use $\Delta t = 0.01 T_1$ in all cases. The measurement inefficiencies are symmetric ($\eta_3 = \eta = \eta_4)$, and are $\eta = 98\%$ (a,b), $\eta = 95\%$ (c,d), and $\eta = 75\%$ (e,f). The ability of this homodyne measurement to generate any entanglement at all is contingent on having $\eta > 50\%$ \cite{LongFlor2019,Leigh2019}. Below $\eta = 50\%$, no feedback based on LOCC can remedy the fact that measurement is incapable of generating entanglement. We see the pronounced degrading effect of the measurement inefficiency on both feedback schemes, and that the quasi--deterministic dynamics of the ideal case (see Fig.~\ref{fig-homfeed}) are lost. The curves for the ideal case without $\pi$-pulses (dash--dotted green), and with flips (dashed magenta) are shown for reference. We additionally show curves representing the average concurrence from the case without any feedback in dotted red (which follow $\bar{\mathcal{C}}(t) = 2(2\eta-1)e^{-\gamma t}(1-e^{-\gamma t})$, \cite{LongFlor2019}). By comparing the average concurrence from the present simulation (solid blue) to these other references we see that our modified scheme outperforms both the no--feedback average for the comparable efficiency (dotted red), and the \emph{ideal} Hamiltonian feedback without the extra flips we have introduced (dash--dotted green), after longer evolution times $t \agt 3T_1$. } 
\label{fig-HomEtaFB}
\end{figure*}

\par We begin with inefficient photodetection; simulations of our feedback scheme with symmetric ($\eta_3 = \eta = \eta_4$) and less than ideal $\eta<1$ photon counting measurements, and subsequent feedback, are shown in Fig.~\ref{fig-PDetaFB}. 
We find that, without additional modifications to our feedback scheme, the addition of measurement inefficiency leads to substantial degradation of the preserved concurrence. 
This is not especially surprising, since the maximum concurrence achievable by the bare measurement before feedback is bounded by a decaying solution \cite{LongFlor2019}
\be \label{CmaxPDeta}
\mathcal{C}_{max}^\eta(t) = \frac{1}{(1-\eta) e^{\gamma t} + \eta},
\ee
where $\eta_3 = \eta = \eta_4$.
In the long time limit, our modified scheme does still achieve some steady--state concurrence, which is still an advantage over the case without feedback, in the longer--time limit.
It is possible that a more complex feedback protocol may be able to further mitigate the undesirable effects of measurement inefficiency, but ultimately, if too much information is lost to the environment without being measured, other schemes which demand additional resources (e.g.~extra long-lived energy levels) for storing entanglement \cite{BarrettKok, Santos2012, Leigh2019} are likely to be more successful. 
As our scheme does not use e.g.~additional transitions to effectively turn off the decay interaction with the environment after it has allowed us to generate entanglement, it is most effective when that lone transition is monitored efficiently.

%%%%%%%%%%%%%%%%%%%%%%%%%%%%%%%%%%%%%%%%%%%%%%%%%%%%%%%%%%%%%%%%%%%%%%%%%%%%%%%%%%%%%%%%%%%%%%%%%%%
\subsection{Inefficient Homodyne Detection}

We may perform the comparable test for the homodyne--based variant on the scheme of \cite{Leigh2019}. The only modification we make to the operator \eqref{homfeedU}, which was optimal in the ideal case, is to scale the readouts by a factor $\sqrt{\eta}$, such that $\hat{\mathcal{U}}_\eta = $
\be \label{homfeedU-eta} 
e^{\displaystyle \tfrac{ i \:\Delta t\: \sqrt{\rho_{00}}}{\sqrt{\rho_{00}}+\sqrt{\rho_{33}}} \sqrt{\tfrac{\gamma}{2}} \left(\sqrt{\eta_3} \,r_3 (\hat{\sigma}_y^A+\hat{\sigma}_y^B) + \sqrt{\eta_4}\, r_4(\hat{\sigma}_x^B-\hat{\sigma}_x^A) \right)},
\ee
where $\rho_{00} = \bra{ee}\rho\ket{ee} = {\color{coquelicot} \blacktriangle}$ and $\rho_{33} = \bra{gg} \rho \ket{gg} = {\color{electricviolet} \blacktriangledown} $ (see \eqref{colorrho} regarding notation).  
We have shown elsewhere \cite{LongFlor2019} that the homodyne measurement under consideration (without feedback) is unable to generate entanglement for $\eta \leq 50\%$. 
Since local unitary operations cannot change the concurrence of the two--qubit state, it not possible for any local feedback protocol to remedy this.
In Fig.~\ref{fig-HomEtaFB}, we simulate the effect of measurement and feedback \eqref{homfeedU-eta} for efficiencies (with $\eta_3 = \eta = \eta_4$) $\eta = 98\%$, $\eta = 95\%$, and $\eta = 75\%$, both without and then with the interjection of qubit flips, as in previous sections. 
We use $\Delta t = 0.01 T_1$ in all instances there.
In broad strokes, we see that the quasi--deterministic nature of the dynamics we had in the ideal case is eroded by the measurement inefficiency. The average entanglement yield suffers from this as expected (consistent with Martin and Whaley's results \cite{Leigh2019}).
The stability of the scheme, at the level of individual trajectories, is quite adversely affected by the measurement inefficiency and the return of some stochasticity to the dynamics.
We do see however, that the effect of our qubit flips on the concurrence is still a net positive at longer times, allowing us to stabilize a large fraction of the entanglement generated by the measurement, on average. 

%%%%%%%%%%%%%%%%%%%%%%%%%%%%%%%%%%%%%%%%%%%%%%%%%%%%%%%%%%%%%%%%%%%%%%%%%%%%%%%%%%%%%%%%%%%%%%%%%%%
%%%%%%%%%%%%%%%%%%%%%%%%%%%%%%%%%%%%%%%%%%%%%%%%%%%%%%%%%%%%%%%%%%%%%%%%%%%%%%%%%%%%%%%%%%%%%%%%%%%
%%%%%%%%%%%%%%%%%%%%%%%%%%%%%%%%%%%%%%%%%%%%%%%%%%%%%%%%%%%%%%%%%%%%%%%%%%%%%%%%%%%%%%%%%%%%%%%%%%%
\section{Discussion \label{sec-conclusions}}

We have proposed a pair of feedback protocols which involve interjecting $\pi$--pulses between measurements (or supplementing an existing feedback control protocol \cite{Leigh2019} with such operations). Our schemes are based on the devices illustrated in Fig.~\ref{fig-exp}, with which we obtain quantum trajectories from continuously measuring the spontaneous emission of two qubits, and then implement local control operations in response to the real--time measurement outcomes.
The devices we consider are set up such that the joint measurements of the qubits may generate entanglement between them \cite{LongFlor2019}, and the aim of our feedback protocols is to increase the yield and/or lifetime of the entanglement generated by the device. 
We have shown that $\pi$--pulse--based  control, in conjunction with continuous photodetection, allows us to implement a measurement reversal procedure, which can protect any two--qubit state against the $T_1$ decay dynamics.
Combining the same methods with a Hamiltonian control protocol \cite{Leigh2019}, for the case of homodyne detection and diffusive quantum trajectories, allows us to create a stable limit cycle about a Bell state, again protecting concurrence from erosion via the qubits' natural decay channel. 
Although both schemes are negatively affected by measurement inefficiency, we are able to demonstrate that carrying them out still results in some net gain in entanglement yield and/or lifetime, compared with not carrying them out, across a wide variety of situations.
The schemes we have considered are grounded in existing experimental protocols; quantum trajectories obtained from measurements of spontaneous emission have been realized on single superconducting qubits \cite{Campagne-Ibarcq2016, naghiloo2015fluores, Mahdi2016, Mahdi2017Qtherm, PCI-2016-2, Tan2017, Ficheux2018}, could be implemented on other quantum information platforms, and single qubit unitary operations can generally be performed with high fidelity.

Entanglement is an important part of many emerging applications drawing broad scientific interest, such as quantum computing or quantum communication, and is also of foundational interest (e.g.~in connection with Bell tests \cite{HansonLoopholeFree}).
Decay due to spontaneous emission is, in many quantum--information systems, one of the important sources of errors.
Protecting entanglement against such errors is consequently of great practical interest.
The protocols we describe above offer a novel approach to this task, based on tools which are realistic extensions of existing devices and experiments. 
Moreover, the novel combination of continuous measurement and feedback with BB--like controls to achieve a measurement reversal suggests a new approach for correcting a wide range of errors on quantum systems that occur through a measureable channel to the environment.

\begin{acknowledgements}
We are grateful to Leigh S.~Martin for helpful correspondence and discussion regarding his work \cite{Leigh2019}. We thank Lorenza Viola for pointing out a few pertinent references after reading v1 of our pre-print. We acknowledge funding from NSF grant no.~DMR-1809343, and US Army Research Office grant no.~W911NF-18-10178. PL acknowledges support from the US Department of Education grant No.~GR506598 as a GAANN fellow.
\end{acknowledgements}

%\emph{Conflicts of Interest}---The authors declare that there are no competing interests.

%\emph{Author Contributions}---PL conceptualized the approach to the main result, in conversations with CE and ANJ. All authors contributed to derivations of the analytic results. PL prepared all figures and the underlying numerical codes to generate them. All authors contributed to the preparation of the manuscript. ANJ supervised the project.

%\emph{Data Availability}---Python codes written to generate most figures and run all underlying numerical methods can be made available upon request. There are no other relevant data, as this is a theoretical work. 

\appendix

%%%%%%%%%%%%%%%%%%%%%%%%%%%%%%%%%%%%%%%%%%%%%%%%%%%%%%%%%%%%%%%%%%%%%%%%%%%%%%%%%%%%%%%%%%%%%%%%%%%
%%%%%%%%%%%%%%%%%%%%%%%%%%%%%%%%%%%%%%%%%%%%%%%%%%%%%%%%%%%%%%%%%%%%%%%%%%%%%%%%%%%%%%%%%%%%%%%%%%%
%%%%%%%%%%%%%%%%%%%%%%%%%%%%%%%%%%%%%%%%%%%%%%%%%%%%%%%%%%%%%%%%%%%%%%%%%%%%%%%%%%%%%%%%%%%%%%%%%%%
\section{Additional Plots \label{sec-extraplots}}

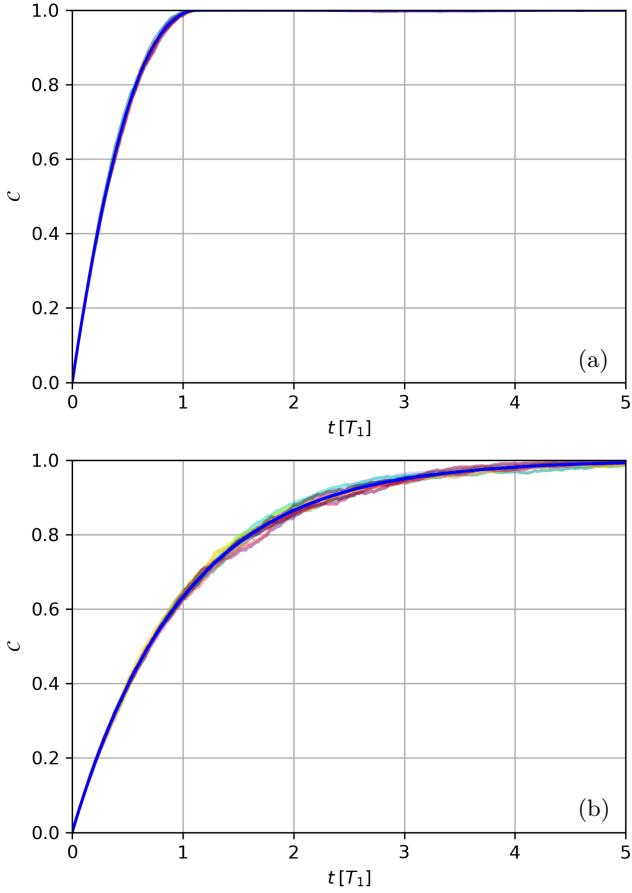
\begin{figure}
    \centering
    \includegraphics[width=\columnwidth,trim = {5 0 30 25}, clip]{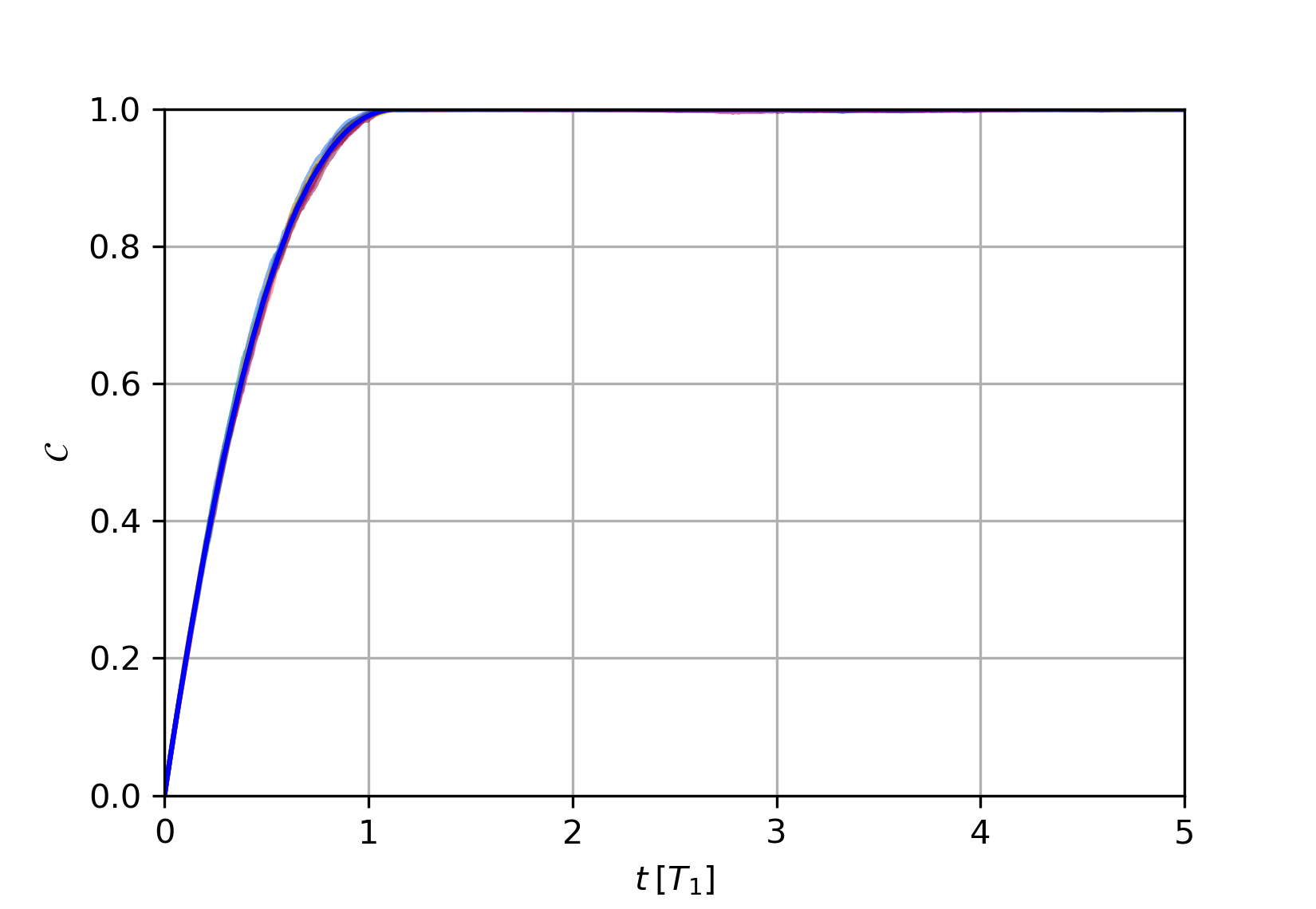}\\
    \includegraphics[width=\columnwidth,trim = {5 0 30 25}, clip]{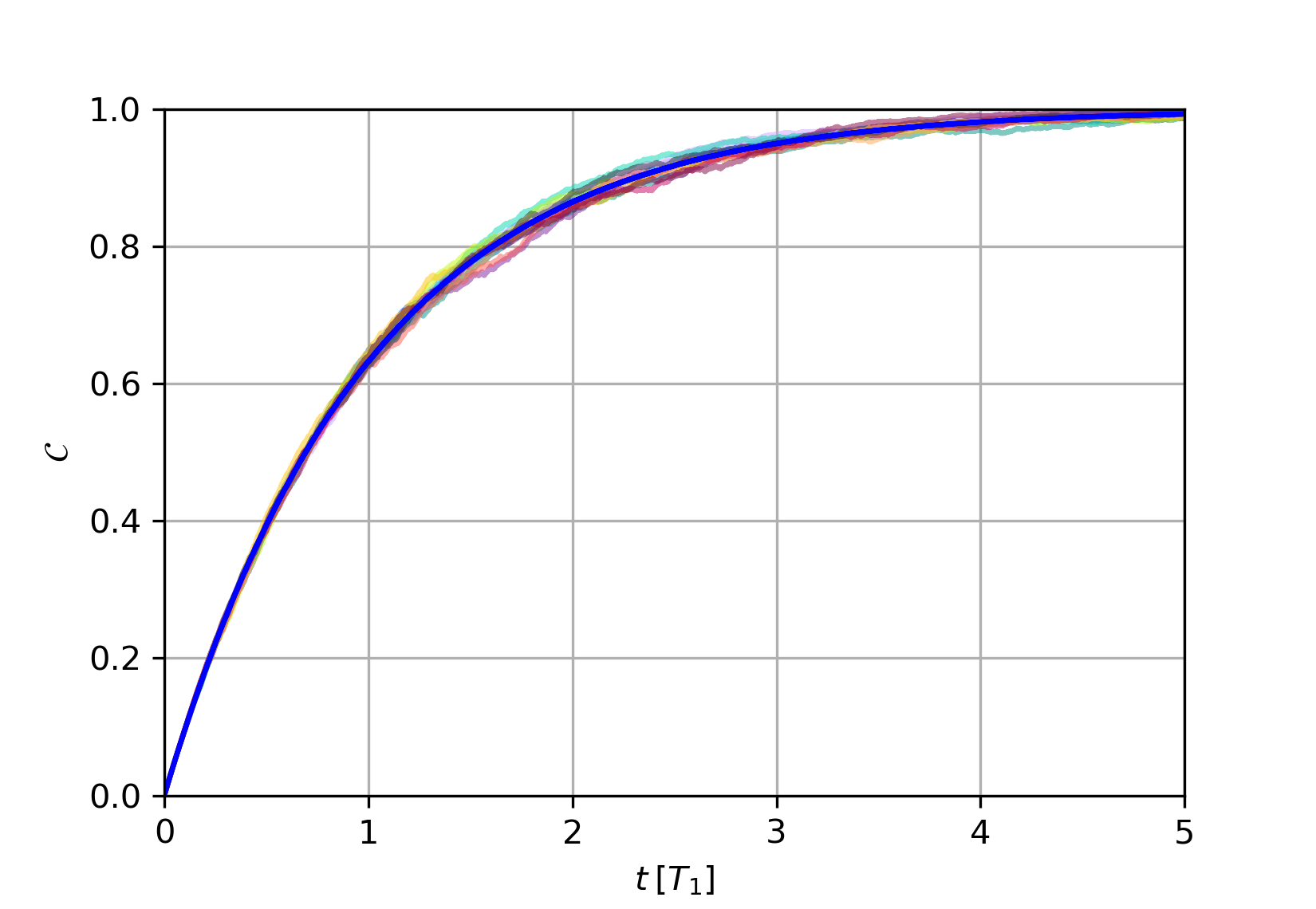}\\
    \begin{picture}(0,0)
    \put(100,210){(a)}
    \put(100,40){(b)}
    \end{picture} \vspace{-10pt}
    \caption{We repeat Fig.~\ref{fig-homfeed} with $\Delta t$ one order of magnitude smaller ($\epsilon = 10^{-3}$ here). While interjecting $\pi$--pulses that fast may no longer be realistic, by comparing to Fig.~\ref{fig-homfeed} we see that deviations from deterministic dynamics are suppressed as we take a step towards the time--continuum limit. As in Fig.~\ref{fig-homfeed}, we begin adding $\pi$--pulses after maximal concurrence is generated at $t_e = 1.13T_1$ in (a), while in (b) we see that we asymptotically approach maximal concurrence if we run the $\pi$--pulses over the entire duration; this serves to confirm that the coarser time--step of Fig.~\ref{fig-homfeed} was adequate to capture the main features of the dynamics, despite the more pronounced stochasticity we had there, on account of operating further from the time--continuum limit.}
    \label{fig-homfeed2}
\end{figure}

We include some additional figures which further support secondary claims we make in the main text. 
In Fig.~\ref{fig-homfeed2} we essentially reproduce the simulation of Fig.~\ref{fig-homfeed}, but this time with a smaller timestep. 
While spacing $\pi$--pulses so closely (every $T_1/1000$) may be less realistic in practice, Fig.~\ref{fig-homfeed2} serves to confirm that as we approach the time--continuum limit $\epsilon\rightarrow 0$, we recover the deterministic dynamics described by Martin and Whaley \cite{Leigh2019}; we see that deviations from deterministic dynamics are suppressed in Fig.~\ref{fig-homfeed2} as compared with the more realistic Fig.~\ref{fig-homfeed}.
Together, these two figures illustrate that 1) there is a tradeoff between the practical necessity of having a modest $\Delta t$, and acheiving exact deterministic evolution from \eqref{homfeedU} promised in the continuum limit, but 2) that this tradeoff is not a limiting factor for the overall effectiveness of our scheme.

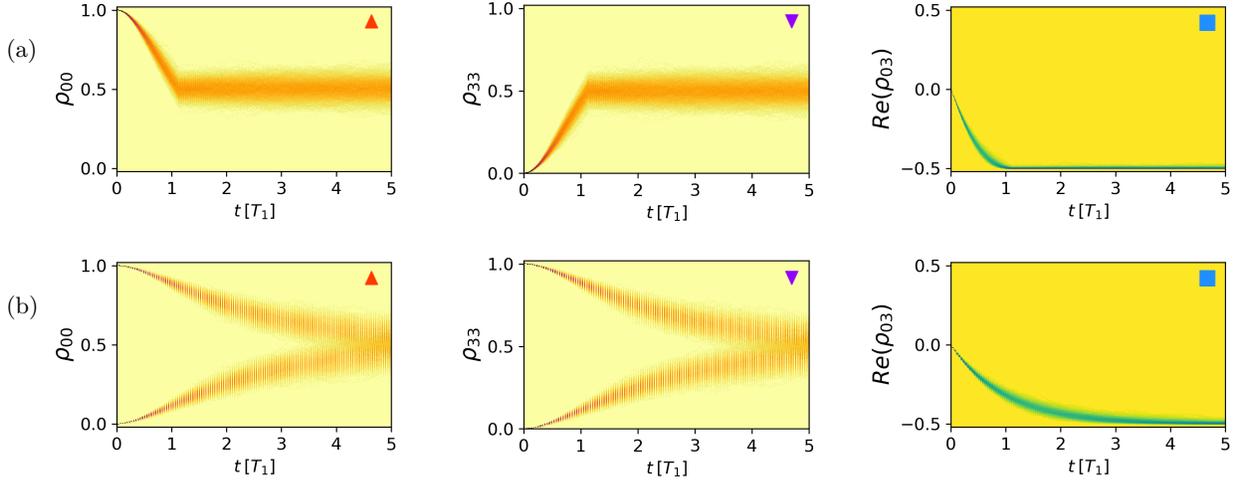
\begin{figure*}
    \centering
    \includegraphics[width = .3\textwidth]{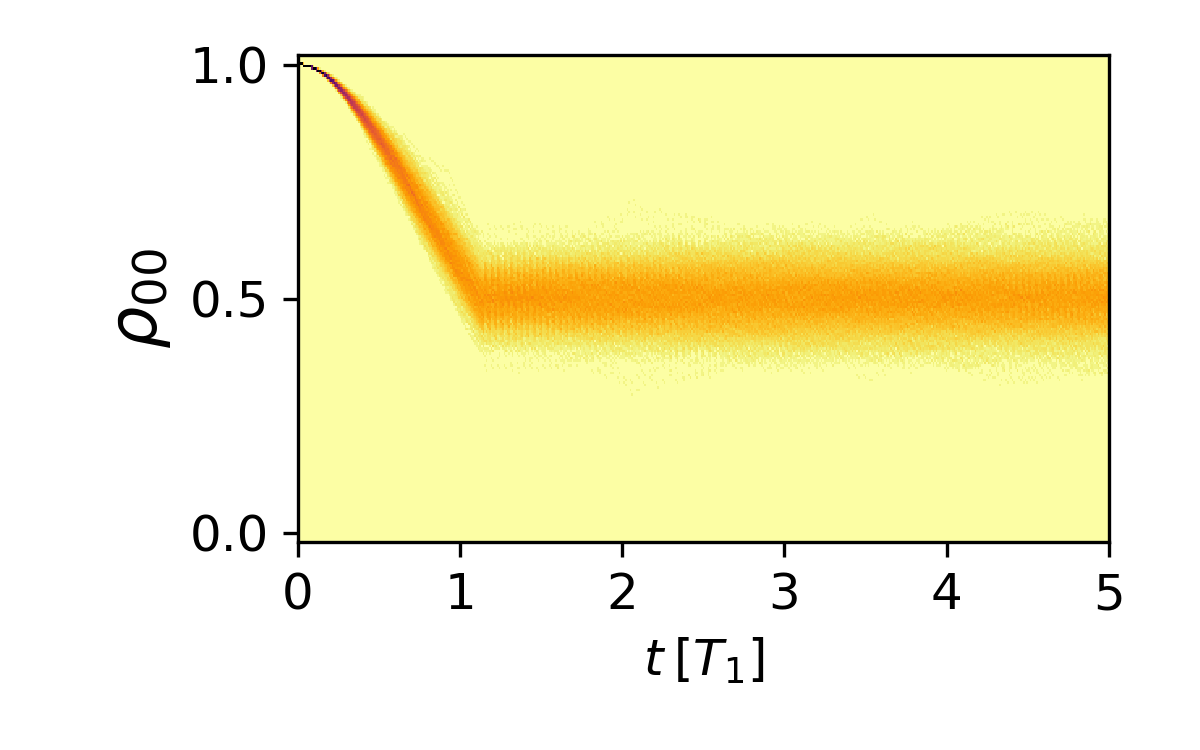}
    \includegraphics[width = .3\textwidth]{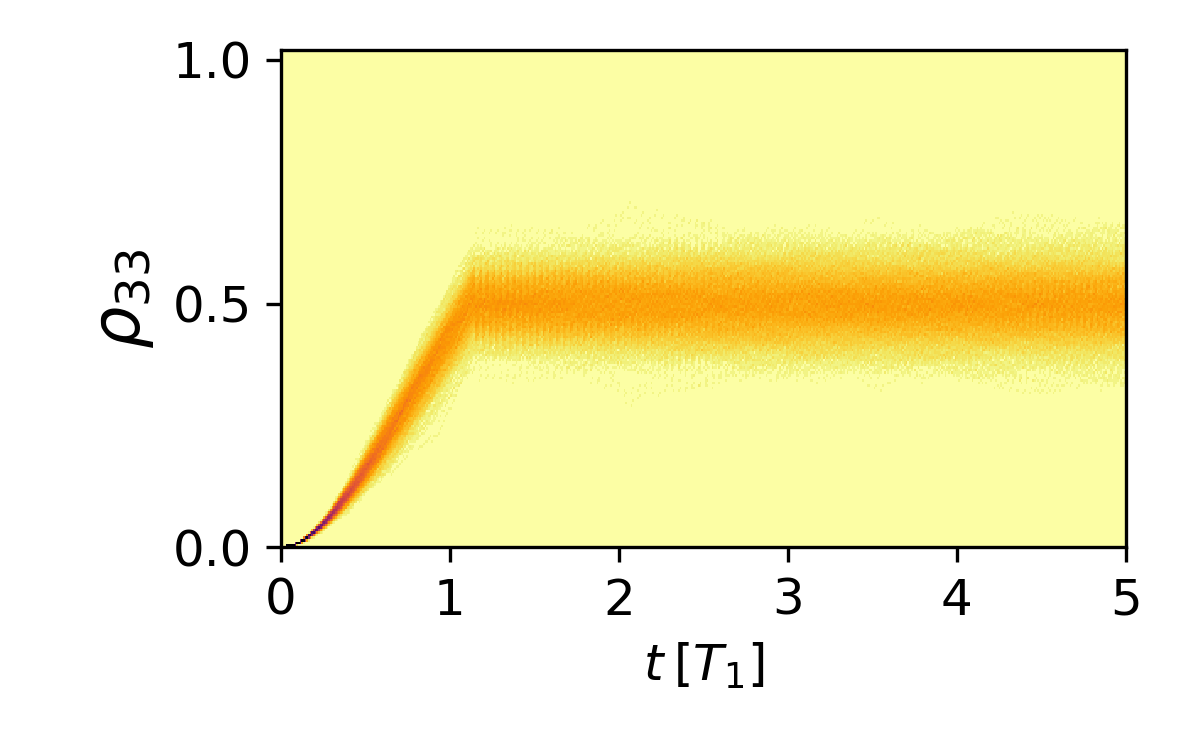}
    \includegraphics[width = .3\textwidth]{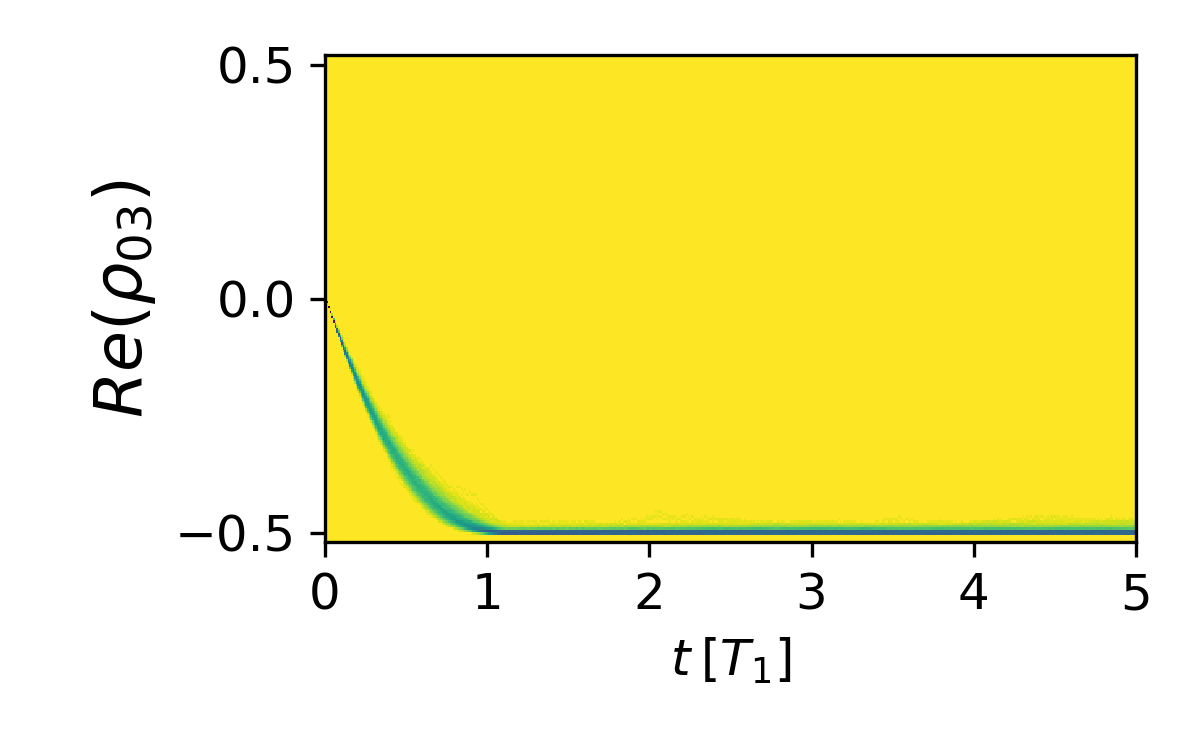} \\
    \includegraphics[width = .3\textwidth]{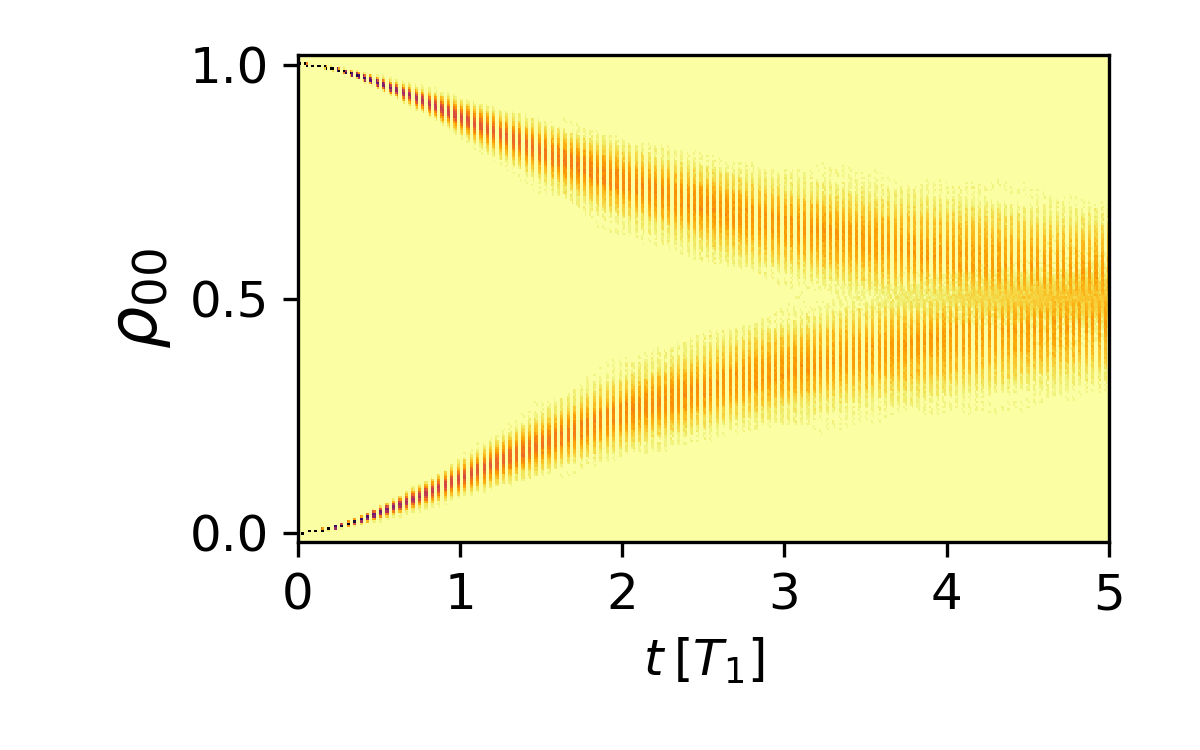}
    \includegraphics[width = .3\textwidth]{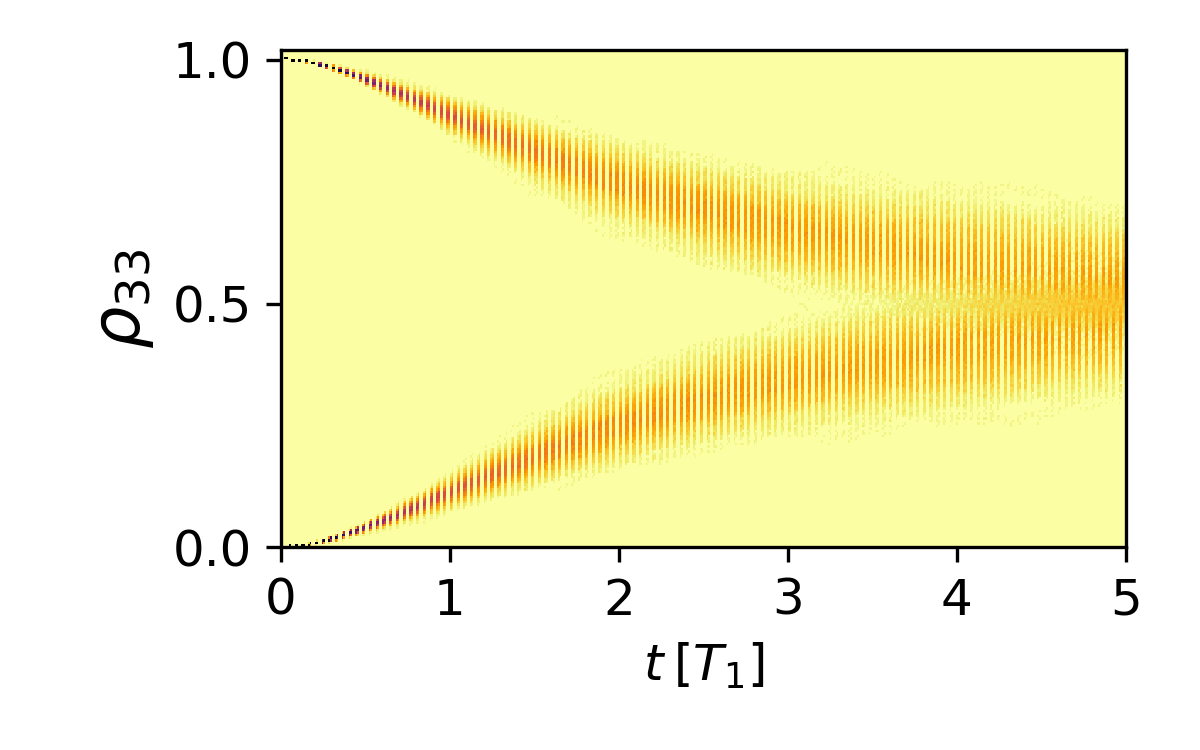}
    \includegraphics[width = .3\textwidth]{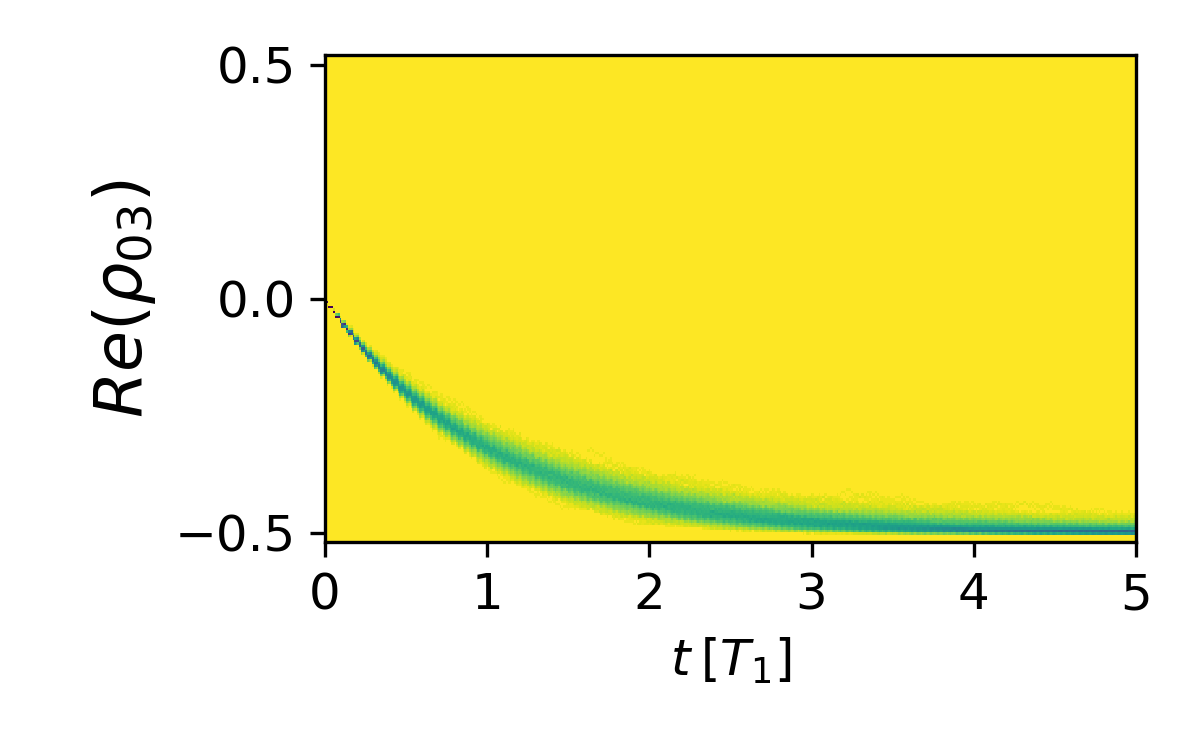} \\
    \begin{picture}(0,0)
    \put(-235,177){(a)}
    \put(-235,80){(b)}
    \put(-100,188){\color{coquelicot} $\blacktriangle$}
    \put(-100,91){\color{coquelicot} $\blacktriangle$}
    \put(59,188){\color{electricviolet} $\blacktriangledown$}
    \put(59,91){\color{electricviolet} $\blacktriangledown$}
    \put(216,187){\color{dodgerblue} $\blacksquare$}
    \put(216,90){\color{dodgerblue} $\blacksquare$}
    \end{picture} \vspace{-20pt}
    \caption{We show an element of the two--qubit density matrix in each panel; within each we plot the density of an ensemble of stochastic trajectories. In row (a) we plot elements corresponding to the case of Fig.~\ref{fig-homfeed}(b), wherein we add our $\pi$--pulse modification to the scheme of \cite{Leigh2019} only after entanglement is already established, whereas in row (b) we plot elements corresponding to the case of Fig.~\ref{fig-homfeed}(a), wherein the $\pi$--pulse modification is present over the entire evolution. }
    \label{fig-homfeedelements}
\end{figure*}

In Fig.~\ref{fig-homfeedelements} we plot the density of stochastic trajectories in the simulated ensemble of Fig.~\ref{fig-homfeed}, represted with selected elements of the density matrix. 
The symbolic / color scheme for notating density matrix elements goes like
\be \label{colorrho}
\rho = \left(\begin{array}{cccc} {\color{coquelicot} \blacktriangle} & 
{\color{electricgreen} \blacksquare}-\text{\colorbox{black}{${\color{white}i}{\color{electricgreen}\blacklozenge}$}} &  
{\color{mint} \blacksquare}-\text{\colorbox{black}{${\color{white}i}{\color{mint}\blacklozenge}$}}
& {\color{dodgerblue} \blacksquare}-\text{\colorbox{black}{${\color{white}i}{\color{dodgerblue}\blacklozenge}$}} \\ 
{\color{electricgreen} \blacksquare}+\text{\colorbox{black}{${\color{white}i}{\color{electricgreen}\blacklozenge}$}} & 
{\color{crimsonglory} \blacktriangleleft} & 
{\color{lincolngreen} \blacksquare}-\text{\colorbox{black}{${\color{white}i}{\color{lincolngreen}\blacklozenge}$}} & 
{\color{persianblue} \blacksquare}-\text{\colorbox{black}{${\color{white}i}{\color{persianblue}\blacklozenge}$}}\\ 
{\color{mint} \blacksquare}+\text{\colorbox{black}{${\color{white}i}{\color{mint}\blacklozenge}$}} & 
{\color{lincolngreen} \blacksquare}+\text{\colorbox{black}{${\color{white}i}{\color{lincolngreen}\blacklozenge}$}}& 
{\color{deeppink} \blacktriangleright}& 
 {\color{patriarch} \blacksquare}-\text{\colorbox{black}{${\color{white}i}{\color{patriarch}\blacklozenge}$}}\\ 
{\color{dodgerblue} \blacksquare}+\text{\colorbox{black}{${\color{white}i}{\color{dodgerblue}\blacklozenge}$}}
& {\color{persianblue} \blacksquare}+\text{\colorbox{black}{${\color{white}i}{\color{persianblue}\blacklozenge}$}}
& {\color{patriarch} \blacksquare}+\text{\colorbox{black}{${\color{white}i}{\color{patriarch}\blacklozenge}$}}
& {\color{electricviolet} \blacktriangledown}\end{array} \right),
\ee
where the basis is such that e.g.~${\color{coquelicot} \blacktriangle}$ represents the population in $\ket{ee}$, ${\color{electricviolet} \blacktriangledown}$ represents the population in $\ket{gg}$, and ${\color{dodgerblue} \blacksquare}$ represents the real part of the coherence $\ket{ee}\bra{gg}$ between them. %${\color{coquelicot} \blacktriangle}$ and ${\color{electricviolet} \blacktriangledown}$.
The full basis, used here and elsewhere in the manuscript assumes pure states notated according to 
\be 
\ket{\psi} = \left( \begin{array}{c}
\mathsf{a} \\ \mathsf{b} \\ \mathsf{c} \\ \mathsf{d} 
\end{array}\right) \quad\sim\quad 
{\color{codegray}
\left\lbrace \begin{array}{c}
\ket{ee} \\ \ket{eg} \\ \ket{ge} \\ \ket{gg}
\end{array}\right\rbrace.
}\ee

%%%%%%%%%%%%%%%%%%%%%%%%%%%%%%%%%%%%%%%%%%%%%%%%%%%%%%%%%%%%%%%%%%%%%%%%%%%%%%%%%%%%%%%%%%%%%%%%%%%
%%%%%%%%%%%%%%%%%%%%%%%%%%%%%%%%%%%%%%%%%%%%%%%%%%%%%%%%%%%%%%%%%%%%%%%%%%%%%%%%%%%%%%%%%%%%%%%%%%%
%%%%%%%%%%%%%%%%%%%%%%%%%%%%%%%%%%%%%%%%%%%%%%%%%%%%%%%%%%%%%%%%%%%%%%%%%%%%%%%%%%%%%%%%%%%%%%%%%%%
\section{Summary of Fluorescence Measurement Formalism \label{sec-kraus-review}}

We review our Kraus operators, used throughout the main text, for completeness. Everything included in this section in brief is explained in far greater detail in \cite{FlorTeach2019} (the one--qubit case), and \cite{LongFlor2019} (the two--qubit case). 
Refer to Fig.~\ref{fig-exp} for a sketch of the relevant apparatus.
We begin with the matrix 
\be \label{Mmat}
\mathcal{M} = \left( \begin{array}{cccc}
1- \epsilon & 0 & 0 & 0 \\
\sqrt{\epsilon(1-\epsilon)} \hat{a}_2^\dag & \sqrt{1-\epsilon} & 0 & 0 \\
\sqrt{\epsilon(1-\epsilon)} \hat{a}_1^\dag & 0 & \sqrt{1-\epsilon} & 0 \\
\epsilon \hat{a}_1^\dag \hat{a}_2^\dag & \sqrt{\epsilon} \hat{a}_1^\dag & \sqrt{\epsilon}\hat{a}_2^\dag & 1
\end{array} \right),
\ee
which may be used to update the joint state of the qubits and optical modes 1 \& 2 they emit into, over a short time $\Delta t \ll T_1$ (equivalently, $\epsilon = \gamma \: \Delta t \ll 1$). 
We assume that both qubit--cavity systems have the same emission rate $\gamma = 1/T_1$ for simplicity.
The operators $\hat{a}^\dag_1$ and $\hat{a}^\dag_2$ are creation operators for photons in ports (modes) 1 and 2, respectively.
The effect of the beamsplitter may be modeled by the unitary transformation
\be \label{bs}
\hat{a}_1^\dag = \tfrac{1}{\sqrt{2}}\left( \hat{a}_3^\dag e^{i\phi} + \hat{a}_4^\dag e^{i\varphi} \right), \: \hat{a}_2^\dag = \tfrac{1}{\sqrt{2}}\left( \hat{a}_3^\dag e^{i\phi} - \hat{a}_4^\dag e^{i\varphi} \right), 
\ee
which mixes the modes 1 \& 2 in order to obtain the measured modes 3 \& 4.
This 50/50 beamsplitter plays an important role in concealing information about which qubit emitted a signal; erasure of this which--path information is a key condition in allowing subsequent measurements to be entangling.

\par In order to obtain a Kraus operator which acts on the qubits alone, it is necessary to select the initial and final states of the optical modes. We will assume that the modes are in vacuum at the start of each measurement interval $\Delta t$, such that the initial state of modes 3 \& 4 is $\ket{0_3 0_4}$ (which implies the same for 1 \& 2).
The final state of the output modes is determined by the type of measurement that is performed. 
For example, photodetection at outputs 3 and 4 leads to outcomes in the Fock basis, and a Kraus operator
\be 
\hat{\mathcal{M}}_{n_3,n_4} = \bra{n_3 n_4} \mathcal{M} \ket{0_3 0_4};
\ee
This generates a set of five operators, one for each of the five outcomes $\lbrace n_3, n_4 \rbrace = \lbrace 0,0 \rbrace, \lbrace 1,0 \rbrace, \lbrace 0,1 \rbrace, \lbrace 2,0 \rbrace, \lbrace 0,2 \rbrace$ allowed in any step $\Delta t$ (which form a complete set of POVM elements). 
Likewise, homodyne detection at both outputs leads to projection onto eigenstates of a quadrature operator, i.e.~for $\ket{X}$ an eigenstate of $\hat{X} = (\hat{a}_3^\dag + \hat{a}_3)/\sqrt{2}$ and $\ket{Y}$ an eigenstate of $\hat{Y} =  (\hat{a}_4^\dag + \hat{a}_4)/\sqrt{2}$, the Kraus operator is obtained from 
\be 
\hat{\mathcal{M}}_{XY} = \bra{X Y} \mathcal{M} \ket{0_3 0_4},
\ee
which reduces to \eqref{M-hom} for the phase choices $\phi = 0$ and $\varphi = 90^\circ$.

\par Measurement inefficiency is most--straightforwardly modeled with an additional set of unbalanced beamsplitters, as shown in Fig.~\ref{fig-exp}.
The effect of these is to split modes 3 and 4 into a ``signal portion'', which goes to the relevant (otherwise still ideal) detector with probability $\eta$, and a ``lost portion''. 
Algebraically, this is expressed the transformations
\be \begin{split}
&\hat{a}_3^\dag \rightarrow \sqrt{\eta_3} \: \hat{a}_{3s}^\dag + \sqrt{1-\eta_3} \: \hat{a}_{3\ell}^\dag \quad\text{ and } \\
&\hat{a}_4^\dag \rightarrow \sqrt{\eta_4} \: \hat{a}_{4s}^\dag + \sqrt{1-\eta_4} \: \hat{a}_{4\ell}^\dag
\end{split} \ee
which can be carried out inside of $\mathcal{M}$ to obtain $\mathcal{M}_\eta$.
While this could be used to model a situation in which four measurements are made, our interest is to use measurement outcomes at the signal ports only, while tracing out all of the possible (but unknown) outcomes which could have occurred in the lost ports.
For example, for inefficient photodetection with the outcome $\lbrace 0,0\rbrace$ at the signal ports, we would have a four--output Kraus operator
\be 
\hat{\mathcal{M}}_{0 0 n_3^\ell n_4^\ell} = \bra{0_{3}^s 0_{4}^s n_3^\ell n_4^\ell}\mathcal{M}_{\eta} \ket{0000}
\ee 
(assuming that the paired extra input modes, required by the unitarity of the transformation, are in vacuum), and the state update equation
\begin{widetext}\be 
\rho(t+\Delta t) = \frac{\hat{\mathcal{M}}_{0000}\rho(t)\hat{\mathcal{M}}^\dag_{0000}+ \hat{\mathcal{M}}_{0010}\rho(t)\hat{\mathcal{M}}^\dag_{0010}+\hat{\mathcal{M}}_{0001}\rho(t)\hat{\mathcal{M}}^\dag_{0001}+\hat{\mathcal{M}}_{0020}\rho(t)\hat{\mathcal{M}}^\dag_{0020}+\hat{\mathcal{M}}_{0002}\rho(t)\hat{\mathcal{M}}^\dag_{0002}}{\text{tr}\left( \hat{\mathcal{M}}_{0000}\rho(t)\hat{\mathcal{M}}^\dag_{0000}+ \hat{\mathcal{M}}_{0010}\rho(t)\hat{\mathcal{M}}^\dag_{0010}+\hat{\mathcal{M}}_{0001}\rho(t)\hat{\mathcal{M}}^\dag_{0001}+\hat{\mathcal{M}}_{0020}\rho(t)\hat{\mathcal{M}}^\dag_{0020}+\hat{\mathcal{M}}_{0002}\rho(t)\hat{\mathcal{M}}^\dag_{0002}\right)},
\ee
which includes the trace over all possible lost--mode states that are consistent with having recieved the outcome $\lbrace 0 ,0\rbrace$.
For such an update with finite measurement efficiency, the basis in which we do the trace over the outcomes in the lost mode does not matter, as long as it represents a complete set of outcomes.
By that token, inefficient homodyne detection is best--modeled by an operator
\be  
\hat{\mathcal{M}}_{X Y n_3^\ell n_4^\ell} = \bra{X Y n_3^\ell n_4^\ell}\mathcal{M}_{\eta} \ket{0000},
\ee 
which can be used with the state update
\be 
\rho' = \frac{\hat{\mathcal{M}}_{XY00}\rho\hat{\mathcal{M}}^\dag_{XY00}+ \hat{\mathcal{M}}_{XY10}\rho\hat{\mathcal{M}}^\dag_{XY10}+\hat{\mathcal{M}}_{XY01}\rho\hat{\mathcal{M}}^\dag_{XY01}+\hat{\mathcal{M}}_{XY20}\rho\hat{\mathcal{M}}^\dag_{XY20}+\hat{\mathcal{M}}_{XY02}\rho\hat{\mathcal{M}}^\dag_{XY02}}{\text{tr}\left( \hat{\mathcal{M}}_{XY00}\rho\hat{\mathcal{M}}^\dag_{XY00}+ \hat{\mathcal{M}}_{XY10}\rho\hat{\mathcal{M}}^\dag_{XY10}+\hat{\mathcal{M}}_{XY01}\rho\hat{\mathcal{M}}^\dag_{XY01}+\hat{\mathcal{M}}_{XY20}\rho\hat{\mathcal{M}}^\dag_{XY20}+\hat{\mathcal{M}}_{XY02}\rho\hat{\mathcal{M}}^\dag_{XY02}\right)},
\ee\end{widetext}
for $\rho' = \rho(t+\Delta t)$ and $\rho = \rho(t)$; summing over the lost modes in the discrete Fock basis is computationally simpler than integrating out another pair of continuous--valued homodyne (quadrature basis) outcomes, although the latter would give an equivalent state update.

\bibliography{refs}

\begin{thebibliography}{100}%
\makeatletter
\providecommand \@ifxundefined [1]{%
 \@ifx{#1\undefined}
}%
\providecommand \@ifnum [1]{%
 \ifnum #1\expandafter \@firstoftwo
 \else \expandafter \@secondoftwo
 \fi
}%
\providecommand \@ifx [1]{%
 \ifx #1\expandafter \@firstoftwo
 \else \expandafter \@secondoftwo
 \fi
}%
\providecommand \natexlab [1]{#1}%
\providecommand \enquote  [1]{``#1''}%
\providecommand \bibnamefont  [1]{#1}%
\providecommand \bibfnamefont [1]{#1}%
\providecommand \citenamefont [1]{#1}%
\providecommand \href@noop [0]{\@secondoftwo}%
\providecommand \href [0]{\begingroup \@sanitize@url \@href}%
\providecommand \@href[1]{\@@startlink{#1}\@@href}%
\providecommand \@@href[1]{\endgroup#1\@@endlink}%
\providecommand \@sanitize@url [0]{\catcode `\\12\catcode `\$12\catcode
  `\&12\catcode `\#12\catcode `\^12\catcode `\_12\catcode `\%12\relax}%
\providecommand \@@startlink[1]{}%
\providecommand \@@endlink[0]{}%
\providecommand \url  [0]{\begingroup\@sanitize@url \@url }%
\providecommand \@url [1]{\endgroup\@href {#1}{\urlprefix }}%
\providecommand \urlprefix  [0]{URL }%
\providecommand \Eprint [0]{\href }%
\providecommand \doibase [0]{http://dx.doi.org/}%
\providecommand \selectlanguage [0]{\@gobble}%
\providecommand \bibinfo  [0]{\@secondoftwo}%
\providecommand \bibfield  [0]{\@secondoftwo}%
\providecommand \translation [1]{[#1]}%
\providecommand \BibitemOpen [0]{}%
\providecommand \bibitemStop [0]{}%
\providecommand \bibitemNoStop [0]{.\EOS\space}%
\providecommand \EOS [0]{\spacefactor3000\relax}%
\providecommand \BibitemShut  [1]{\csname bibitem#1\endcsname}%
\let\auto@bib@innerbib\@empty
%</preamble>
\bibitem [{\citenamefont {Yu}\ and\ \citenamefont
  {Eberly}(2004)}]{YuEberly_2004}%
  \BibitemOpen
  \bibfield  {author} {\bibinfo {author} {\bibfnamefont {Ting}\ \bibnamefont
  {Yu}}\ and\ \bibinfo {author} {\bibfnamefont {J.~H.}\ \bibnamefont
  {Eberly}},\ }\bibfield  {title} {\enquote {\bibinfo {title} {Finite-time
  disentanglement via spontaneous emission},}\ }\href {\doibase
  10.1103/PhysRevLett.93.140404} {\bibfield  {journal} {\bibinfo  {journal}
  {Phys. Rev. Lett.}\ }\textbf {\bibinfo {volume} {93}},\ \bibinfo {pages}
  {140404} (\bibinfo {year} {2004})}\BibitemShut {NoStop}%
\bibitem [{\citenamefont {Cabrillo}\ \emph {et~al.}(1999)\citenamefont
  {Cabrillo}, \citenamefont {Cirac}, \citenamefont {Garc\'{\i}a-Fern\'andez},\
  and\ \citenamefont {Zoller}}]{Cabrillo1998}%
  \BibitemOpen
  \bibfield  {author} {\bibinfo {author} {\bibfnamefont {C.}~\bibnamefont
  {Cabrillo}}, \bibinfo {author} {\bibfnamefont {J.~I.}\ \bibnamefont {Cirac}},
  \bibinfo {author} {\bibfnamefont {P.}~\bibnamefont
  {Garc\'{\i}a-Fern\'andez}}, \ and\ \bibinfo {author} {\bibfnamefont
  {P.}~\bibnamefont {Zoller}},\ }\bibfield  {title} {\enquote {\bibinfo {title}
  {Creation of entangled states of distant atoms by interference},}\ }\href
  {\doibase 10.1103/PhysRevA.59.1025} {\bibfield  {journal} {\bibinfo
  {journal} {Phys. Rev. A}\ }\textbf {\bibinfo {volume} {59}},\ \bibinfo
  {pages} {1025--1033} (\bibinfo {year} {1999})}\BibitemShut {NoStop}%
\bibitem [{\citenamefont {Bose}\ \emph {et~al.}(1999)\citenamefont {Bose},
  \citenamefont {Knight}, \citenamefont {Plenio},\ and\ \citenamefont
  {Vedral}}]{Bose1999}%
  \BibitemOpen
  \bibfield  {author} {\bibinfo {author} {\bibfnamefont {S.}~\bibnamefont
  {Bose}}, \bibinfo {author} {\bibfnamefont {P.~L.}\ \bibnamefont {Knight}},
  \bibinfo {author} {\bibfnamefont {M.~B.}\ \bibnamefont {Plenio}}, \ and\
  \bibinfo {author} {\bibfnamefont {V.}~\bibnamefont {Vedral}},\ }\bibfield
  {title} {\enquote {\bibinfo {title} {Proposal for teleportation of an atomic
  state via cavity decay},}\ }\href {\doibase 10.1103/PhysRevLett.83.5158}
  {\bibfield  {journal} {\bibinfo  {journal} {Phys. Rev. Lett.}\ }\textbf
  {\bibinfo {volume} {83}},\ \bibinfo {pages} {5158--5161} (\bibinfo {year}
  {1999})}\BibitemShut {NoStop}%
\bibitem [{\citenamefont {Plenio}\ \emph {et~al.}(1999)\citenamefont {Plenio},
  \citenamefont {Huelga}, \citenamefont {Beige},\ and\ \citenamefont
  {Knight}}]{Plenio1999}%
  \BibitemOpen
  \bibfield  {author} {\bibinfo {author} {\bibfnamefont {M.~B.}\ \bibnamefont
  {Plenio}}, \bibinfo {author} {\bibfnamefont {S.~F.}\ \bibnamefont {Huelga}},
  \bibinfo {author} {\bibfnamefont {A.}~\bibnamefont {Beige}}, \ and\ \bibinfo
  {author} {\bibfnamefont {P.~L.}\ \bibnamefont {Knight}},\ }\bibfield  {title}
  {\enquote {\bibinfo {title} {Cavity-loss-induced generation of entangled
  atoms},}\ }\href {\doibase 10.1103/PhysRevA.59.2468} {\bibfield  {journal}
  {\bibinfo  {journal} {Phys. Rev. A}\ }\textbf {\bibinfo {volume} {59}},\
  \bibinfo {pages} {2468--2475} (\bibinfo {year} {1999})}\BibitemShut {NoStop}%
\bibitem [{\citenamefont {Barrett}\ and\ \citenamefont
  {Kok}(2005)}]{BarrettKok}%
  \BibitemOpen
  \bibfield  {author} {\bibinfo {author} {\bibfnamefont {Sean~D.}\ \bibnamefont
  {Barrett}}\ and\ \bibinfo {author} {\bibfnamefont {Pieter}\ \bibnamefont
  {Kok}},\ }\bibfield  {title} {\enquote {\bibinfo {title} {Efficient
  high-fidelity quantum computation using matter qubits and linear optics},}\
  }\href {\doibase 10.1103/PhysRevA.71.060310} {\bibfield  {journal} {\bibinfo
  {journal} {Phys. Rev. A}\ }\textbf {\bibinfo {volume} {71}},\ \bibinfo
  {pages} {060310(R)} (\bibinfo {year} {2005})}\BibitemShut {NoStop}%
\bibitem [{\citenamefont {Duan}\ and\ \citenamefont
  {Kimble}(2003)}]{DuanKimble2003}%
  \BibitemOpen
  \bibfield  {author} {\bibinfo {author} {\bibfnamefont {L.-M.}\ \bibnamefont
  {Duan}}\ and\ \bibinfo {author} {\bibfnamefont {H.~J.}\ \bibnamefont
  {Kimble}},\ }\bibfield  {title} {\enquote {\bibinfo {title} {Efficient
  engineering of multiatom entanglement through single-photon detections},}\
  }\href {\doibase 10.1103/PhysRevLett.90.253601} {\bibfield  {journal}
  {\bibinfo  {journal} {Phys. Rev. Lett.}\ }\textbf {\bibinfo {volume} {90}},\
  \bibinfo {pages} {253601} (\bibinfo {year} {2003})}\BibitemShut {NoStop}%
\bibitem [{\citenamefont {Browne}\ \emph {et~al.}(2003)\citenamefont {Browne},
  \citenamefont {Plenio},\ and\ \citenamefont {Huelga}}]{Browne2003}%
  \BibitemOpen
  \bibfield  {author} {\bibinfo {author} {\bibfnamefont {Daniel~E.}\
  \bibnamefont {Browne}}, \bibinfo {author} {\bibfnamefont {Martin~B.}\
  \bibnamefont {Plenio}}, \ and\ \bibinfo {author} {\bibfnamefont {Susana~F.}\
  \bibnamefont {Huelga}},\ }\bibfield  {title} {\enquote {\bibinfo {title}
  {Robust creation of entanglement between ions in spatially separate
  cavities},}\ }\href {\doibase 10.1103/PhysRevLett.91.067901} {\bibfield
  {journal} {\bibinfo  {journal} {Phys. Rev. Lett.}\ }\textbf {\bibinfo
  {volume} {91}},\ \bibinfo {pages} {067901} (\bibinfo {year}
  {2003})}\BibitemShut {NoStop}%
\bibitem [{\citenamefont {Simon}\ and\ \citenamefont
  {Irvine}(2003)}]{Simon2003}%
  \BibitemOpen
  \bibfield  {author} {\bibinfo {author} {\bibfnamefont {Christoph}\
  \bibnamefont {Simon}}\ and\ \bibinfo {author} {\bibfnamefont {William T.~M.}\
  \bibnamefont {Irvine}},\ }\bibfield  {title} {\enquote {\bibinfo {title}
  {Robust long-distance entanglement and a loophole-free {B}ell test with ions
  and photons},}\ }\href {\doibase 10.1103/PhysRevLett.91.110405} {\bibfield
  {journal} {\bibinfo  {journal} {Phys. Rev. Lett.}\ }\textbf {\bibinfo
  {volume} {91}},\ \bibinfo {pages} {110405} (\bibinfo {year}
  {2003})}\BibitemShut {NoStop}%
\bibitem [{\citenamefont {Lim}\ \emph {et~al.}(2005)\citenamefont {Lim},
  \citenamefont {Beige},\ and\ \citenamefont {Kwek}}]{Lim2005}%
  \BibitemOpen
  \bibfield  {author} {\bibinfo {author} {\bibfnamefont {Yuan~Liang}\
  \bibnamefont {Lim}}, \bibinfo {author} {\bibfnamefont {Almut}\ \bibnamefont
  {Beige}}, \ and\ \bibinfo {author} {\bibfnamefont {Leong~Chuan}\ \bibnamefont
  {Kwek}},\ }\bibfield  {title} {\enquote {\bibinfo {title}
  {Repeat-until-success linear optics distributed quantum computing},}\ }\href
  {\doibase 10.1103/PhysRevLett.95.030505} {\bibfield  {journal} {\bibinfo
  {journal} {Phys. Rev. Lett.}\ }\textbf {\bibinfo {volume} {95}},\ \bibinfo
  {pages} {030505} (\bibinfo {year} {2005})}\BibitemShut {NoStop}%
\bibitem [{\citenamefont {Moehring}\ \emph {et~al.}(2007)\citenamefont
  {Moehring}, \citenamefont {Maunz}, \citenamefont {Olmschenk}, \citenamefont
  {Younge}, \citenamefont {Matsukevich}, \citenamefont {Duan},\ and\
  \citenamefont {Monroe}}]{Moehring2007}%
  \BibitemOpen
  \bibfield  {author} {\bibinfo {author} {\bibfnamefont {D.~L.}\ \bibnamefont
  {Moehring}}, \bibinfo {author} {\bibfnamefont {P.}~\bibnamefont {Maunz}},
  \bibinfo {author} {\bibfnamefont {S.}~\bibnamefont {Olmschenk}}, \bibinfo
  {author} {\bibfnamefont {K.~C.}\ \bibnamefont {Younge}}, \bibinfo {author}
  {\bibfnamefont {D.~N.}\ \bibnamefont {Matsukevich}}, \bibinfo {author}
  {\bibfnamefont {L.-M.}\ \bibnamefont {Duan}}, \ and\ \bibinfo {author}
  {\bibfnamefont {C.}~\bibnamefont {Monroe}},\ }\bibfield  {title} {\enquote
  {\bibinfo {title} {Entanglement of single-atom quantum bits at a distance},}\
  }\href {https://www.nature.com/articles/nature06118} {\bibfield  {journal}
  {\bibinfo  {journal} {Nature}\ }\textbf {\bibinfo {volume} {449}},\ \bibinfo
  {pages} {68} (\bibinfo {year} {2007})}\BibitemShut {NoStop}%
\bibitem [{\citenamefont {Maunz}\ \emph {et~al.}(2009)\citenamefont {Maunz},
  \citenamefont {Olmschenk}, \citenamefont {Hayes}, \citenamefont
  {Matsukevich}, \citenamefont {Duan},\ and\ \citenamefont
  {Monroe}}]{Maunz2009}%
  \BibitemOpen
  \bibfield  {author} {\bibinfo {author} {\bibfnamefont {P.}~\bibnamefont
  {Maunz}}, \bibinfo {author} {\bibfnamefont {S.}~\bibnamefont {Olmschenk}},
  \bibinfo {author} {\bibfnamefont {D.}~\bibnamefont {Hayes}}, \bibinfo
  {author} {\bibfnamefont {D.~N.}\ \bibnamefont {Matsukevich}}, \bibinfo
  {author} {\bibfnamefont {L.-M.}\ \bibnamefont {Duan}}, \ and\ \bibinfo
  {author} {\bibfnamefont {C.}~\bibnamefont {Monroe}},\ }\bibfield  {title}
  {\enquote {\bibinfo {title} {Heralded quantum gate between remote quantum
  memories},}\ }\href {https://link.aps.org/doi/10.1103/PhysRevLett.102.250502}
  {\bibfield  {journal} {\bibinfo  {journal} {Phys. Rev. Lett.}\ }\textbf
  {\bibinfo {volume} {102}},\ \bibinfo {pages} {250502} (\bibinfo {year}
  {2009})}\BibitemShut {NoStop}%
\bibitem [{\citenamefont {Hofmann}\ \emph {et~al.}(2012)\citenamefont
  {Hofmann}, \citenamefont {Krug}, \citenamefont {Ortegel}, \citenamefont
  {G{\'e}rard}, \citenamefont {Weber}, \citenamefont {Rosenfeld},\ and\
  \citenamefont {Weinfurter}}]{Hofmann2012}%
  \BibitemOpen
  \bibfield  {author} {\bibinfo {author} {\bibfnamefont {Julian}\ \bibnamefont
  {Hofmann}}, \bibinfo {author} {\bibfnamefont {Michael}\ \bibnamefont {Krug}},
  \bibinfo {author} {\bibfnamefont {Norbert}\ \bibnamefont {Ortegel}}, \bibinfo
  {author} {\bibfnamefont {Lea}\ \bibnamefont {G{\'e}rard}}, \bibinfo {author}
  {\bibfnamefont {Markus}\ \bibnamefont {Weber}}, \bibinfo {author}
  {\bibfnamefont {Wenjamin}\ \bibnamefont {Rosenfeld}}, \ and\ \bibinfo
  {author} {\bibfnamefont {Harald}\ \bibnamefont {Weinfurter}},\ }\bibfield
  {title} {\enquote {\bibinfo {title} {Heralded entanglement between widely
  separated atoms},}\ }\href {\doibase 10.1126/science.1221856} {\bibfield
  {journal} {\bibinfo  {journal} {Science}\ }\textbf {\bibinfo {volume}
  {337}},\ \bibinfo {pages} {72--75} (\bibinfo {year} {2012})}\BibitemShut
  {NoStop}%
\bibitem [{\citenamefont {Bernien}\ \emph {et~al.}(2013)\citenamefont
  {Bernien}, \citenamefont {Hensen}, \citenamefont {Pfaff}, \citenamefont
  {Koolstra}, \citenamefont {Blok}, \citenamefont {Robledo}, \citenamefont
  {Taminiau}, \citenamefont {Markham}, \citenamefont {Twitchen}, \citenamefont
  {Childress},\ and\ \citenamefont {Hanson}}]{Hanson2013Heralded}%
  \BibitemOpen
  \bibfield  {author} {\bibinfo {author} {\bibfnamefont {H.}~\bibnamefont
  {Bernien}}, \bibinfo {author} {\bibfnamefont {B.}~\bibnamefont {Hensen}},
  \bibinfo {author} {\bibfnamefont {W.}~\bibnamefont {Pfaff}}, \bibinfo
  {author} {\bibfnamefont {G.}~\bibnamefont {Koolstra}}, \bibinfo {author}
  {\bibfnamefont {M.~S.}\ \bibnamefont {Blok}}, \bibinfo {author}
  {\bibfnamefont {L.}~\bibnamefont {Robledo}}, \bibinfo {author} {\bibfnamefont
  {T.~H.}\ \bibnamefont {Taminiau}}, \bibinfo {author} {\bibfnamefont
  {M.}~\bibnamefont {Markham}}, \bibinfo {author} {\bibfnamefont {D.~J.}\
  \bibnamefont {Twitchen}}, \bibinfo {author} {\bibfnamefont {L.}~\bibnamefont
  {Childress}}, \ and\ \bibinfo {author} {\bibfnamefont {R.}~\bibnamefont
  {Hanson}},\ }\bibfield  {title} {\enquote {\bibinfo {title} {Heralded
  entanglement between solid-state qubits separated by three metres},}\ }\href
  {https://www.nature.com/articles/nature12016} {\bibfield  {journal} {\bibinfo
   {journal} {Nature}\ }\textbf {\bibinfo {volume} {497}},\ \bibinfo {pages}
  {86} (\bibinfo {year} {2013})}\BibitemShut {NoStop}%
\bibitem [{\citenamefont {Hensen}\ \emph {et~al.}(2015)\citenamefont {Hensen},
  \citenamefont {Bernien}, \citenamefont {Dr\'{e}au}, \citenamefont {Reiserer},
  \citenamefont {Kalb}, \citenamefont {Blok}, \citenamefont {Ruitenberg},
  \citenamefont {Vermeulen}, \citenamefont {Schouten}, \citenamefont
  {Abell\'{a}n}, \citenamefont {Amaya}, \citenamefont {Pruneri}, \citenamefont
  {Mitchell}, \citenamefont {Markham}, \citenamefont {Twitchen}, \citenamefont
  {Elkouss}, \citenamefont {Wehner}, \citenamefont {Taminiau},\ and\
  \citenamefont {Hanson}}]{HansonLoopholeFree}%
  \BibitemOpen
  \bibfield  {author} {\bibinfo {author} {\bibfnamefont {B.}~\bibnamefont
  {Hensen}}, \bibinfo {author} {\bibfnamefont {H.}~\bibnamefont {Bernien}},
  \bibinfo {author} {\bibfnamefont {A.~E.}\ \bibnamefont {Dr\'{e}au}}, \bibinfo
  {author} {\bibfnamefont {A.}~\bibnamefont {Reiserer}}, \bibinfo {author}
  {\bibfnamefont {N.}~\bibnamefont {Kalb}}, \bibinfo {author} {\bibfnamefont
  {M.~S.}\ \bibnamefont {Blok}}, \bibinfo {author} {\bibfnamefont
  {J.}~\bibnamefont {Ruitenberg}}, \bibinfo {author} {\bibfnamefont {R.~F.~L.}\
  \bibnamefont {Vermeulen}}, \bibinfo {author} {\bibfnamefont {R.~N.}\
  \bibnamefont {Schouten}}, \bibinfo {author} {\bibfnamefont {C.}~\bibnamefont
  {Abell\'{a}n}}, \bibinfo {author} {\bibfnamefont {W.}~\bibnamefont {Amaya}},
  \bibinfo {author} {\bibfnamefont {V.}~\bibnamefont {Pruneri}}, \bibinfo
  {author} {\bibfnamefont {M.~W.}\ \bibnamefont {Mitchell}}, \bibinfo {author}
  {\bibfnamefont {M.}~\bibnamefont {Markham}}, \bibinfo {author} {\bibfnamefont
  {D.~J.}\ \bibnamefont {Twitchen}}, \bibinfo {author} {\bibfnamefont
  {D.}~\bibnamefont {Elkouss}}, \bibinfo {author} {\bibfnamefont
  {S.}~\bibnamefont {Wehner}}, \bibinfo {author} {\bibfnamefont {T.~H.}\
  \bibnamefont {Taminiau}}, \ and\ \bibinfo {author} {\bibfnamefont
  {R.}~\bibnamefont {Hanson}},\ }\bibfield  {title} {\enquote {\bibinfo {title}
  {{Loophole-free Bell inequality violation using electron spins separated by
  1.3 kilometres}},}\ }\href {https://www.nature.com/articles/nature15759}
  {\bibfield  {journal} {\bibinfo  {journal} {Nature}\ }\textbf {\bibinfo
  {volume} {526}},\ \bibinfo {pages} {682} (\bibinfo {year}
  {2015})}\BibitemShut {NoStop}%
\bibitem [{\citenamefont {Ohm}\ and\ \citenamefont {Hassler}(2017)}]{Ohm2017}%
  \BibitemOpen
  \bibfield  {author} {\bibinfo {author} {\bibfnamefont {Christoph}\
  \bibnamefont {Ohm}}\ and\ \bibinfo {author} {\bibfnamefont {Fabian}\
  \bibnamefont {Hassler}},\ }\bibfield  {title} {\enquote {\bibinfo {title}
  {Measurement-induced entanglement of two transmon qubits by a single
  photon},}\ }\href {\doibase 10.1088/1367-2630/aa6d46} {\bibfield  {journal}
  {\bibinfo  {journal} {New Journal of Physics}\ }\textbf {\bibinfo {volume}
  {19}},\ \bibinfo {pages} {053018} (\bibinfo {year} {2017})}\BibitemShut
  {NoStop}%
\bibitem [{\citenamefont {Mintert}\ \emph {et~al.}(2005)\citenamefont
  {Mintert}, \citenamefont {Carvalho}, \citenamefont {Ku\'{s}},\ and\
  \citenamefont {Buchleitner}}]{Mintert2005}%
  \BibitemOpen
  \bibfield  {author} {\bibinfo {author} {\bibfnamefont {Florian}\ \bibnamefont
  {Mintert}}, \bibinfo {author} {\bibfnamefont {Andr\'{e} R.~R.}\ \bibnamefont
  {Carvalho}}, \bibinfo {author} {\bibfnamefont {Marek}\ \bibnamefont
  {Ku\'{s}}}, \ and\ \bibinfo {author} {\bibfnamefont {Andreas}\ \bibnamefont
  {Buchleitner}},\ }\bibfield  {title} {\enquote {\bibinfo {title} {Measures
  and dynamics of entangled states},}\ }\href {\doibase
  https://doi.org/10.1016/j.physrep.2005.04.006} {\bibfield  {journal}
  {\bibinfo  {journal} {Physics Reports}\ }\textbf {\bibinfo {volume} {415}},\
  \bibinfo {pages} {207 -- 259} (\bibinfo {year} {2005})}\BibitemShut {NoStop}%
\bibitem [{\citenamefont {Carvalho}\ \emph {et~al.}(2007)\citenamefont
  {Carvalho}, \citenamefont {Busse}, \citenamefont {Brodier}, \citenamefont
  {Viviescas},\ and\ \citenamefont {Buchleitner}}]{Carvalho2007}%
  \BibitemOpen
  \bibfield  {author} {\bibinfo {author} {\bibfnamefont {Andr\'e R.~R.}\
  \bibnamefont {Carvalho}}, \bibinfo {author} {\bibfnamefont {Marc}\
  \bibnamefont {Busse}}, \bibinfo {author} {\bibfnamefont {Olivier}\
  \bibnamefont {Brodier}}, \bibinfo {author} {\bibfnamefont {Carlos}\
  \bibnamefont {Viviescas}}, \ and\ \bibinfo {author} {\bibfnamefont {Andreas}\
  \bibnamefont {Buchleitner}},\ }\bibfield  {title} {\enquote {\bibinfo {title}
  {Optimal dynamical characterization of entanglement},}\ }\href
  {https://link.aps.org/doi/10.1103/PhysRevLett.98.190501} {\bibfield
  {journal} {\bibinfo  {journal} {Phys. Rev. Lett.}\ }\textbf {\bibinfo
  {volume} {98}},\ \bibinfo {pages} {190501} (\bibinfo {year}
  {2007})}\BibitemShut {NoStop}%
\bibitem [{\citenamefont {Viviescas}\ \emph {et~al.}(2010)\citenamefont
  {Viviescas}, \citenamefont {Guevara}, \citenamefont {Carvalho}, \citenamefont
  {Busse},\ and\ \citenamefont {Buchleitner}}]{Viviescas2010}%
  \BibitemOpen
  \bibfield  {author} {\bibinfo {author} {\bibfnamefont {Carlos}\ \bibnamefont
  {Viviescas}}, \bibinfo {author} {\bibfnamefont {Ivonne}\ \bibnamefont
  {Guevara}}, \bibinfo {author} {\bibfnamefont {Andr\'e R.~R.}\ \bibnamefont
  {Carvalho}}, \bibinfo {author} {\bibfnamefont {Marc}\ \bibnamefont {Busse}},
  \ and\ \bibinfo {author} {\bibfnamefont {Andreas}\ \bibnamefont
  {Buchleitner}},\ }\bibfield  {title} {\enquote {\bibinfo {title}
  {Entanglement dynamics in open two-qubit systems via diffusive quantum
  trajectories},}\ }\href {\doibase 10.1103/PhysRevLett.105.210502} {\bibfield
  {journal} {\bibinfo  {journal} {Phys. Rev. Lett.}\ }\textbf {\bibinfo
  {volume} {105}},\ \bibinfo {pages} {210502} (\bibinfo {year}
  {2010})}\BibitemShut {NoStop}%
\bibitem [{\citenamefont {Mascarenhas}\ \emph {et~al.}(2010)\citenamefont
  {Mascarenhas}, \citenamefont {Marques}, \citenamefont {Cunha},\ and\
  \citenamefont {Santos}}]{Mascarenhas2010}%
  \BibitemOpen
  \bibfield  {author} {\bibinfo {author} {\bibfnamefont {Eduardo}\ \bibnamefont
  {Mascarenhas}}, \bibinfo {author} {\bibfnamefont {Breno}\ \bibnamefont
  {Marques}}, \bibinfo {author} {\bibfnamefont {Marcelo~Terra}\ \bibnamefont
  {Cunha}}, \ and\ \bibinfo {author} {\bibfnamefont {Marcelo~Fran\c{c}a}\
  \bibnamefont {Santos}},\ }\bibfield  {title} {\enquote {\bibinfo {title}
  {Continuous quantum error correction through local operations},}\ }\href
  {\doibase 10.1103/PhysRevA.82.032327} {\bibfield  {journal} {\bibinfo
  {journal} {Phys. Rev. A}\ }\textbf {\bibinfo {volume} {82}},\ \bibinfo
  {pages} {032327} (\bibinfo {year} {2010})}\BibitemShut {NoStop}%
\bibitem [{\citenamefont {Mascarenhas}\ \emph {et~al.}(2011)\citenamefont
  {Mascarenhas}, \citenamefont {Cavalcanti}, \citenamefont {Vedral},\ and\
  \citenamefont {Santos}}]{Mascarenhas2011}%
  \BibitemOpen
  \bibfield  {author} {\bibinfo {author} {\bibfnamefont {Eduardo}\ \bibnamefont
  {Mascarenhas}}, \bibinfo {author} {\bibfnamefont {Daniel}\ \bibnamefont
  {Cavalcanti}}, \bibinfo {author} {\bibfnamefont {Vlatko}\ \bibnamefont
  {Vedral}}, \ and\ \bibinfo {author} {\bibfnamefont {Marcelo~Fran\c{c}a}\
  \bibnamefont {Santos}},\ }\bibfield  {title} {\enquote {\bibinfo {title}
  {Physically realizable entanglement by local continuous measurements},}\
  }\href {\doibase 10.1103/PhysRevA.83.022311} {\bibfield  {journal} {\bibinfo
  {journal} {Phys. Rev. A}\ }\textbf {\bibinfo {volume} {83}},\ \bibinfo
  {pages} {022311} (\bibinfo {year} {2011})}\BibitemShut {NoStop}%
\bibitem [{\citenamefont {Carvalho}\ and\ \citenamefont
  {Santos}(2011)}]{Carvalho2011}%
  \BibitemOpen
  \bibfield  {author} {\bibinfo {author} {\bibfnamefont {Andr\'{e} R.~R.}\
  \bibnamefont {Carvalho}}\ and\ \bibinfo {author} {\bibfnamefont
  {Marcelo~Fran\c{c}a}\ \bibnamefont {Santos}},\ }\bibfield  {title} {\enquote
  {\bibinfo {title} {Distant entanglement protected through artificially
  increased local temperature},}\ }\href {\doibase
  10.1088/1367-2630/13/1/013010} {\bibfield  {journal} {\bibinfo  {journal}
  {New Journal of Physics}\ }\textbf {\bibinfo {volume} {13}},\ \bibinfo
  {pages} {013010} (\bibinfo {year} {2011})}\BibitemShut {NoStop}%
\bibitem [{\citenamefont {Santos}\ \emph {et~al.}(2012)\citenamefont {Santos},
  \citenamefont {Terra~Cunha}, \citenamefont {Chaves},\ and\ \citenamefont
  {Carvalho}}]{Santos2012}%
  \BibitemOpen
  \bibfield  {author} {\bibinfo {author} {\bibfnamefont {M.~F.}\ \bibnamefont
  {Santos}}, \bibinfo {author} {\bibfnamefont {M.}~\bibnamefont {Terra~Cunha}},
  \bibinfo {author} {\bibfnamefont {R.}~\bibnamefont {Chaves}}, \ and\ \bibinfo
  {author} {\bibfnamefont {A.~R.~R.}\ \bibnamefont {Carvalho}},\ }\bibfield
  {title} {\enquote {\bibinfo {title} {Quantum computing with incoherent
  resources and quantum jumps},}\ }\href {\doibase
  10.1103/PhysRevLett.108.170501} {\bibfield  {journal} {\bibinfo  {journal}
  {Phys. Rev. Lett.}\ }\textbf {\bibinfo {volume} {108}},\ \bibinfo {pages}
  {170501} (\bibinfo {year} {2012})}\BibitemShut {NoStop}%
\bibitem [{\citenamefont {Lewalle}\ \emph {et~al.}(2020)\citenamefont
  {Lewalle}, \citenamefont {Manikandan}, \citenamefont {Elouard},\ and\
  \citenamefont {Jordan}}]{FlorTeach2019}%
  \BibitemOpen
  \bibfield  {author} {\bibinfo {author} {\bibfnamefont {Philippe}\
  \bibnamefont {Lewalle}}, \bibinfo {author} {\bibfnamefont {Sreenath~K.}\
  \bibnamefont {Manikandan}}, \bibinfo {author} {\bibfnamefont {Cyril}\
  \bibnamefont {Elouard}}, \ and\ \bibinfo {author} {\bibfnamefont {Andrew~N.}\
  \bibnamefont {Jordan}},\ }\bibfield  {title} {\enquote {\bibinfo {title}
  {Measuring fluorescence to track a quantum emitter's state: A theory
  review},}\ }\href {https://doi.org/10.1080/00107514.2020.1747201} {\bibfield
  {journal} {\bibinfo  {journal} {Contemporary Physics}\ }\textbf {\bibinfo
  {volume} {61}},\ \bibinfo {pages} {26--50} (\bibinfo {year}
  {2020})}\BibitemShut {NoStop}%
\bibitem [{\citenamefont {Lewalle}\ \emph {et~al.}(2019)\citenamefont
  {Lewalle}, \citenamefont {Elouard}, \citenamefont {Manikandan}, \citenamefont
  {Qian}, \citenamefont {Eberly},\ and\ \citenamefont {Jordan}}]{LongFlor2019}%
  \BibitemOpen
  \bibfield  {author} {\bibinfo {author} {\bibfnamefont {Philippe}\
  \bibnamefont {Lewalle}}, \bibinfo {author} {\bibfnamefont {Cyril}\
  \bibnamefont {Elouard}}, \bibinfo {author} {\bibfnamefont {Sreenath~K.}\
  \bibnamefont {Manikandan}}, \bibinfo {author} {\bibfnamefont {Xiao-Feng}\
  \bibnamefont {Qian}}, \bibinfo {author} {\bibfnamefont {Joseph~H.}\
  \bibnamefont {Eberly}}, \ and\ \bibinfo {author} {\bibfnamefont {Andrew~N.}\
  \bibnamefont {Jordan}},\ }\bibfield  {title} {\enquote {\bibinfo {title}
  {Entanglement of a pair of quantum emitters under continuous fluorescence
  measurements},}\ }\href {https://arxiv.org/abs/1910.01206} {\bibfield
  {journal} {\bibinfo  {journal} {arXiv:1910.01206}\ } (\bibinfo {year}
  {2019})}\BibitemShut {NoStop}%
\bibitem [{\citenamefont {Martin}\ and\ \citenamefont
  {Whaley}(2019)}]{Leigh2019}%
  \BibitemOpen
  \bibfield  {author} {\bibinfo {author} {\bibfnamefont {Leigh~S.}\
  \bibnamefont {Martin}}\ and\ \bibinfo {author} {\bibfnamefont {K.~Birgitta}\
  \bibnamefont {Whaley}},\ }\bibfield  {title} {\enquote {\bibinfo {title}
  {Single-shot deterministic entanglement between non-interacting systems with
  linear optics},}\ }\href {https://arxiv.org/abs/1912.00067} {\bibfield
  {journal} {\bibinfo  {journal} {arXiv:1912.00067}\ } (\bibinfo {year}
  {2019})}\BibitemShut {NoStop}%
\bibitem [{\citenamefont {Carmichael}(1993)}]{BookCarmichael}%
  \BibitemOpen
  \bibfield  {author} {\bibinfo {author} {\bibfnamefont {H.~J.}\ \bibnamefont
  {Carmichael}},\ }\href@noop {} {\emph {\bibinfo {title} {An Open Systems
  Approach to Quantum Optics}}}\ (\bibinfo  {publisher} {Springer, Berlin},\
  \bibinfo {year} {1993})\BibitemShut {NoStop}%
\bibitem [{\citenamefont {Percival}(1998)}]{BookPercival}%
  \BibitemOpen
  \bibfield  {author} {\bibinfo {author} {\bibfnamefont {I.~C.}\ \bibnamefont
  {Percival}},\ }\href {https://books.google.com/books?id=AlXSmZTHxtwC} {\emph
  {\bibinfo {title} {Quantum State Diffusion}}}\ (\bibinfo  {publisher}
  {Cambridge University Press},\ \bibinfo {year} {1998})\BibitemShut {NoStop}%
\bibitem [{\citenamefont {Gardiner}\ and\ \citenamefont
  {Zoller}(2004)}]{BookGardiner}%
  \BibitemOpen
  \bibfield  {author} {\bibinfo {author} {\bibfnamefont {C.~W.}\ \bibnamefont
  {Gardiner}}\ and\ \bibinfo {author} {\bibfnamefont {P.}~\bibnamefont
  {Zoller}},\ }\href@noop {} {\emph {\bibinfo {title} {{Quantum Noise: A
  Handbook of Markovian and Non-Markovian Quantum Stochastic Methods with
  Applications to Quantum Optics}}}}\ (\bibinfo  {publisher} {Springer},\
  \bibinfo {year} {2004})\BibitemShut {NoStop}%
\bibitem [{\citenamefont {Wiseman}\ and\ \citenamefont
  {Milburn}(2010)}]{BookWiseman}%
  \BibitemOpen
  \bibfield  {author} {\bibinfo {author} {\bibfnamefont {H.~M.}\ \bibnamefont
  {Wiseman}}\ and\ \bibinfo {author} {\bibfnamefont {G.~J.}\ \bibnamefont
  {Milburn}},\ }\href@noop {} {\emph {\bibinfo {title} {Quantum measurement and
  control}}}\ (\bibinfo  {publisher} {Cambridge University Press},\ \bibinfo
  {year} {2010})\BibitemShut {NoStop}%
\bibitem [{\citenamefont {Barchielli}\ and\ \citenamefont
  {Gregoratti}(2009)}]{BookBarchielli}%
  \BibitemOpen
  \bibfield  {author} {\bibinfo {author} {\bibfnamefont {A.}~\bibnamefont
  {Barchielli}}\ and\ \bibinfo {author} {\bibfnamefont {M.}~\bibnamefont
  {Gregoratti}},\ }\href@noop {} {\emph {\bibinfo {title} {Quantum trajectories
  and measurements in continuous time}}}\ (\bibinfo  {publisher}
  {Springer-Verlag Berlin Heidelberg},\ \bibinfo {year} {2009})\BibitemShut
  {NoStop}%
\bibitem [{\citenamefont {Jacobs}(2014)}]{BookJacobs}%
  \BibitemOpen
  \bibfield  {author} {\bibinfo {author} {\bibfnamefont {Kurt}\ \bibnamefont
  {Jacobs}},\ }\href@noop {} {\emph {\bibinfo {title} {{Quantum Measurement
  Theory and its Applications}}}}\ (\bibinfo  {publisher} {Cambridge University
  Press},\ \bibinfo {year} {2014})\BibitemShut {NoStop}%
\bibitem [{\citenamefont {Wiseman}(1996)}]{Wiseman1996Review}%
  \BibitemOpen
  \bibfield  {author} {\bibinfo {author} {\bibfnamefont {H.~M.}\ \bibnamefont
  {Wiseman}},\ }\bibfield  {title} {\enquote {\bibinfo {title} {Quantum
  trajectories and quantum measurement theory},}\ }\href
  {https://iopscience.iop.org/article/10.1088/1355-5111/8/1/015} {\bibfield
  {journal} {\bibinfo  {journal} {Quantum and Semiclassical Optics: Journal of
  the European Optical Society Part B}\ }\textbf {\bibinfo {volume} {8}},\
  \bibinfo {pages} {205--222} (\bibinfo {year} {1996})}\BibitemShut {NoStop}%
\bibitem [{\citenamefont {Brun}({2002})}]{Brun2001Teach}%
  \BibitemOpen
  \bibfield  {author} {\bibinfo {author} {\bibfnamefont {Todd~A.}\ \bibnamefont
  {Brun}},\ }\bibfield  {title} {\enquote {\bibinfo {title} {A simple model of
  quantum trajectories},}\ }\href {\doibase 10.1119/1.1475328} {\bibfield
  {journal} {\bibinfo  {journal} {{American Journal of Physics}}\ }\textbf
  {\bibinfo {volume} {70}},\ \bibinfo {pages} {719} (\bibinfo {year}
  {{2002}})}\BibitemShut {NoStop}%
\bibitem [{\citenamefont {Jacobs}\ and\ \citenamefont
  {Steck}(2006)}]{Jacobs2006}%
  \BibitemOpen
  \bibfield  {author} {\bibinfo {author} {\bibfnamefont {Kurt}\ \bibnamefont
  {Jacobs}}\ and\ \bibinfo {author} {\bibfnamefont {Daniel~A.}\ \bibnamefont
  {Steck}},\ }\bibfield  {title} {\enquote {\bibinfo {title} {A straightforward
  introduction to continuous quantum measurement},}\ }\href {\doibase
  10.1080/00107510601101934} {\bibfield  {journal} {\bibinfo  {journal}
  {Contemporary Physics}\ }\textbf {\bibinfo {volume} {47}},\ \bibinfo {pages}
  {279--303} (\bibinfo {year} {2006})}\BibitemShut {NoStop}%
\bibitem [{\citenamefont {Zhang}\ \emph {et~al.}(2017)\citenamefont {Zhang},
  \citenamefont {xi~Liu}, \citenamefont {Wu}, \citenamefont {Jacobs},\ and\
  \citenamefont {Nori}}]{Zhang2017}%
  \BibitemOpen
  \bibfield  {author} {\bibinfo {author} {\bibfnamefont {Jing}\ \bibnamefont
  {Zhang}}, \bibinfo {author} {\bibfnamefont {Yu}~\bibnamefont {xi~Liu}},
  \bibinfo {author} {\bibfnamefont {Re-Bing}\ \bibnamefont {Wu}}, \bibinfo
  {author} {\bibfnamefont {Kurt}\ \bibnamefont {Jacobs}}, \ and\ \bibinfo
  {author} {\bibfnamefont {Franco}\ \bibnamefont {Nori}},\ }\bibfield  {title}
  {\enquote {\bibinfo {title} {Quantum feedback: Theory, experiments, and
  applications},}\ }\href {\doibase
  http://doi.org/10.1016/j.physrep.2017.02.003} {\bibfield  {journal} {\bibinfo
   {journal} {Physics Reports}\ }\textbf {\bibinfo {volume} {679}},\ \bibinfo
  {pages} {1 -- 60} (\bibinfo {year} {2017})}\BibitemShut {NoStop}%
\bibitem [{\citenamefont {Wiseman}(1994)}]{Wiseman1994}%
  \BibitemOpen
  \bibfield  {author} {\bibinfo {author} {\bibfnamefont {H.~M.}\ \bibnamefont
  {Wiseman}},\ }\bibfield  {title} {\enquote {\bibinfo {title} {Quantum theory
  of continuous feedback},}\ }\href {\doibase 10.1103/PhysRevA.49.2133}
  {\bibfield  {journal} {\bibinfo  {journal} {Phys. Rev. A}\ }\textbf {\bibinfo
  {volume} {49}},\ \bibinfo {pages} {2133--2150} (\bibinfo {year}
  {1994})}\BibitemShut {NoStop}%
\bibitem [{\citenamefont {Ahn}\ \emph {et~al.}(2002)\citenamefont {Ahn},
  \citenamefont {Doherty},\ and\ \citenamefont {Landahl}}]{Ahn2002}%
  \BibitemOpen
  \bibfield  {author} {\bibinfo {author} {\bibfnamefont {Charlene}\
  \bibnamefont {Ahn}}, \bibinfo {author} {\bibfnamefont {Andrew~C.}\
  \bibnamefont {Doherty}}, \ and\ \bibinfo {author} {\bibfnamefont {Andrew~J.}\
  \bibnamefont {Landahl}},\ }\bibfield  {title} {\enquote {\bibinfo {title}
  {Continuous quantum error correction via quantum feedback control},}\ }\href
  {\doibase 10.1103/PhysRevA.65.042301} {\bibfield  {journal} {\bibinfo
  {journal} {Phys. Rev. A}\ }\textbf {\bibinfo {volume} {65}},\ \bibinfo
  {pages} {042301} (\bibinfo {year} {2002})}\BibitemShut {NoStop}%
\bibitem [{\citenamefont {Ahn}\ \emph {et~al.}(2003{\natexlab{a}})\citenamefont
  {Ahn}, \citenamefont {Wiseman},\ and\ \citenamefont {Milburn}}]{Ahn2003}%
  \BibitemOpen
  \bibfield  {author} {\bibinfo {author} {\bibfnamefont {C.}~\bibnamefont
  {Ahn}}, \bibinfo {author} {\bibfnamefont {H.~M.}\ \bibnamefont {Wiseman}}, \
  and\ \bibinfo {author} {\bibfnamefont {G.~J.}\ \bibnamefont {Milburn}},\
  }\bibfield  {title} {\enquote {\bibinfo {title} {Quantum error correction for
  continuously detected errors},}\ }in\ \href {\doibase
  10.1109/PHYCON.2003.1237011} {\emph {\bibinfo {booktitle} {2003 IEEE
  International Workshop on Workload Characterization (IEEE Cat.
  No.03EX775)}}},\ Vol.~\bibinfo {volume} {3}\ (\bibinfo {year} {2003})\ pp.\
  \bibinfo {pages} {834--839 vol.3}\BibitemShut {NoStop}%
\bibitem [{\citenamefont {Ahn}\ \emph {et~al.}(2003{\natexlab{b}})\citenamefont
  {Ahn}, \citenamefont {Wiseman},\ and\ \citenamefont {Milburn}}]{Ahn2003-2}%
  \BibitemOpen
  \bibfield  {author} {\bibinfo {author} {\bibfnamefont {C.}~\bibnamefont
  {Ahn}}, \bibinfo {author} {\bibfnamefont {H.M.}\ \bibnamefont {Wiseman}}, \
  and\ \bibinfo {author} {\bibfnamefont {G.J.}\ \bibnamefont {Milburn}},\
  }\bibfield  {title} {\enquote {\bibinfo {title} {Quantum control and quantum
  entanglement},}\ }\href {\doibase https://doi.org/10.3166/ejc.9.279-284}
  {\bibfield  {journal} {\bibinfo  {journal} {European Journal of Control}\
  }\textbf {\bibinfo {volume} {9}},\ \bibinfo {pages} {279 -- 284} (\bibinfo
  {year} {2003}{\natexlab{b}})}\BibitemShut {NoStop}%
\bibitem [{\citenamefont {Ahn}\ \emph {et~al.}(2003{\natexlab{c}})\citenamefont
  {Ahn}, \citenamefont {Wiseman},\ and\ \citenamefont {Milburn}}]{Ahn2003-3}%
  \BibitemOpen
  \bibfield  {author} {\bibinfo {author} {\bibfnamefont {Charlene}\
  \bibnamefont {Ahn}}, \bibinfo {author} {\bibfnamefont {H.~M.}\ \bibnamefont
  {Wiseman}}, \ and\ \bibinfo {author} {\bibfnamefont {G.~J.}\ \bibnamefont
  {Milburn}},\ }\bibfield  {title} {\enquote {\bibinfo {title} {Quantum error
  correction for continuously detected errors},}\ }\href {\doibase
  10.1103/PhysRevA.67.052310} {\bibfield  {journal} {\bibinfo  {journal} {Phys.
  Rev. A}\ }\textbf {\bibinfo {volume} {67}},\ \bibinfo {pages} {052310}
  (\bibinfo {year} {2003}{\natexlab{c}})}\BibitemShut {NoStop}%
\bibitem [{\citenamefont {Sarovar}\ \emph {et~al.}(2004)\citenamefont
  {Sarovar}, \citenamefont {Ahn}, \citenamefont {Jacobs},\ and\ \citenamefont
  {Milburn}}]{Sarovar2004}%
  \BibitemOpen
  \bibfield  {author} {\bibinfo {author} {\bibfnamefont {Mohan}\ \bibnamefont
  {Sarovar}}, \bibinfo {author} {\bibfnamefont {Charlene}\ \bibnamefont {Ahn}},
  \bibinfo {author} {\bibfnamefont {Kurt}\ \bibnamefont {Jacobs}}, \ and\
  \bibinfo {author} {\bibfnamefont {Gerard~J.}\ \bibnamefont {Milburn}},\
  }\bibfield  {title} {\enquote {\bibinfo {title} {Practical scheme for error
  control using feedback},}\ }\href {\doibase 10.1103/PhysRevA.69.052324}
  {\bibfield  {journal} {\bibinfo  {journal} {Phys. Rev. A}\ }\textbf {\bibinfo
  {volume} {69}},\ \bibinfo {pages} {052324} (\bibinfo {year}
  {2004})}\BibitemShut {NoStop}%
\bibitem [{\citenamefont {Mancini}\ and\ \citenamefont
  {Wiseman}(2007)}]{Mancini2007}%
  \BibitemOpen
  \bibfield  {author} {\bibinfo {author} {\bibfnamefont {Stefano}\ \bibnamefont
  {Mancini}}\ and\ \bibinfo {author} {\bibfnamefont {Howard~M.}\ \bibnamefont
  {Wiseman}},\ }\bibfield  {title} {\enquote {\bibinfo {title} {Optimal control
  of entanglement via quantum feedback},}\ }\href {\doibase
  10.1103/PhysRevA.75.012330} {\bibfield  {journal} {\bibinfo  {journal} {Phys.
  Rev. A}\ }\textbf {\bibinfo {volume} {75}},\ \bibinfo {pages} {012330}
  (\bibinfo {year} {2007})}\BibitemShut {NoStop}%
\bibitem [{\citenamefont {Carvalho}\ and\ \citenamefont
  {Hope}(2007)}]{Carvalho2007-2}%
  \BibitemOpen
  \bibfield  {author} {\bibinfo {author} {\bibfnamefont {A.~R.~R.}\
  \bibnamefont {Carvalho}}\ and\ \bibinfo {author} {\bibfnamefont {J.~J.}\
  \bibnamefont {Hope}},\ }\bibfield  {title} {\enquote {\bibinfo {title}
  {Stabilizing entanglement by quantum-jump-based feedback},}\ }\href {\doibase
  10.1103/PhysRevA.76.010301} {\bibfield  {journal} {\bibinfo  {journal} {Phys.
  Rev. A}\ }\textbf {\bibinfo {volume} {76}},\ \bibinfo {pages} {010301(R)}
  (\bibinfo {year} {2007})}\BibitemShut {NoStop}%
\bibitem [{\citenamefont {Carvalho}\ \emph {et~al.}(2008)\citenamefont
  {Carvalho}, \citenamefont {Reid},\ and\ \citenamefont {Hope}}]{Carvalho2008}%
  \BibitemOpen
  \bibfield  {author} {\bibinfo {author} {\bibfnamefont {A.~R.~R.}\
  \bibnamefont {Carvalho}}, \bibinfo {author} {\bibfnamefont {A.~J.~S.}\
  \bibnamefont {Reid}}, \ and\ \bibinfo {author} {\bibfnamefont {J.~J.}\
  \bibnamefont {Hope}},\ }\bibfield  {title} {\enquote {\bibinfo {title}
  {Controlling entanglement by direct quantum feedback},}\ }\href {\doibase
  10.1103/PhysRevA.78.012334} {\bibfield  {journal} {\bibinfo  {journal} {Phys.
  Rev. A}\ }\textbf {\bibinfo {volume} {78}},\ \bibinfo {pages} {012334}
  (\bibinfo {year} {2008})}\BibitemShut {NoStop}%
\bibitem [{\citenamefont {Hill}\ and\ \citenamefont {Ralph}(2008)}]{Hill2008}%
  \BibitemOpen
  \bibfield  {author} {\bibinfo {author} {\bibfnamefont {Charles}\ \bibnamefont
  {Hill}}\ and\ \bibinfo {author} {\bibfnamefont {Jason}\ \bibnamefont
  {Ralph}},\ }\bibfield  {title} {\enquote {\bibinfo {title} {Weak measurement
  and control of entanglement generation},}\ }\href {\doibase
  10.1103/PhysRevA.77.014305} {\bibfield  {journal} {\bibinfo  {journal} {Phys.
  Rev. A}\ }\textbf {\bibinfo {volume} {77}},\ \bibinfo {pages} {014305}
  (\bibinfo {year} {2008})}\BibitemShut {NoStop}%
\bibitem [{\citenamefont {Serafini}\ and\ \citenamefont
  {Mancini}(2010)}]{Serafini2010}%
  \BibitemOpen
  \bibfield  {author} {\bibinfo {author} {\bibfnamefont {Alessio}\ \bibnamefont
  {Serafini}}\ and\ \bibinfo {author} {\bibfnamefont {Stefano}\ \bibnamefont
  {Mancini}},\ }\bibfield  {title} {\enquote {\bibinfo {title} {Determination
  of maximal {G}aussian entanglement achievable by feedback-controlled
  dynamics},}\ }\href {\doibase 10.1103/PhysRevLett.104.220501} {\bibfield
  {journal} {\bibinfo  {journal} {Phys. Rev. Lett.}\ }\textbf {\bibinfo
  {volume} {104}},\ \bibinfo {pages} {220501} (\bibinfo {year}
  {2010})}\BibitemShut {NoStop}%
\bibitem [{\citenamefont {Vijay}\ \emph {et~al.}(2012)\citenamefont {Vijay},
  \citenamefont {Macklin}, \citenamefont {Slichter}, \citenamefont {Weber},
  \citenamefont {Murch}, \citenamefont {Naik}, \citenamefont {Korotkov},\ and\
  \citenamefont {Siddiqi}}]{Vijay2012}%
  \BibitemOpen
  \bibfield  {author} {\bibinfo {author} {\bibfnamefont {R.}~\bibnamefont
  {Vijay}}, \bibinfo {author} {\bibfnamefont {C.}~\bibnamefont {Macklin}},
  \bibinfo {author} {\bibfnamefont {D.~H.}\ \bibnamefont {Slichter}}, \bibinfo
  {author} {\bibfnamefont {S.~J.}\ \bibnamefont {Weber}}, \bibinfo {author}
  {\bibfnamefont {K.~W.}\ \bibnamefont {Murch}}, \bibinfo {author}
  {\bibfnamefont {R.}~\bibnamefont {Naik}}, \bibinfo {author} {\bibfnamefont
  {A.~N.}\ \bibnamefont {Korotkov}}, \ and\ \bibinfo {author} {\bibfnamefont
  {I.}~\bibnamefont {Siddiqi}},\ }\bibfield  {title} {\enquote {\bibinfo
  {title} {Stabilizing {R}abi oscillations in a superconducting qubit using
  quantum feedback},}\ }\href {\doibase 10.1038/nature11505} {\bibfield
  {journal} {\bibinfo  {journal} {Nature}\ }\textbf {\bibinfo {volume} {490}},\
  \bibinfo {pages} {77} (\bibinfo {year} {2012})}\BibitemShut {NoStop}%
\bibitem [{\citenamefont {Balouchi}\ and\ \citenamefont
  {Jacobs}(2014)}]{Balouchi_2014}%
  \BibitemOpen
  \bibfield  {author} {\bibinfo {author} {\bibfnamefont {Ashkan}\ \bibnamefont
  {Balouchi}}\ and\ \bibinfo {author} {\bibfnamefont {Kurt}\ \bibnamefont
  {Jacobs}},\ }\bibfield  {title} {\enquote {\bibinfo {title} {Optimal
  measurement-based feedback control for a single qubit: a candidate
  protocol},}\ }\href {\doibase 10.1088/1367-2630/16/9/093059} {\bibfield
  {journal} {\bibinfo  {journal} {New Journal of Physics}\ }\textbf {\bibinfo
  {volume} {16}},\ \bibinfo {pages} {093059} (\bibinfo {year}
  {2014})}\BibitemShut {NoStop}%
\bibitem [{\citenamefont {Meyer~zu Rheda}\ \emph {et~al.}(2014)\citenamefont
  {Meyer~zu Rheda}, \citenamefont {Haack},\ and\ \citenamefont
  {Romito}}]{Rheda2014}%
  \BibitemOpen
  \bibfield  {author} {\bibinfo {author} {\bibfnamefont {Clemens}\ \bibnamefont
  {Meyer~zu Rheda}}, \bibinfo {author} {\bibfnamefont {G\'eraldine}\
  \bibnamefont {Haack}}, \ and\ \bibinfo {author} {\bibfnamefont {Alessandro}\
  \bibnamefont {Romito}},\ }\bibfield  {title} {\enquote {\bibinfo {title}
  {On-demand maximally entangled states with a parity meter and continuous
  feedback},}\ }\href {\doibase 10.1103/PhysRevB.90.155438} {\bibfield
  {journal} {\bibinfo  {journal} {Phys. Rev. B}\ }\textbf {\bibinfo {volume}
  {90}},\ \bibinfo {pages} {155438} (\bibinfo {year} {2014})}\BibitemShut
  {NoStop}%
\bibitem [{\citenamefont {Szigeti}\ \emph {et~al.}(2014)\citenamefont
  {Szigeti}, \citenamefont {Carvalho}, \citenamefont {Morley},\ and\
  \citenamefont {Hush}}]{Szigeti2014}%
  \BibitemOpen
  \bibfield  {author} {\bibinfo {author} {\bibfnamefont {Stuart~S.}\
  \bibnamefont {Szigeti}}, \bibinfo {author} {\bibfnamefont {Andre R.~R.}\
  \bibnamefont {Carvalho}}, \bibinfo {author} {\bibfnamefont {James~G.}\
  \bibnamefont {Morley}}, \ and\ \bibinfo {author} {\bibfnamefont {Michael~R.}\
  \bibnamefont {Hush}},\ }\bibfield  {title} {\enquote {\bibinfo {title}
  {Ignorance is bliss: General and robust cancellation of decoherence via
  no-knowledge quantum feedback},}\ }\href {\doibase
  10.1103/PhysRevLett.113.020407} {\bibfield  {journal} {\bibinfo  {journal}
  {Phys. Rev. Lett.}\ }\textbf {\bibinfo {volume} {113}},\ \bibinfo {pages}
  {020407} (\bibinfo {year} {2014})}\BibitemShut {NoStop}%
\bibitem [{\citenamefont {Hsu}\ and\ \citenamefont {Brun}(2016)}]{Hsu2016}%
  \BibitemOpen
  \bibfield  {author} {\bibinfo {author} {\bibfnamefont {Kung-Chuan}\
  \bibnamefont {Hsu}}\ and\ \bibinfo {author} {\bibfnamefont {Todd~A.}\
  \bibnamefont {Brun}},\ }\bibfield  {title} {\enquote {\bibinfo {title}
  {Method for quantum-jump continuous-time quantum error correction},}\ }\href
  {\doibase 10.1103/PhysRevA.93.022321} {\bibfield  {journal} {\bibinfo
  {journal} {Phys. Rev. A}\ }\textbf {\bibinfo {volume} {93}},\ \bibinfo
  {pages} {022321} (\bibinfo {year} {2016})}\BibitemShut {NoStop}%
\bibitem [{\citenamefont {Magazz{\`{u}}}\ \emph {et~al.}(2018)\citenamefont
  {Magazz{\`{u}}}, \citenamefont {Jaramillo}, \citenamefont {Talkner},\ and\
  \citenamefont {H\"{a}nggi}}]{Magazzu2018}%
  \BibitemOpen
  \bibfield  {author} {\bibinfo {author} {\bibfnamefont {L}~\bibnamefont
  {Magazz{\`{u}}}}, \bibinfo {author} {\bibfnamefont {J.~D.}\ \bibnamefont
  {Jaramillo}}, \bibinfo {author} {\bibfnamefont {P.}~\bibnamefont {Talkner}},
  \ and\ \bibinfo {author} {\bibfnamefont {P.}~\bibnamefont {H\"{a}nggi}},\
  }\bibfield  {title} {\enquote {\bibinfo {title} {{Generation and
  stabilization of Bell states via repeated projective measurements on a driven
  ancilla qubit}},}\ }\href {\doibase 10.1088/1402-4896/aabeb8} {\bibfield
  {journal} {\bibinfo  {journal} {Physica Scripta}\ }\textbf {\bibinfo {volume}
  {93}},\ \bibinfo {pages} {064001} (\bibinfo {year} {2018})}\BibitemShut
  {NoStop}%
\bibitem [{\citenamefont {Zhang}\ \emph {et~al.}(2020)\citenamefont {Zhang},
  \citenamefont {Martin},\ and\ \citenamefont {Whaley}}]{Zhang2018}%
  \BibitemOpen
  \bibfield  {author} {\bibinfo {author} {\bibfnamefont {Song}\ \bibnamefont
  {Zhang}}, \bibinfo {author} {\bibfnamefont {Leigh~S.}\ \bibnamefont
  {Martin}}, \ and\ \bibinfo {author} {\bibfnamefont {K.~Birgitta}\
  \bibnamefont {Whaley}},\ }\bibfield  {title} {\enquote {\bibinfo {title}
  {Locally optimal measurement-based quantum feedback with application to
  multiqubit entanglement generation},}\ }\href {\doibase
  10.1103/PhysRevA.102.062418} {\bibfield  {journal} {\bibinfo  {journal}
  {Phys. Rev. A}\ }\textbf {\bibinfo {volume} {102}},\ \bibinfo {pages}
  {062418} (\bibinfo {year} {2020})}\BibitemShut {NoStop}%
\bibitem [{\citenamefont {Hacohen-Gourgy}\ \emph {et~al.}(2018)\citenamefont
  {Hacohen-Gourgy}, \citenamefont {Garc\'{\i}a-Pintos}, \citenamefont {Martin},
  \citenamefont {Dressel},\ and\ \citenamefont {Siddiqi}}]{Gourgy2018}%
  \BibitemOpen
  \bibfield  {author} {\bibinfo {author} {\bibfnamefont {S.}~\bibnamefont
  {Hacohen-Gourgy}}, \bibinfo {author} {\bibfnamefont {L.~P.}\ \bibnamefont
  {Garc\'{\i}a-Pintos}}, \bibinfo {author} {\bibfnamefont {L.~S.}\ \bibnamefont
  {Martin}}, \bibinfo {author} {\bibfnamefont {J.}~\bibnamefont {Dressel}}, \
  and\ \bibinfo {author} {\bibfnamefont {I.}~\bibnamefont {Siddiqi}},\
  }\bibfield  {title} {\enquote {\bibinfo {title} {{Incoherent Qubit Control
  Using the Quantum Zeno Effect}},}\ }\href {\doibase
  10.1103/PhysRevLett.120.020505} {\bibfield  {journal} {\bibinfo  {journal}
  {Phys. Rev. Lett.}\ }\textbf {\bibinfo {volume} {120}},\ \bibinfo {pages}
  {020505} (\bibinfo {year} {2018})}\BibitemShut {NoStop}%
\bibitem [{\citenamefont {Minev}\ \emph {et~al.}(2019)\citenamefont {Minev},
  \citenamefont {Mundhada}, \citenamefont {Shankar}, \citenamefont {Reinhold},
  \citenamefont {Gutierrez-Jauregui}, \citenamefont {Schoelkopf}, \citenamefont
  {Mirrahimi}, \citenamefont {Carmichael},\ and\ \citenamefont
  {Devoret}}]{Minev2018}%
  \BibitemOpen
  \bibfield  {author} {\bibinfo {author} {\bibfnamefont {Z.~K.}\ \bibnamefont
  {Minev}}, \bibinfo {author} {\bibfnamefont {S.~O.}\ \bibnamefont {Mundhada}},
  \bibinfo {author} {\bibfnamefont {S.}~\bibnamefont {Shankar}}, \bibinfo
  {author} {\bibfnamefont {P.}~\bibnamefont {Reinhold}}, \bibinfo {author}
  {\bibfnamefont {R.}~\bibnamefont {Gutierrez-Jauregui}}, \bibinfo {author}
  {\bibfnamefont {R.~J.}\ \bibnamefont {Schoelkopf}}, \bibinfo {author}
  {\bibfnamefont {M.}~\bibnamefont {Mirrahimi}}, \bibinfo {author}
  {\bibfnamefont {H.~J.}\ \bibnamefont {Carmichael}}, \ and\ \bibinfo {author}
  {\bibfnamefont {M.~H.}\ \bibnamefont {Devoret}},\ }\bibfield  {title}
  {\enquote {\bibinfo {title} {To catch and reverse a quantum jump
  mid-flight},}\ }\href {https://www.nature.com/articles/s41586-019-1287-z}
  {\bibfield  {journal} {\bibinfo  {journal} {Nature}\ }\textbf {\bibinfo
  {volume} {570}},\ \bibinfo {pages} {200} (\bibinfo {year}
  {2019})}\BibitemShut {NoStop}%
\bibitem [{\citenamefont {Martin}\ \emph {et~al.}(2019)\citenamefont {Martin},
  \citenamefont {Livingston}, \citenamefont {Hacohen-Gourgy}, \citenamefont
  {Wiseman},\ and\ \citenamefont {Siddiqi}}]{Martin2019}%
  \BibitemOpen
  \bibfield  {author} {\bibinfo {author} {\bibfnamefont {Leigh~S.}\
  \bibnamefont {Martin}}, \bibinfo {author} {\bibfnamefont {William~P.}\
  \bibnamefont {Livingston}}, \bibinfo {author} {\bibfnamefont {Shay}\
  \bibnamefont {Hacohen-Gourgy}}, \bibinfo {author} {\bibfnamefont {Howard~M.}\
  \bibnamefont {Wiseman}}, \ and\ \bibinfo {author} {\bibfnamefont {Irfan}\
  \bibnamefont {Siddiqi}},\ }\bibfield  {title} {\enquote {\bibinfo {title}
  {Implementation of a canonical phase measurement with quantum feedback},}\
  }\href {https://arxiv.org/abs/1906.07274} {\bibfield  {journal} {\bibinfo
  {journal} {arXiv:1906.07274}\ } (\bibinfo {year} {2019})}\BibitemShut
  {NoStop}%
\bibitem [{\citenamefont {Cardona}\ \emph {et~al.}(2019)\citenamefont
  {Cardona}, \citenamefont {Sarlette},\ and\ \citenamefont
  {Rouchon}}]{Cardona2019}%
  \BibitemOpen
  \bibfield  {author} {\bibinfo {author} {\bibfnamefont {Gerardo}\ \bibnamefont
  {Cardona}}, \bibinfo {author} {\bibfnamefont {Alain}\ \bibnamefont
  {Sarlette}}, \ and\ \bibinfo {author} {\bibfnamefont {Pierre}\ \bibnamefont
  {Rouchon}},\ }\bibfield  {title} {\enquote {\bibinfo {title}
  {Continuous--time quantum error correction with noise-assisted quantum
  feedback},}\ }\href {https://arxiv.org/abs/1902.00115} {\bibfield  {journal}
  {\bibinfo  {journal} {arXiv:1902.00115}\ } (\bibinfo {year}
  {2019})}\BibitemShut {NoStop}%
\bibitem [{\citenamefont {Mohseninia}\ \emph {et~al.}(2020)\citenamefont
  {Mohseninia}, \citenamefont {Yang}, \citenamefont {Siddiqi}, \citenamefont
  {Jordan},\ and\ \citenamefont {Dressel}}]{Mohseninia2019}%
  \BibitemOpen
  \bibfield  {author} {\bibinfo {author} {\bibfnamefont {Razieh}\ \bibnamefont
  {Mohseninia}}, \bibinfo {author} {\bibfnamefont {Jing}\ \bibnamefont {Yang}},
  \bibinfo {author} {\bibfnamefont {Irfan}\ \bibnamefont {Siddiqi}}, \bibinfo
  {author} {\bibfnamefont {Andrew~N.}\ \bibnamefont {Jordan}}, \ and\ \bibinfo
  {author} {\bibfnamefont {Justin}\ \bibnamefont {Dressel}},\ }\bibfield
  {title} {\enquote {\bibinfo {title} {Always-{O}n {Q}uantum {E}rror {T}racking
  with {C}ontinuous {P}arity {M}easurements},}\ }\href {\doibase
  10.22331/q-2020-11-04-358} {\bibfield  {journal} {\bibinfo  {journal}
  {{Quantum}}\ }\textbf {\bibinfo {volume} {4}},\ \bibinfo {pages} {358}
  (\bibinfo {year} {2020})}\BibitemShut {NoStop}%
\bibitem [{\citenamefont {Bolund}\ and\ \citenamefont
  {M\o{}lmer}(2014)}]{Bolund2014}%
  \BibitemOpen
  \bibfield  {author} {\bibinfo {author} {\bibfnamefont {Anders}\ \bibnamefont
  {Bolund}}\ and\ \bibinfo {author} {\bibfnamefont {Klaus}\ \bibnamefont
  {M\o{}lmer}},\ }\bibfield  {title} {\enquote {\bibinfo {title} {Stochastic
  excitation during the decay of a two-level emitter subject to homodyne and
  heterodyne detection},}\ }\href {\doibase 10.1103/PhysRevA.89.023827}
  {\bibfield  {journal} {\bibinfo  {journal} {Phys. Rev. A}\ }\textbf {\bibinfo
  {volume} {89}},\ \bibinfo {pages} {023827} (\bibinfo {year}
  {2014})}\BibitemShut {NoStop}%
\bibitem [{\citenamefont {Jordan}\ \emph {et~al.}(2015)\citenamefont {Jordan},
  \citenamefont {Chantasri}, \citenamefont {Rouchon},\ and\ \citenamefont
  {Huard}}]{Jordan2015flor}%
  \BibitemOpen
  \bibfield  {author} {\bibinfo {author} {\bibfnamefont {Andrew~N.}\
  \bibnamefont {Jordan}}, \bibinfo {author} {\bibfnamefont {Areeya}\
  \bibnamefont {Chantasri}}, \bibinfo {author} {\bibfnamefont {Pierre}\
  \bibnamefont {Rouchon}}, \ and\ \bibinfo {author} {\bibfnamefont {Benjamin}\
  \bibnamefont {Huard}},\ }\bibfield  {title} {\enquote {\bibinfo {title}
  {Anatomy of fluorescence: Quantum trajectory statistics from continuously
  measuring spontaneous emission},}\ }\href {\doibase
  10.1007/s40509-016-0075-9} {\bibfield  {journal} {\bibinfo  {journal}
  {Quantum Studies: Math. and Found.}\ }\textbf {\bibinfo {volume} {3}},\
  \bibinfo {pages} {237} (\bibinfo {year} {2015})}\BibitemShut {NoStop}%
\bibitem [{\citenamefont {Campagne-Ibarcq}\ \emph
  {et~al.}(2016{\natexlab{a}})\citenamefont {Campagne-Ibarcq}, \citenamefont
  {Six}, \citenamefont {Bretheau}, \citenamefont {Sarlette}, \citenamefont
  {Mirrahimi}, \citenamefont {Rouchon},\ and\ \citenamefont
  {Huard}}]{Campagne-Ibarcq2016}%
  \BibitemOpen
  \bibfield  {author} {\bibinfo {author} {\bibfnamefont {P.}~\bibnamefont
  {Campagne-Ibarcq}}, \bibinfo {author} {\bibfnamefont {P.}~\bibnamefont
  {Six}}, \bibinfo {author} {\bibfnamefont {L.}~\bibnamefont {Bretheau}},
  \bibinfo {author} {\bibfnamefont {A.}~\bibnamefont {Sarlette}}, \bibinfo
  {author} {\bibfnamefont {M.}~\bibnamefont {Mirrahimi}}, \bibinfo {author}
  {\bibfnamefont {P.}~\bibnamefont {Rouchon}}, \ and\ \bibinfo {author}
  {\bibfnamefont {B.}~\bibnamefont {Huard}},\ }\bibfield  {title} {\enquote
  {\bibinfo {title} {Observing quantum state diffusion by heterodyne detection
  of fluorescence},}\ }\href {\doibase 10.1103/PhysRevX.6.011002} {\bibfield
  {journal} {\bibinfo  {journal} {Phys. Rev. X}\ }\textbf {\bibinfo {volume}
  {6}},\ \bibinfo {pages} {011002} (\bibinfo {year}
  {2016}{\natexlab{a}})}\BibitemShut {NoStop}%
\bibitem [{\citenamefont {Naghiloo}\ \emph {et~al.}(2016)\citenamefont
  {Naghiloo}, \citenamefont {Foroozani}, \citenamefont {Tan}, \citenamefont
  {Jadbabaie},\ and\ \citenamefont {Murch}}]{naghiloo2015fluores}%
  \BibitemOpen
  \bibfield  {author} {\bibinfo {author} {\bibfnamefont {M.}~\bibnamefont
  {Naghiloo}}, \bibinfo {author} {\bibfnamefont {N.}~\bibnamefont {Foroozani}},
  \bibinfo {author} {\bibfnamefont {D.}~\bibnamefont {Tan}}, \bibinfo {author}
  {\bibfnamefont {A.}~\bibnamefont {Jadbabaie}}, \ and\ \bibinfo {author}
  {\bibfnamefont {K.~W.}\ \bibnamefont {Murch}},\ }\bibfield  {title} {\enquote
  {\bibinfo {title} {Mapping quantum state dynamics in spontaneous emission},}\
  }\href {https://www.nature.com/articles/ncomms11527} {\bibfield  {journal}
  {\bibinfo  {journal} {Nature Communications}\ }\textbf {\bibinfo {volume}
  {7}},\ \bibinfo {pages} {11527} (\bibinfo {year} {2016})}\BibitemShut
  {NoStop}%
\bibitem [{\citenamefont {Naghiloo}\ \emph {et~al.}(2017)\citenamefont
  {Naghiloo}, \citenamefont {Tan}, \citenamefont {Harrington}, \citenamefont
  {Lewalle}, \citenamefont {Jordan},\ and\ \citenamefont {Murch}}]{Mahdi2016}%
  \BibitemOpen
  \bibfield  {author} {\bibinfo {author} {\bibfnamefont {M.}~\bibnamefont
  {Naghiloo}}, \bibinfo {author} {\bibfnamefont {D.}~\bibnamefont {Tan}},
  \bibinfo {author} {\bibfnamefont {P.~M.}\ \bibnamefont {Harrington}},
  \bibinfo {author} {\bibfnamefont {P.}~\bibnamefont {Lewalle}}, \bibinfo
  {author} {\bibfnamefont {A.~N.}\ \bibnamefont {Jordan}}, \ and\ \bibinfo
  {author} {\bibfnamefont {K.~W.}\ \bibnamefont {Murch}},\ }\bibfield  {title}
  {\enquote {\bibinfo {title} {Quantum caustics in resonance-fluorescence
  trajectories},}\ }\href {\doibase 10.1103/PhysRevA.96.053807} {\bibfield
  {journal} {\bibinfo  {journal} {Phys. Rev. A}\ }\textbf {\bibinfo {volume}
  {96}},\ \bibinfo {pages} {053807} (\bibinfo {year} {2017})}\BibitemShut
  {NoStop}%
\bibitem [{\citenamefont {Tan}\ \emph {et~al.}(2017)\citenamefont {Tan},
  \citenamefont {Foroozani}, \citenamefont {Naghiloo}, \citenamefont
  {Kiilerich}, \citenamefont {M\o{}lmer},\ and\ \citenamefont
  {Murch}}]{Tan2017}%
  \BibitemOpen
  \bibfield  {author} {\bibinfo {author} {\bibfnamefont {D.}~\bibnamefont
  {Tan}}, \bibinfo {author} {\bibfnamefont {N.}~\bibnamefont {Foroozani}},
  \bibinfo {author} {\bibfnamefont {M.}~\bibnamefont {Naghiloo}}, \bibinfo
  {author} {\bibfnamefont {A.~H.}\ \bibnamefont {Kiilerich}}, \bibinfo {author}
  {\bibfnamefont {K.}~\bibnamefont {M\o{}lmer}}, \ and\ \bibinfo {author}
  {\bibfnamefont {K.~W.}\ \bibnamefont {Murch}},\ }\bibfield  {title} {\enquote
  {\bibinfo {title} {Homodyne monitoring of postselected decay},}\ }\href
  {\doibase 10.1103/PhysRevA.96.022104} {\bibfield  {journal} {\bibinfo
  {journal} {Phys. Rev. A}\ }\textbf {\bibinfo {volume} {96}},\ \bibinfo
  {pages} {022104} (\bibinfo {year} {2017})}\BibitemShut {NoStop}%
\bibitem [{\citenamefont {Ficheux}\ \emph {et~al.}(2018)\citenamefont
  {Ficheux}, \citenamefont {Jezouin}, \citenamefont {Leghtas},\ and\
  \citenamefont {Huard}}]{Ficheux2018}%
  \BibitemOpen
  \bibfield  {author} {\bibinfo {author} {\bibfnamefont {Q.}~\bibnamefont
  {Ficheux}}, \bibinfo {author} {\bibfnamefont {S.}~\bibnamefont {Jezouin}},
  \bibinfo {author} {\bibfnamefont {Z.}~\bibnamefont {Leghtas}}, \ and\
  \bibinfo {author} {\bibfnamefont {B.}~\bibnamefont {Huard}},\ }\bibfield
  {title} {\enquote {\bibinfo {title} {Dynamics of a qubit while simultaneously
  monitoring its relaxation and dephasing},}\ }\href
  {https://www.nature.com/articles/s41467-018-04372-9} {\bibfield  {journal}
  {\bibinfo  {journal} {Nat. Comm.}\ }\textbf {\bibinfo {volume} {9}},\
  \bibinfo {pages} {1926} (\bibinfo {year} {2018})}\BibitemShut {NoStop}%
\bibitem [{\citenamefont {Campagne-Ibarcq}\ \emph
  {et~al.}(2016{\natexlab{b}})\citenamefont {Campagne-Ibarcq}, \citenamefont
  {Jezouin}, \citenamefont {Cottet}, \citenamefont {Six}, \citenamefont
  {Bretheau}, \citenamefont {Mallet}, \citenamefont {Sarlette}, \citenamefont
  {Rouchon},\ and\ \citenamefont {Huard}}]{PCI-2016-2}%
  \BibitemOpen
  \bibfield  {author} {\bibinfo {author} {\bibfnamefont {P.}~\bibnamefont
  {Campagne-Ibarcq}}, \bibinfo {author} {\bibfnamefont {S.}~\bibnamefont
  {Jezouin}}, \bibinfo {author} {\bibfnamefont {N.}~\bibnamefont {Cottet}},
  \bibinfo {author} {\bibfnamefont {P.}~\bibnamefont {Six}}, \bibinfo {author}
  {\bibfnamefont {L.}~\bibnamefont {Bretheau}}, \bibinfo {author}
  {\bibfnamefont {F.}~\bibnamefont {Mallet}}, \bibinfo {author} {\bibfnamefont
  {A.}~\bibnamefont {Sarlette}}, \bibinfo {author} {\bibfnamefont
  {P.}~\bibnamefont {Rouchon}}, \ and\ \bibinfo {author} {\bibfnamefont
  {B.}~\bibnamefont {Huard}},\ }\bibfield  {title} {\enquote {\bibinfo {title}
  {Using spontaneous emission of a qubit as a resource for feedback control},}\
  }\href {\doibase 10.1103/PhysRevLett.117.060502} {\bibfield  {journal}
  {\bibinfo  {journal} {Phys. Rev. Lett.}\ }\textbf {\bibinfo {volume} {117}},\
  \bibinfo {pages} {060502} (\bibinfo {year} {2016}{\natexlab{b}})}\BibitemShut
  {NoStop}%
\bibitem [{\citenamefont {Naghiloo}\ \emph {et~al.}(2020)\citenamefont
  {Naghiloo}, \citenamefont {Tan}, \citenamefont {Harrington}, \citenamefont
  {Alonso}, \citenamefont {Lutz}, \citenamefont {Romito},\ and\ \citenamefont
  {Murch}}]{Mahdi2017Qtherm}%
  \BibitemOpen
  \bibfield  {author} {\bibinfo {author} {\bibfnamefont {M.}~\bibnamefont
  {Naghiloo}}, \bibinfo {author} {\bibfnamefont {D.}~\bibnamefont {Tan}},
  \bibinfo {author} {\bibfnamefont {P.~M.}\ \bibnamefont {Harrington}},
  \bibinfo {author} {\bibfnamefont {J.~J.}\ \bibnamefont {Alonso}}, \bibinfo
  {author} {\bibfnamefont {E.}~\bibnamefont {Lutz}}, \bibinfo {author}
  {\bibfnamefont {A.}~\bibnamefont {Romito}}, \ and\ \bibinfo {author}
  {\bibfnamefont {K.~W.}\ \bibnamefont {Murch}},\ }\bibfield  {title} {\enquote
  {\bibinfo {title} {Heat and work along individual trajectories of a quantum
  bit},}\ }\href {\doibase 10.1103/PhysRevLett.124.110604} {\bibfield
  {journal} {\bibinfo  {journal} {Phys. Rev. Lett.}\ }\textbf {\bibinfo
  {volume} {124}},\ \bibinfo {pages} {110604} (\bibinfo {year}
  {2020})}\BibitemShut {NoStop}%
\bibitem [{\citenamefont {Plenio}\ and\ \citenamefont
  {Shashank}(2007)}]{Plenio07}%
  \BibitemOpen
  \bibfield  {author} {\bibinfo {author} {\bibfnamefont {Martin~B.}\
  \bibnamefont {Plenio}}\ and\ \bibinfo {author} {\bibfnamefont {Virmani}\
  \bibnamefont {Shashank}},\ }\bibfield  {title} {\enquote {\bibinfo {title}
  {{An introduction to entanglement measures}},}\ }\href
  {https://dl.acm.org/doi/10.5555/2011706.2011707} {\bibfield  {journal}
  {\bibinfo  {journal} {Quantum Inf. Comput.}\ } (\bibinfo {year}
  {2007})}\BibitemShut {NoStop}%
\bibitem [{\citenamefont {Horodecki}\ \emph {et~al.}(2009)\citenamefont
  {Horodecki}, \citenamefont {Horodecki}, \citenamefont {Horodecki},\ and\
  \citenamefont {Horodecki}}]{Horodecki09}%
  \BibitemOpen
  \bibfield  {author} {\bibinfo {author} {\bibfnamefont {Ryszard}\ \bibnamefont
  {Horodecki}}, \bibinfo {author} {\bibfnamefont {Pawe{\l}}\ \bibnamefont
  {Horodecki}}, \bibinfo {author} {\bibfnamefont {Micha{\l}}\ \bibnamefont
  {Horodecki}}, \ and\ \bibinfo {author} {\bibfnamefont {Karol}\ \bibnamefont
  {Horodecki}},\ }\bibfield  {title} {\enquote {\bibinfo {title} {{Quantum
  entanglement}},}\ }\href {\doibase 10.1103/RevModPhys.81.865} {\bibfield
  {journal} {\bibinfo  {journal} {Rev. Mod. Phys.}\ }\textbf {\bibinfo {volume}
  {81}},\ \bibinfo {pages} {865--942} (\bibinfo {year} {2009})}\BibitemShut
  {NoStop}%
\bibitem [{\citenamefont {Hahn}(1950)}]{HahnSpinEcho}%
  \BibitemOpen
  \bibfield  {author} {\bibinfo {author} {\bibfnamefont {E.~L.}\ \bibnamefont
  {Hahn}},\ }\bibfield  {title} {\enquote {\bibinfo {title} {Spin echoes},}\
  }\href {\doibase 10.1103/PhysRev.80.580} {\bibfield  {journal} {\bibinfo
  {journal} {Phys. Rev.}\ }\textbf {\bibinfo {volume} {80}},\ \bibinfo {pages}
  {580--594} (\bibinfo {year} {1950})}\BibitemShut {NoStop}%
\bibitem [{\citenamefont {Viola}\ and\ \citenamefont
  {Lloyd}(1998)}]{ViolaBangBang}%
  \BibitemOpen
  \bibfield  {author} {\bibinfo {author} {\bibfnamefont {Lorenza}\ \bibnamefont
  {Viola}}\ and\ \bibinfo {author} {\bibfnamefont {Seth}\ \bibnamefont
  {Lloyd}},\ }\bibfield  {title} {\enquote {\bibinfo {title} {Dynamical
  suppression of decoherence in two-state quantum systems},}\ }\href {\doibase
  10.1103/PhysRevA.58.2733} {\bibfield  {journal} {\bibinfo  {journal} {Phys.
  Rev. A}\ }\textbf {\bibinfo {volume} {58}},\ \bibinfo {pages} {2733--2744}
  (\bibinfo {year} {1998})}\BibitemShut {NoStop}%
\bibitem [{\citenamefont {Viola}\ \emph
  {et~al.}(1999{\natexlab{a}})\citenamefont {Viola}, \citenamefont {Knill},\
  and\ \citenamefont {Lloyd}}]{ViolaDecoupling}%
  \BibitemOpen
  \bibfield  {author} {\bibinfo {author} {\bibfnamefont {Lorenza}\ \bibnamefont
  {Viola}}, \bibinfo {author} {\bibfnamefont {Emanuel}\ \bibnamefont {Knill}},
  \ and\ \bibinfo {author} {\bibfnamefont {Seth}\ \bibnamefont {Lloyd}},\
  }\bibfield  {title} {\enquote {\bibinfo {title} {Dynamical decoupling of open
  quantum systems},}\ }\href {\doibase 10.1103/PhysRevLett.82.2417} {\bibfield
  {journal} {\bibinfo  {journal} {Phys. Rev. Lett.}\ }\textbf {\bibinfo
  {volume} {82}},\ \bibinfo {pages} {2417--2421} (\bibinfo {year}
  {1999}{\natexlab{a}})}\BibitemShut {NoStop}%
\bibitem [{\citenamefont {Viola}\ \emph
  {et~al.}(1999{\natexlab{b}})\citenamefont {Viola}, \citenamefont {Lloyd},\
  and\ \citenamefont {Knill}}]{Viola1999}%
  \BibitemOpen
  \bibfield  {author} {\bibinfo {author} {\bibfnamefont {Lorenza}\ \bibnamefont
  {Viola}}, \bibinfo {author} {\bibfnamefont {Seth}\ \bibnamefont {Lloyd}}, \
  and\ \bibinfo {author} {\bibfnamefont {Emanuel}\ \bibnamefont {Knill}},\
  }\bibfield  {title} {\enquote {\bibinfo {title} {Universal control of
  decoupled quantum systems},}\ }\href {\doibase 10.1103/PhysRevLett.83.4888}
  {\bibfield  {journal} {\bibinfo  {journal} {Phys. Rev. Lett.}\ }\textbf
  {\bibinfo {volume} {83}},\ \bibinfo {pages} {4888--4891} (\bibinfo {year}
  {1999}{\natexlab{b}})}\BibitemShut {NoStop}%
\bibitem [{\citenamefont {Viola}\ \emph {et~al.}(2000)\citenamefont {Viola},
  \citenamefont {Knill},\ and\ \citenamefont {Lloyd}}]{Viola2000}%
  \BibitemOpen
  \bibfield  {author} {\bibinfo {author} {\bibfnamefont {Lorenza}\ \bibnamefont
  {Viola}}, \bibinfo {author} {\bibfnamefont {Emanuel}\ \bibnamefont {Knill}},
  \ and\ \bibinfo {author} {\bibfnamefont {Seth}\ \bibnamefont {Lloyd}},\
  }\bibfield  {title} {\enquote {\bibinfo {title} {Dynamical generation of
  noiseless quantum subsystems},}\ }\href {\doibase
  10.1103/PhysRevLett.85.3520} {\bibfield  {journal} {\bibinfo  {journal}
  {Phys. Rev. Lett.}\ }\textbf {\bibinfo {volume} {85}},\ \bibinfo {pages}
  {3520--3523} (\bibinfo {year} {2000})}\BibitemShut {NoStop}%
\bibitem [{\citenamefont {Byrd}\ and\ \citenamefont {Lidar}(2002)}]{Byrd2002}%
  \BibitemOpen
  \bibfield  {author} {\bibinfo {author} {\bibfnamefont {Mark~S.}\ \bibnamefont
  {Byrd}}\ and\ \bibinfo {author} {\bibfnamefont {Daniel~A.}\ \bibnamefont
  {Lidar}},\ }\bibfield  {title} {\enquote {\bibinfo {title} {Comprehensive
  encoding and decoupling solution to problems of decoherence and design in
  solid-state quantum computing},}\ }\href {\doibase
  10.1103/PhysRevLett.89.047901} {\bibfield  {journal} {\bibinfo  {journal}
  {Phys. Rev. Lett.}\ }\textbf {\bibinfo {volume} {89}},\ \bibinfo {pages}
  {047901} (\bibinfo {year} {2002})}\BibitemShut {NoStop}%
\bibitem [{\citenamefont {Viola}\ and\ \citenamefont
  {Knill}(2003)}]{Viola2003}%
  \BibitemOpen
  \bibfield  {author} {\bibinfo {author} {\bibfnamefont {Lorenza}\ \bibnamefont
  {Viola}}\ and\ \bibinfo {author} {\bibfnamefont {Emanuel}\ \bibnamefont
  {Knill}},\ }\bibfield  {title} {\enquote {\bibinfo {title} {Robust dynamical
  decoupling of quantum systems with bounded controls},}\ }\href {\doibase
  10.1103/PhysRevLett.90.037901} {\bibfield  {journal} {\bibinfo  {journal}
  {Phys. Rev. Lett.}\ }\textbf {\bibinfo {volume} {90}},\ \bibinfo {pages}
  {037901} (\bibinfo {year} {2003})}\BibitemShut {NoStop}%
\bibitem [{\citenamefont {Byrd}\ and\ \citenamefont {Lidar}(2003)}]{Byrd2003}%
  \BibitemOpen
  \bibfield  {author} {\bibinfo {author} {\bibfnamefont {Mark~S.}\ \bibnamefont
  {Byrd}}\ and\ \bibinfo {author} {\bibfnamefont {Daniel~A.}\ \bibnamefont
  {Lidar}},\ }\bibfield  {title} {\enquote {\bibinfo {title} {Combined error
  correction techniques for quantum computing architectures},}\ }\href
  {\doibase 10.1080/09500340308235203} {\bibfield  {journal} {\bibinfo
  {journal} {Journal of Modern Optics}\ }\textbf {\bibinfo {volume} {50}},\
  \bibinfo {pages} {1285--1297} (\bibinfo {year} {2003})}\BibitemShut {NoStop}%
\bibitem [{\citenamefont {Viola}(2004)}]{Viola2004}%
  \BibitemOpen
  \bibfield  {author} {\bibinfo {author} {\bibfnamefont {Lorenza}\ \bibnamefont
  {Viola}},\ }\bibfield  {title} {\enquote {\bibinfo {title} {Advances in
  decoherence control},}\ }\href {\doibase 10.1080/09500340408231795}
  {\bibfield  {journal} {\bibinfo  {journal} {Journal of Modern Optics}\
  }\textbf {\bibinfo {volume} {51}},\ \bibinfo {pages} {2357--2367} (\bibinfo
  {year} {2004})}\BibitemShut {NoStop}%
\bibitem [{\citenamefont {Byrd}\ \emph {et~al.}(2004)\citenamefont {Byrd},
  \citenamefont {Wu},\ and\ \citenamefont {Lidar}}]{Byrd2004}%
  \BibitemOpen
  \bibfield  {author} {\bibinfo {author} {\bibfnamefont {Mark~S.}\ \bibnamefont
  {Byrd}}, \bibinfo {author} {\bibfnamefont {Lian-Ao}\ \bibnamefont {Wu}}, \
  and\ \bibinfo {author} {\bibfnamefont {Daniel~A.}\ \bibnamefont {Lidar}},\
  }\bibfield  {title} {\enquote {\bibinfo {title} {Overview of quantum error
  prevention and leakage elimination},}\ }\href {\doibase
  10.1080/09500340408231803} {\bibfield  {journal} {\bibinfo  {journal}
  {Journal of Modern Optics}\ }\textbf {\bibinfo {volume} {51}},\ \bibinfo
  {pages} {2449--2460} (\bibinfo {year} {2004})}\BibitemShut {NoStop}%
\bibitem [{\citenamefont {Facchi}\ \emph {et~al.}(2004)\citenamefont {Facchi},
  \citenamefont {Lidar},\ and\ \citenamefont {Pascazio}}]{Facchi2004}%
  \BibitemOpen
  \bibfield  {author} {\bibinfo {author} {\bibfnamefont {P.}~\bibnamefont
  {Facchi}}, \bibinfo {author} {\bibfnamefont {D.~A.}\ \bibnamefont {Lidar}}, \
  and\ \bibinfo {author} {\bibfnamefont {S.}~\bibnamefont {Pascazio}},\
  }\bibfield  {title} {\enquote {\bibinfo {title} {{Unification of dynamical
  decoupling and the quantum Zeno effect}},}\ }\href {\doibase
  10.1103/PhysRevA.69.032314} {\bibfield  {journal} {\bibinfo  {journal} {Phys.
  Rev. A}\ }\textbf {\bibinfo {volume} {69}},\ \bibinfo {pages} {032314}
  (\bibinfo {year} {2004})}\BibitemShut {NoStop}%
\bibitem [{\citenamefont {Facchi}\ \emph {et~al.}(2005)\citenamefont {Facchi},
  \citenamefont {Tasaki}, \citenamefont {Pascazio}, \citenamefont {Nakazato},
  \citenamefont {Tokuse},\ and\ \citenamefont {Lidar}}]{Facchi2005}%
  \BibitemOpen
  \bibfield  {author} {\bibinfo {author} {\bibfnamefont {P.}~\bibnamefont
  {Facchi}}, \bibinfo {author} {\bibfnamefont {S.}~\bibnamefont {Tasaki}},
  \bibinfo {author} {\bibfnamefont {S.}~\bibnamefont {Pascazio}}, \bibinfo
  {author} {\bibfnamefont {H.}~\bibnamefont {Nakazato}}, \bibinfo {author}
  {\bibfnamefont {A.}~\bibnamefont {Tokuse}}, \ and\ \bibinfo {author}
  {\bibfnamefont {D.~A.}\ \bibnamefont {Lidar}},\ }\bibfield  {title} {\enquote
  {\bibinfo {title} {Control of decoherence: Analysis and comparison of three
  different strategies},}\ }\href {\doibase 10.1103/PhysRevA.71.022302}
  {\bibfield  {journal} {\bibinfo  {journal} {Phys. Rev. A}\ }\textbf {\bibinfo
  {volume} {71}},\ \bibinfo {pages} {022302} (\bibinfo {year}
  {2005})}\BibitemShut {NoStop}%
\bibitem [{\citenamefont {Khodjasteh}\ and\ \citenamefont
  {Lidar}(2005)}]{Khodjasteh2005}%
  \BibitemOpen
  \bibfield  {author} {\bibinfo {author} {\bibfnamefont {K.}~\bibnamefont
  {Khodjasteh}}\ and\ \bibinfo {author} {\bibfnamefont {D.~A.}\ \bibnamefont
  {Lidar}},\ }\bibfield  {title} {\enquote {\bibinfo {title} {Fault-tolerant
  quantum dynamical decoupling},}\ }\href {\doibase
  10.1103/PhysRevLett.95.180501} {\bibfield  {journal} {\bibinfo  {journal}
  {Phys. Rev. Lett.}\ }\textbf {\bibinfo {volume} {95}},\ \bibinfo {pages}
  {180501} (\bibinfo {year} {2005})}\BibitemShut {NoStop}%
\bibitem [{\citenamefont {Morton}\ \emph {et~al.}(2006)\citenamefont {Morton},
  \citenamefont {Tyryshkin}, \citenamefont {Ardavan}, \citenamefont {Benjamin},
  \citenamefont {Porfyrakis}, \citenamefont {Lyon},\ and\ \citenamefont
  {Briggs}}]{Morton2006}%
  \BibitemOpen
  \bibfield  {author} {\bibinfo {author} {\bibfnamefont {John J.~L.}\
  \bibnamefont {Morton}}, \bibinfo {author} {\bibfnamefont {Alexei~M.}\
  \bibnamefont {Tyryshkin}}, \bibinfo {author} {\bibfnamefont {Arzhang}\
  \bibnamefont {Ardavan}}, \bibinfo {author} {\bibfnamefont {Simon~C.}\
  \bibnamefont {Benjamin}}, \bibinfo {author} {\bibfnamefont {Kyriakos}\
  \bibnamefont {Porfyrakis}}, \bibinfo {author} {\bibfnamefont {S.~A.}\
  \bibnamefont {Lyon}}, \ and\ \bibinfo {author} {\bibfnamefont {G.~Andrew~D.}\
  \bibnamefont {Briggs}},\ }\bibfield  {title} {\enquote {\bibinfo {title}
  {Bang–bang control of fullerene qubits using ultrafast phase gates},}\
  }\href {https://www.nature.com/articles/nphys192} {\bibfield  {journal}
  {\bibinfo  {journal} {Nature Physics}\ }\textbf {\bibinfo {volume} {2}},\
  \bibinfo {pages} {40} (\bibinfo {year} {2006})}\BibitemShut {NoStop}%
\bibitem [{\citenamefont {Pryadko}\ and\ \citenamefont
  {Quiroz}(2009)}]{Pryadko2009}%
  \BibitemOpen
  \bibfield  {author} {\bibinfo {author} {\bibfnamefont {Leonid~P.}\
  \bibnamefont {Pryadko}}\ and\ \bibinfo {author} {\bibfnamefont {Gregory}\
  \bibnamefont {Quiroz}},\ }\bibfield  {title} {\enquote {\bibinfo {title}
  {{Soft-pulse dynamical decoupling with Markovian decoherence}},}\ }\href
  {\doibase 10.1103/PhysRevA.80.042317} {\bibfield  {journal} {\bibinfo
  {journal} {Phys. Rev. A}\ }\textbf {\bibinfo {volume} {80}},\ \bibinfo
  {pages} {042317} (\bibinfo {year} {2009})}\BibitemShut {NoStop}%
\bibitem [{\citenamefont {Damodarakurup}\ \emph {et~al.}(2009)\citenamefont
  {Damodarakurup}, \citenamefont {Lucamarini}, \citenamefont {Di~Giuseppe},
  \citenamefont {Vitali},\ and\ \citenamefont {Tombesi}}]{Damodarakurup2009}%
  \BibitemOpen
  \bibfield  {author} {\bibinfo {author} {\bibfnamefont {S.}~\bibnamefont
  {Damodarakurup}}, \bibinfo {author} {\bibfnamefont {M.}~\bibnamefont
  {Lucamarini}}, \bibinfo {author} {\bibfnamefont {G.}~\bibnamefont
  {Di~Giuseppe}}, \bibinfo {author} {\bibfnamefont {D.}~\bibnamefont {Vitali}},
  \ and\ \bibinfo {author} {\bibfnamefont {P.}~\bibnamefont {Tombesi}},\
  }\bibfield  {title} {\enquote {\bibinfo {title} {Experimental inhibition of
  decoherence on flying qubits via ``bang-bang'' control},}\ }\href {\doibase
  10.1103/PhysRevLett.103.040502} {\bibfield  {journal} {\bibinfo  {journal}
  {Phys. Rev. Lett.}\ }\textbf {\bibinfo {volume} {103}},\ \bibinfo {pages}
  {040502} (\bibinfo {year} {2009})}\BibitemShut {NoStop}%
\bibitem [{\citenamefont {Wang}\ \emph {et~al.}(2012)\citenamefont {Wang},
  \citenamefont {Zhang}, \citenamefont {Tyryshkin}, \citenamefont {Lyon},
  \citenamefont {Ager}, \citenamefont {Haller},\ and\ \citenamefont
  {Dobrovitski}}]{Wang2012}%
  \BibitemOpen
  \bibfield  {author} {\bibinfo {author} {\bibfnamefont {Zhi-Hui}\ \bibnamefont
  {Wang}}, \bibinfo {author} {\bibfnamefont {Wenxian}\ \bibnamefont {Zhang}},
  \bibinfo {author} {\bibfnamefont {A.~M.}\ \bibnamefont {Tyryshkin}}, \bibinfo
  {author} {\bibfnamefont {S.~A.}\ \bibnamefont {Lyon}}, \bibinfo {author}
  {\bibfnamefont {J.~W.}\ \bibnamefont {Ager}}, \bibinfo {author}
  {\bibfnamefont {E.~E.}\ \bibnamefont {Haller}}, \ and\ \bibinfo {author}
  {\bibfnamefont {V.~V.}\ \bibnamefont {Dobrovitski}},\ }\bibfield  {title}
  {\enquote {\bibinfo {title} {Effect of pulse error accumulation on dynamical
  decoupling of the electron spins of phosphorus donors in silicon},}\ }\href
  {\doibase 10.1103/PhysRevB.85.085206} {\bibfield  {journal} {\bibinfo
  {journal} {Phys. Rev. B}\ }\textbf {\bibinfo {volume} {85}},\ \bibinfo
  {pages} {085206} (\bibinfo {year} {2012})}\BibitemShut {NoStop}%
\bibitem [{\citenamefont {Xu}\ and\ \citenamefont {bo~Xu}(2012)}]{Xu2012}%
  \BibitemOpen
  \bibfield  {author} {\bibinfo {author} {\bibfnamefont {Hang-Shi}\
  \bibnamefont {Xu}}\ and\ \bibinfo {author} {\bibfnamefont {Jing}\
  \bibnamefont {bo~Xu}},\ }\bibfield  {title} {\enquote {\bibinfo {title}
  {{Protecting quantum correlations of two qubits in independent non-Markovian
  environments by bang-bang pulses}},}\ }\href {\doibase
  10.1364/JOSAB.29.002074} {\bibfield  {journal} {\bibinfo  {journal} {J. Opt.
  Soc. Am. B}\ }\textbf {\bibinfo {volume} {29}},\ \bibinfo {pages}
  {2074--2079} (\bibinfo {year} {2012})}\BibitemShut {NoStop}%
\bibitem [{\citenamefont {Bhole}\ \emph {et~al.}(2016)\citenamefont {Bhole},
  \citenamefont {Anjusha},\ and\ \citenamefont {Mahesh}}]{Bhole2016}%
  \BibitemOpen
  \bibfield  {author} {\bibinfo {author} {\bibfnamefont {Gaurav}\ \bibnamefont
  {Bhole}}, \bibinfo {author} {\bibfnamefont {V.~S.}\ \bibnamefont {Anjusha}},
  \ and\ \bibinfo {author} {\bibfnamefont {T.~S.}\ \bibnamefont {Mahesh}},\
  }\bibfield  {title} {\enquote {\bibinfo {title} {Steering quantum dynamics
  via bang-bang control: Implementing optimal fixed-point quantum search
  algorithm},}\ }\href {\doibase 10.1103/PhysRevA.93.042339} {\bibfield
  {journal} {\bibinfo  {journal} {Phys. Rev. A}\ }\textbf {\bibinfo {volume}
  {93}},\ \bibinfo {pages} {042339} (\bibinfo {year} {2016})}\BibitemShut
  {NoStop}%
\bibitem [{\citenamefont {Ticozzi}\ and\ \citenamefont
  {Viola}(2006)}]{Ticozzi2006}%
  \BibitemOpen
  \bibfield  {author} {\bibinfo {author} {\bibfnamefont {Francesco}\
  \bibnamefont {Ticozzi}}\ and\ \bibinfo {author} {\bibfnamefont {Lorenza}\
  \bibnamefont {Viola}},\ }\bibfield  {title} {\enquote {\bibinfo {title}
  {Single-bit feedback and quantum-dynamical decoupling},}\ }\href {\doibase
  10.1103/PhysRevA.74.052328} {\bibfield  {journal} {\bibinfo  {journal} {Phys.
  Rev. A}\ }\textbf {\bibinfo {volume} {74}},\ \bibinfo {pages} {052328}
  (\bibinfo {year} {2006})}\BibitemShut {NoStop}%
\bibitem [{\citenamefont {Gong}\ and\ \citenamefont {Yao}(2013)}]{Gong2013}%
  \BibitemOpen
  \bibfield  {author} {\bibinfo {author} {\bibfnamefont {Z.~R.}\ \bibnamefont
  {Gong}}\ and\ \bibinfo {author} {\bibfnamefont {Wang}\ \bibnamefont {Yao}},\
  }\bibfield  {title} {\enquote {\bibinfo {title} {Protecting dissipative
  quantum state preparation via dynamical decoupling},}\ }\href {\doibase
  10.1103/PhysRevA.87.032314} {\bibfield  {journal} {\bibinfo  {journal} {Phys.
  Rev. A}\ }\textbf {\bibinfo {volume} {87}},\ \bibinfo {pages} {032314}
  (\bibinfo {year} {2013})}\BibitemShut {NoStop}%
\bibitem [{\citenamefont {Sun}\ \emph {et~al.}(2010)\citenamefont {Sun},
  \citenamefont {Al-Amri}, \citenamefont {Davidovich},\ and\ \citenamefont
  {Zubairy}}]{Sun2010}%
  \BibitemOpen
  \bibfield  {author} {\bibinfo {author} {\bibfnamefont {Qingqing}\
  \bibnamefont {Sun}}, \bibinfo {author} {\bibfnamefont {M.}~\bibnamefont
  {Al-Amri}}, \bibinfo {author} {\bibfnamefont {Luiz}\ \bibnamefont
  {Davidovich}}, \ and\ \bibinfo {author} {\bibfnamefont {M.~Suhail}\
  \bibnamefont {Zubairy}},\ }\bibfield  {title} {\enquote {\bibinfo {title}
  {Reversing entanglement change by a weak measurement},}\ }\href {\doibase
  10.1103/PhysRevA.82.052323} {\bibfield  {journal} {\bibinfo  {journal} {Phys.
  Rev. A}\ }\textbf {\bibinfo {volume} {82}},\ \bibinfo {pages} {052323}
  (\bibinfo {year} {2010})}\BibitemShut {NoStop}%
\bibitem [{\citenamefont {Korotkov}\ and\ \citenamefont
  {Jordan}(2006)}]{Jordan2006Reversal}%
  \BibitemOpen
  \bibfield  {author} {\bibinfo {author} {\bibfnamefont {Alexander~N.}\
  \bibnamefont {Korotkov}}\ and\ \bibinfo {author} {\bibfnamefont {Andrew~N.}\
  \bibnamefont {Jordan}},\ }\bibfield  {title} {\enquote {\bibinfo {title}
  {Undoing a weak quantum measurement of a solid-state qubit},}\ }\href
  {\doibase 10.1103/PhysRevLett.97.166805} {\bibfield  {journal} {\bibinfo
  {journal} {Phys. Rev. Lett.}\ }\textbf {\bibinfo {volume} {97}},\ \bibinfo
  {pages} {166805} (\bibinfo {year} {2006})}\BibitemShut {NoStop}%
\bibitem [{\citenamefont {Jordan}\ and\ \citenamefont
  {Korotkov}(2010)}]{JordanCPReversal}%
  \BibitemOpen
  \bibfield  {author} {\bibinfo {author} {\bibfnamefont {Andrew~N.}\
  \bibnamefont {Jordan}}\ and\ \bibinfo {author} {\bibfnamefont {Alexander~N.}\
  \bibnamefont {Korotkov}},\ }\bibfield  {title} {\enquote {\bibinfo {title}
  {Uncollapsing the wavefunction by undoing quantum measurements},}\ }\href
  {\doibase 10.1080/00107510903385292} {\bibfield  {journal} {\bibinfo
  {journal} {Contemporary Physics}\ }\textbf {\bibinfo {volume} {51}},\
  \bibinfo {pages} {125--147} (\bibinfo {year} {2010})}\BibitemShut {NoStop}%
\bibitem [{\citenamefont {Katz}\ \emph {et~al.}(2008)\citenamefont {Katz},
  \citenamefont {Neeley}, \citenamefont {Ansmann}, \citenamefont {Bialczak},
  \citenamefont {Hofheinz}, \citenamefont {Lucero}, \citenamefont {O'Connell},
  \citenamefont {Wang}, \citenamefont {Cleland}, \citenamefont {Martinis},\
  and\ \citenamefont {Korotkov}}]{Katz2008}%
  \BibitemOpen
  \bibfield  {author} {\bibinfo {author} {\bibfnamefont {Nadav}\ \bibnamefont
  {Katz}}, \bibinfo {author} {\bibfnamefont {Matthew}\ \bibnamefont {Neeley}},
  \bibinfo {author} {\bibfnamefont {M.}~\bibnamefont {Ansmann}}, \bibinfo
  {author} {\bibfnamefont {Radoslaw~C.}\ \bibnamefont {Bialczak}}, \bibinfo
  {author} {\bibfnamefont {M.}~\bibnamefont {Hofheinz}}, \bibinfo {author}
  {\bibfnamefont {Erik}\ \bibnamefont {Lucero}}, \bibinfo {author}
  {\bibfnamefont {A.}~\bibnamefont {O'Connell}}, \bibinfo {author}
  {\bibfnamefont {H.}~\bibnamefont {Wang}}, \bibinfo {author} {\bibfnamefont
  {A.~N.}\ \bibnamefont {Cleland}}, \bibinfo {author} {\bibfnamefont {John~M.}\
  \bibnamefont {Martinis}}, \ and\ \bibinfo {author} {\bibfnamefont
  {Alexander~N.}\ \bibnamefont {Korotkov}},\ }\bibfield  {title} {\enquote
  {\bibinfo {title} {Reversal of the weak measurement of a quantum state in a
  superconducting phase qubit},}\ }\href {\doibase
  10.1103/PhysRevLett.101.200401} {\bibfield  {journal} {\bibinfo  {journal}
  {Phys. Rev. Lett.}\ }\textbf {\bibinfo {volume} {101}},\ \bibinfo {pages}
  {200401} (\bibinfo {year} {2008})}\BibitemShut {NoStop}%
\bibitem [{\citenamefont {Kim}\ \emph {et~al.}(2009)\citenamefont {Kim},
  \citenamefont {Cho}, \citenamefont {Ra},\ and\ \citenamefont
  {Kim}}]{Kim2009}%
  \BibitemOpen
  \bibfield  {author} {\bibinfo {author} {\bibfnamefont {Yong-Su}\ \bibnamefont
  {Kim}}, \bibinfo {author} {\bibfnamefont {Young-Wook}\ \bibnamefont {Cho}},
  \bibinfo {author} {\bibfnamefont {Young-Sik}\ \bibnamefont {Ra}}, \ and\
  \bibinfo {author} {\bibfnamefont {Yoon-Ho}\ \bibnamefont {Kim}},\ }\bibfield
  {title} {\enquote {\bibinfo {title} {Reversing the weak quantum measurement
  for a photonic qubit},}\ }\href {\doibase 10.1364/OE.17.011978} {\bibfield
  {journal} {\bibinfo  {journal} {Opt. Express}\ }\textbf {\bibinfo {volume}
  {17}},\ \bibinfo {pages} {11978--11985} (\bibinfo {year} {2009})}\BibitemShut
  {NoStop}%
\bibitem [{\citenamefont {Korotkov}\ and\ \citenamefont
  {Keane}(2010)}]{Korotkov2010}%
  \BibitemOpen
  \bibfield  {author} {\bibinfo {author} {\bibfnamefont {Alexander~N.}\
  \bibnamefont {Korotkov}}\ and\ \bibinfo {author} {\bibfnamefont {Kyle}\
  \bibnamefont {Keane}},\ }\bibfield  {title} {\enquote {\bibinfo {title}
  {Decoherence suppression by quantum measurement reversal},}\ }\href {\doibase
  10.1103/PhysRevA.81.040103} {\bibfield  {journal} {\bibinfo  {journal} {Phys.
  Rev. A}\ }\textbf {\bibinfo {volume} {81}},\ \bibinfo {pages} {040103(R)}
  (\bibinfo {year} {2010})}\BibitemShut {NoStop}%
\bibitem [{\citenamefont {Kim}\ \emph {et~al.}(2012)\citenamefont {Kim},
  \citenamefont {Lee}, \citenamefont {Kwon},\ and\ \citenamefont
  {Kim}}]{Kim2012}%
  \BibitemOpen
  \bibfield  {author} {\bibinfo {author} {\bibfnamefont {Yong-Su}\ \bibnamefont
  {Kim}}, \bibinfo {author} {\bibfnamefont {Jong-Chan}\ \bibnamefont {Lee}},
  \bibinfo {author} {\bibfnamefont {Osung}\ \bibnamefont {Kwon}}, \ and\
  \bibinfo {author} {\bibfnamefont {Yoon-Ho}\ \bibnamefont {Kim}},\ }\bibfield
  {title} {\enquote {\bibinfo {title} {Protecting entanglement from decoherence
  using weak measurement and quantum measurement reversal},}\ }\href
  {https://www.nature.com/articles/nphys2178} {\bibfield  {journal} {\bibinfo
  {journal} {Nature Physics}\ }\textbf {\bibinfo {volume} {8}},\ \bibinfo
  {pages} {117} (\bibinfo {year} {2012})}\BibitemShut {NoStop}%
\bibitem [{\citenamefont {Korotkov}(2012)}]{Korotkov2012}%
  \BibitemOpen
  \bibfield  {author} {\bibinfo {author} {\bibfnamefont {Alexander~N.}\
  \bibnamefont {Korotkov}},\ }\bibfield  {title} {\enquote {\bibinfo {title}
  {{The Sleeping Beauty approach}},}\ }\href
  {https://www.nature.com/articles/nphys2209} {\bibfield  {journal} {\bibinfo
  {journal} {Nature Physics}\ }\textbf {\bibinfo {volume} {8}},\ \bibinfo
  {pages} {107} (\bibinfo {year} {2012})}\BibitemShut {NoStop}%
\bibitem [{\citenamefont {Ott}(2002)}]{BookOtt}%
  \BibitemOpen
  \bibfield  {author} {\bibinfo {author} {\bibfnamefont {E.}~\bibnamefont
  {Ott}},\ }\href@noop {} {\emph {\bibinfo {title} {{Chaos in Dynamical
  Systems}}}}\ (\bibinfo  {publisher} {Cambridge University Press},\ \bibinfo
  {address} {Cambridge UK},\ \bibinfo {year} {2002})\BibitemShut {NoStop}%
\bibitem [{\citenamefont {Strogatz}(1994)}]{BookStrogatz}%
  \BibitemOpen
  \bibfield  {author} {\bibinfo {author} {\bibfnamefont {S.~H.}\ \bibnamefont
  {Strogatz}},\ }\href@noop {} {\emph {\bibinfo {title} {{Nonlinear Dynamics
  and Chaos}}}}\ (\bibinfo  {publisher} {Westview Press / Perseus Books,
  Cambridge MA},\ \bibinfo {year} {1994})\BibitemShut {NoStop}%
\end{thebibliography}%
\end{document}